\documentclass[a4paper,11pt]{article}
\usepackage{graphicx}
\usepackage{here}
\usepackage{amsmath,amssymb,amsthm}
\usepackage{mathrsfs}
\usepackage{ascmac}
\usepackage[sort, comma]{natbib}
\usepackage[subrefformat=parens]{subcaption}
\usepackage{url}
\usepackage{tikz}
\usetikzlibrary{intersections,calc}
\usepackage{mathtools}
\usepackage{physics}
\mathtoolsset{showonlyrefs=true}
\usepackage[stable]{footmisc}
\usepackage{colortbl}
\usepackage{hyperref}

\newcommand{\ol}[1]{\overline{#1}}

\title{{\LARGE Pattern formation by advection-diffusion\\ in new economic geography}}
\author{Kensuke Ohtake\thanks{Center for General Education, Shinshu University, Matsumoto, Nagano 390-8621, Japan,
E-mail: k\_ohtake@shinshu-u.ac.jp}}
\date{June 23, 2025}

\begin{document}
\maketitle

\begin{abstract}
This paper studies spatial patterns formed by proximate population migration driven by real wage gradients and other idiosyncratic factors. The model consists of a tractable core-periphery model incorporating a quasi-linear log utility function and an advection-diffusion equation that expresses population migration. It is found that diffusion stabilizes a homogeneous stationary solution when transport costs are sufficiently low, and it also inhibits the monotonic facilitation of agglomeration caused by lower transport costs in some cases. When the homogeneous stationary solution is unstable, numerical simulations show spatial patterns with multiple urban areas. Insights into the relation between agglomeration and control parameters (transport costs and preference for variety of consumers) gained from the large-time behavior of solutions confirm the validity of the analysis of linearized equations.
\end{abstract}

\noindent
{\bf Keywords:\hspace{1mm}}
advection-diffusion;
core-periphery model;
economic agglomeration;
new economic geography;
partial differential equation;
pattern formation


\section{Introduction}

New Economic Geography (NEG) is a paradigm in spatial economics that explains the spatial movement and distribution of firms and consumers using general equilibrium theory based on monopolistic competition. The most critical factors in NEG are increasing return technology of firms, transport costs of goods, and preference for variety of consumers. These three factors work together to generate spatial inequality. Firms are attracted to regions where consumers are agglomerated, i.e., where the market is large, because of increasing returns and transportation costs, and consumers are attracted to regions where many firms are agglomerated, i.e., where a large variety of goods are produced, because of transport costs and preference for variety. \citet{Krug91} has proposed a fundamental general equilibrium model in this field known as the core-periphery (CP) model. A somewhat surprising conclusion of the CP model is that agglomeration, or spatial inequality, occurs when transport costs fall sufficiently. This is because, if transport costs are low enough, agglomeration allows firms to enjoy the benefits of increasing returns without losing demand from non-mobile consumers. Additionally, an enhanced preference for variety draws consumers to agglomerated regions. This is a fundamental property of the CP model. It is carried over to derivative CP models, such as \citet{OttaTabThi}, \citet{ForsOtta}, and \citet{Pfl}, which are more tractable versions of the original model by Krugman.\footnote{Exceptionally, \citet{Ohtake2022agg} has proved that agglomeration occurs for any value of transport costs in the model by \citet{OttaTabThi} when the number of regions is multiples of four.} Since the present study extends and utilizes the model by \citet{Pfl}, the same properties are essentially confirmed in the model proposed in this paper, but there are also some interesting differences.

The CP model is often used in combination with the following replicator equation for population migration dynamics. That is, for a geographic space denoted by $\Omega$, the time evolution of a spatial distribution $\lambda$ of mobile population on $\Omega$ is assumed to be
\begin{equation}\label{repl}
\frac{\partial\lambda}{\partial t}(t,x) = v\left[\omega(t,x)-\tilde{\omega}(t)\right]\lambda(t,x), \hspace{3mm} v > 0
\end{equation}
for time $t\geq 0$ and $x\in \Omega$. Here, $\omega$ and $\tilde{\omega}$ represent real wages and a spatially averaged real wage, respectively. The average real wage is defined as $\tilde{\omega}:=\frac{1}{\Lambda}\int_\Omega\omega(t,x)\lambda(t,x)dx$ where $\Lambda> 0$ is the total number of mobile population. The equation \eqref{repl} states that the population migrates from where real wages are below average to where they are above average. This formulation assumes that the population can perceive the global spatial distribution of real wages and can migrate freely, regardless of distance.

There are cases where this may not be appropriate in the real world. We know that the population often migrates only proximally, and migration would be subject to random influences from individual, accidental, and idiosyncratic factors unrelated to real wages. Based on this idea, \citet{Moss} has proposed a pioneering model in which the migration process of population is expressed by a partial differential equation such as
\begin{equation}\label{MossPDE}
\frac{\partial \lambda}{\partial t}(t,x) + 
\frac{\partial}{\partial x}\left(\lambda(t,x)\frac{\partial \omega}{\partial x}(t,x)\right)
= d\frac{\partial^2 \lambda}{\partial x^2}(t,x), \hspace{3mm} d>0
\end{equation}
in one-dimensional periodic space.\footnote{\citet[p.~430]{Moss}} This type of partial differential equation is commonly referred to as the advection-diffusion equation. Here, the second term on the left-hand side expresses that the population moves locally toward higher real wages. This term is called {\it advection} in physics, and in general, it has an advection coefficient that represents its degree (in the Mossay model, the value is $1$). The right-hand side is called {\it diffusion} in physics, which expresses that the population moves locally from areas of high population to areas of low population. The magnitude of diffusion is controlled by the coefficient $d$, commonly referred to as the diffusion coefficient in physics.\footnote{\citet{Moss} calles it the mobility index.} There are several ways to derive the diffusion term. \citet{Moss} derives the diffusion term from a microscopic perspective in which consumers move randomly according to their idiosyncratic preferences, ignoring real wages.\footnote{Another work that introduces such idiosyncratic preferences of consumers is \citet{AkaTakaIke}.} From a more macroscopic point of view, it is possible to interpret the diffusion term as a result of some congestion costs in populated areas, which causes consumers who dislike them to flee to less populated areas. A macroscopic derivation of the advection-diffusion model, commonly used in physics or mathematical biology, is provided in the Appendix.

This paper aims to contribute to the understanding of the role that advection-diffusion plays in shaping spatial patterns in economic geography. In particular, clarifying the difference between models whose dynamics are given by the replicator equation is essential as a basis for discussing the validity of the NEG model in the real world. However, it is generally challenging to compute time-evolving solutions to the Mossay model. This is because the Mossay model is based on the original CP model; one must solve a nonlinear fixed-point problem for the population distribution $\lambda$ at each time $t\geq 0$ to obtain a market equilibrium solution for computing the real wage $\omega$.\footnote{See \citet{MossMar2004}, \citet{TabaEshiSakaTaka}, and \citet{OhtakeYagi_Asym} for iterative algorithms for solving the fixed-point problem of the CP model.} Even without this problem, the numerical computation for the advection-diffusion process is more intensive than that for the replicator equation because it takes a long time for solutions to reach a stationary state due to diffusion effects. These make it critically difficult to elucidate the large-time behavior of solutions to the Mossay model in detail, even numerically. In this paper, instead of the CP model proposed by \citet{Krug91}, we employ a more computationally tractable model to circumvent difficulties in computing market equilibrium solutions. This allowed us to easily track the large-time behavior of solutions to the NEG model with the advection-diffusion dynamics. The model proposed in this paper enables more comprehensive numerical computations with various parameter values without losing the essential elements of the CP model. \citet{MossMar2004} obtained two agglomerated areas in the non-homogeneous stationary solution. In this study, by varying the value of control parameters, simulation results are obtained in which the number of agglomerated areas is $1$, $2$, $\cdots$, and $6$. It could be easily increased if necessary. The simulation code is available at \url{https://github.com/k-ohtake/advection-diffusion-neg}.

The results of this paper are summarized as follows. First, we decompose the solutions around the homogeneous stationary solution into eigenmodes using a Fourier series and investigate the relationship between the spatial frequencies of the time-growing, and thus unstable, modes and the control parameters. The outcomes of this model generally reproduce those of the standard CP model. That is, the absolute value of the spatial frequency of unstable modes decreases as transport costs decrease and, if transport costs are relatively large, the preference for variety strengthens. Since spatial frequency corresponds to the number of agglomerations that form in the early stages of destabilization, this implies the standard NEG consequence that agglomeration is enhanced as transport costs decrease and preference for variety strengthens. On the other hand, diffusion complicates the relationship between the control parameters and the frequency of unstable modes; i) All modes stabilize when transport costs are sufficiently low, and in some cases, even at intermediate values of transport costs. More interestingly, in some cases, there is even a cascade of instability and stability as transport costs decrease. ii) When preference for variety is sufficiently weak, all modes are stabilized. Second, we numerically compute the large-time behavior of solutions to the model. If the homogeneous stationary solution is unstable, the time-evolving solution eventually converges to a non-homogeneous stationary solution having some urban areas. Although the homogeneous stationary solution can become stable at intermediate values of transport costs, only when it is unstable, the number of urban areas decreases as transport costs decrease, which is a well-known trend in NEG. Where transport costs are somewhat higher, a stronger preference for variety leads to an increase in the number of urban areas. These numerical results show that the analysis of unstable modes, which only captures local solution behavior near the homogeneous stationary solution, is also valid in the large-time behavior of solutions.

This paper clarifies the role of advection-diffusion processes in agglomeration pattern formation by comparing them with the case of replicator equations. First, in the case of the replicator equation, the area where each frequency becomes unstable in the control parameter (transport costs and preference for variety) space expands uniformly with an increase in the absolute value of the frequency, whereas in the case of the advection-diffusion equation, this uniformity breaks down, particularly due to diffusion. In addition, a parameter area appears in which all frequencies stabilize due to diffusion. Second, the shape of the population distribution that is formed asymptotically is spiky, with the population agglomerating at a single point in the case of the replicator equation, whereas in the case of diffusion, the population is dispersed over a wider area, forming a mountainous shape. Considering that actual urban formation is not limited to spikey distributions but can also exhibit mountainous distributions, the results of this study can be interpreted as suggesting the existence of a kind of diffusion process in actual population migration dynamics.

The rest of this paper is organized as follows. Section \ref{sec:relatedworks} reviews related works. Section \ref{sec:model} presents the model we handle. Section \ref{sec:stationarysol} discusses the stability of the homogeneous stationary solution. Section \ref{sec:numerical} gives a numerical scheme for simulations and presents the results. Section \ref{sec:comparison} compares the results of the present study with those of similar models. Section \ref{sec:concdisc} discusses the results of the present study and their significance. Section \ref{sec:app} is Appendix.

\section{Literature review}\label{sec:relatedworks}

This study builds upon the background of numerous previous theoretical studies on the multi-regional CP model. Here, {\it multi-regional} means that it consists of more than three discrete regions or a continuous space. The existing literature on discrete regional models are: \citet{CastCorrMoss}, \citet{IkeAkaKon}, \citet{AkaTaka}, \citet{IkeMuroAkaKoTa}, \citet{TabaEshiSaka2014}, \citet{TabaEshiSaka2015}, \citet{TabaEshiKiyoTaka}, and \citet{TabaEshi2016}. More closely related to the present study is the continuous space model: \citet[Chapter 6]{FujiKrugVenab}, \citet{ChiAsh08}, \citet{TabaEshiSakaTaka}, \citet{TabaEshi_existence}, \citet{TabaEshi_explosion}, \citet{TabaEshi23}, \citet{OhtakeYagi_Asym}, and \citet{OhtakeYagi_point}. These are the literatures that are concerned with the CP model in its most basic settings, and there may be a vast literature on those involving more applied elements.

The tractable model used in this study is based on the following research. \citet{Pfl} has devised a tractable model in which the market equilibrium solution can be obtained analytically by changing the consumer's utility function to a quasi-linear-log type. \citet{AkaTakaIke} and \citet{GasCasCorr} have extended the Pfl{\"u}ger model to a space consisting of multi--egional discrete regions.\footnote{\citet{AkaTakaIke} adopt logit dynamics in evolutionary game theory. Their model includes idiosyncratic preferences for location choice, which would be more similar to diffusion in this study than the replicator dynamics. \citet{GasCasCorr} adopt the replicator dynamics.} Then, \citet{Ohtake2023cont} has extended the Pfl{\"u}ger model over continuous space with the replicator dynamics. The market equilibrium equations adopted in this study are those in \citet{Ohtake2023cont}. Let us refer to these equations as the Quasi-Linear-Log Utility (QLLU) model for convenience. Therefore, the focus of this paper is on a model that combines the QLLU model with advection-diffusion dynamics.

\citet{Moss} is the first to introduce the advection-diffusion equation into the CP model, to the author's knowledge. Then, \citet{MossMar2004} successfully solved the model numerically. Other partial differential equation models in economic geography include \citet{alvarez2006estimation}, \citet{mossay2006stability}, and \citet{picard2010self}. Partial differential equations are often used at the intersection of economic growth theory and economic geography. Here, we mention \citet{CaEnLato2010}, \citet{BoCaFa2013}, and \citet{juchem2015capital}, and \citet{ballestra2024modeling}. Other various partial differential equations in Macroeconomics are reviewed by \citet{achdou2014partial}.

\section{The model}\label{sec:model}
\subsection{Settings}
The economy consists of two sectors: manufacturing and agriculture. As \citet[p.45]{FujiKrugVenab} note, ``manufacturing'' here is just a label for a sector dominated by increasing returns and imperfect competition, while ``agriculture'' is just a label for a sector dominated by constant returns and perfect competition, and they do not necessarily correspond to the sectors in the real world. In manufacturing, one firm produces one variety of an infinite number of varieties of differentiated goods. In agriculture, a single variety of goods is produced. When goods are transported between regions, manufactured goods incur transport costs, whereas agricultural goods do not. For the transport costs, the so-called iceberg transport technology by \citet{Sam1952} is assumed. That is, goods melt away during transportation, depending on the distance they are transported. There are two types of workers: mobile workers, who can migrate through space, and immobile workers, who cannot. The manufacturing sector is modeled using the so-called Footloose Entrepreneur setting by \citet{ForsOtta}, where mobile workers and immobile workers are used as fixed inputs and marginal inputs, respectively. In the agricultural sector, one unit of goods is produced by one unit of immobile workers. In this setting, mobile workers can also be interpreted as entrepreneurs in the manufacturing sector. They have skills specific to the manufacturing sector and migrate in search of higher real wages. On the other hand, (immobile) workers are assumed to be locally rooted and therefore do not migrate.\footnote{The reason for assuming two types of workers is also, technically, to create spatially distributed forces. As \citet[p.~322]{FujiThis} point out, if all production factors were mobile and there were no costs associated with agglomeration, the spatial economy would be dominated solely by agglomeration.}

\subsection{Model equations}
We are concerned with the following system of integral-differential equations
\begin{equation}\label{1}
\left\{
\begin{aligned}
&G(t, x) = \left[\frac{1}{F}\int_{\Omega} \lambda(t, y)T(x, y)^{1-\sigma}dy\right]^{\frac{1}{1-\sigma}},\\
&w(t, x)=\frac{\mu}{\sigma F}\int_{\Omega} \left(\phi(y)+\lambda(t, y)\right)G(t, y)^{\sigma-1}T(x, y)^{1-\sigma}dy,\\ 
&\omega(t, x) = w(t, x)-\mu\ln G(t, x),\\
&\frac{\partial \lambda}{\partial t}(t,x) + 
a\nabla\cdot\left(\lambda(t,x)\nabla\omega(t,x)\right)
= d\Delta\lambda(t,x)
\end{aligned}\right.
\end{equation}
for $(t,x) \in [0,\infty)\times \Omega$ with an initial condition $\lambda(0, x)=\lambda_0(x)$. Here, $\Omega$ is a bounded subset of the $n$-dimensional Euclidean space, where $n\geq 1$. The given function $\phi\geq 0$ which satisfies 
\[
\int_\Omega\phi(x)dx=\Phi>0
\]
denotes the population density of immobile workers. The unknown function $\lambda\geq 0$ which satisfies 
\begin{equation}\label{conlmd}
\int_\Omega\lambda(t,x)dx=\Lambda>0,~\forall t\geq 0
\end{equation}
denotes the population density of mobile workers. The other unknown functions, $w$, $G$, and $\omega$, denote the nominal wage of mobile workers, the price index of manufactured goods, and the real wage of mobile workers, respectively. The iceberg transport technology is represented by the two-variable function $T$, which means that $T(x,y)\geq 1$ units of a variety of manufactured goods must be shipped to deliver one unit of the variety from $x\in\Omega$ to $y\in\Omega$.

The first three equations constitute the QLLU model, and the fourth is the advection-diffusion equation.\footnote{See \citet{Ohtake2023cont} for the derivation of the QLLU model. For the derivation of the advection-diffusion equation, see Subsections \ref{subsec:derivationeqcont} and \ref{subsec:derivationad}.} The second term on the left-hand side of the fourth equation describes local migration along the real wage gradient, i.e., {\it advection}. The coefficient $a\geq 0$ is the advection coefficient; the larger the value of $a$, the greater the degree to which the mobile population migrates in response to a gradient in real wages. The right-hand side of the equation describes the random movement of mobile workers, independent of the gradient or an escape behavior from populated regions that incurs some congestion costs, i.e., {\it diffusion}. The coefficient $d\geq 0$ is the diffusion coefficient; the larger the value of $d$, the greater the degree to which the population moves, ignoring the gradient in real wages. Let us emphasize here the significance of the diffusion term. First, it can be interpreted as mobile workers decide to migrate based on idiosyncratic factors not measured by real wages, such as the characteristics of a particular place or accidental life changes. Second, it can be interpreted as expressing the evacuation behavior of mobile workers to avoid some congestion costs associated with agglomeration, which may include higher land prices, commuting costs, environmental pollution, and so on. Thus, the diffusion term works against the agglomeration formed by the advection term. Suppose the diffusion coefficient is sufficiently large compared to the advection coefficient. In that case, population migration is dominated by either just random movement behavior or anti-agglomeration movement of mobile workers, eventually settling into a spatially uniform state.\footnote{\citet[p.435]{Moss} proves that the uniform state is stable for large values of $d$.} It is, in essence, the same as the phenomenon of randomly flying molecules diffusing uniformly in the atmosphere.

The parameters included in the model are as follows. The degree to which each consumer favors manufactured goods is $\mu\in[0,1)$. The elasticity of substitution between any two varieties of manufactured goods is $\sigma>1$. The closer $\sigma$ is to $1$, the stronger the preference for variety of consumers. According to \citet[p.54]{FujiKrugVenab}, the units of measurement for manufactured output are chosen so that the marginal requirement of immobile workers in manufacturing production is $\frac{\sigma-1}{\sigma}$. The fixed input size in manufacturing production is $F>0$. The strengths of advection and diffusion are represented by the advection coefficient $a\geq 0$ and the diffusion coefficient $d\geq 0$, respectively.

\subsection{Racetrack economy}
In the following, we consider the so-called racetrack economy,\footnote{\citet[Chapter 6]{FujiKrugVenab}} which is, a model on the one-dimensional circumference of radius $\rho>0$ denoted by $S$. For this purpose, we consider the model \eqref{1} with the periodic boundary condition on the interval $[-\rho\pi, \rho\pi]\subset\mathbb{R}$, and a point $x\in S$ is identified with a real number $x\in[-\rho\pi, \rho\pi]$. Therefore, the integration of a periodic function $h$ over $S$ is calculated by
\[
\int_S h(x)dx = \int_{-\rho\pi}^{\rho\pi} h(x)dx.
\]
By transforming the variable by
\begin{equation}\label{xrtheta}
x = \rho r
\end{equation}
with an angle $r\in[-\pi,\pi]$, we see that
\begin{equation}\label{Intrtheta}
\int_S h(x)dx = \int_{-\pi}^\pi h(\rho r)\rho dr. 
\end{equation}
The fourth equation in \eqref{1} is then
\begin{equation}\label{PDErtheta}
\frac{\partial \lambda}{\partial t}(t,\rho r) + 
\frac{a}{\rho^2}\frac{\partial}{\partial r}\left(\lambda(t,\rho r)\frac{\partial \omega}{\partial r}(t,\rho r)\right)
= \frac{d}{\rho^2}\frac{\partial^2 \lambda}{\partial r^2}(t,\rho r).
\end{equation}
We specify a function representing the iceberg transport cost as
\begin{equation}\label{Txyexp}
T(x,y) = e^{\tau D(x,y)}
\end{equation}
with $\tau\geq 0$. Here, $D(x,y)$ stands for the shorter distance between $x\in S$ and $y\in S$ along $S$. If $x=\rho r$ and $y=\rho s$ with $r\in[-\pi,\pi]$ and $s\in[-\pi,\pi]$, then 
\[
D\left(x,y\right) (= D\left(\rho r, \rho s\right)) = \rho\min\left\{|r-s|, 2\pi-|r-s|\right\}.
\] 
Then, it is convenient to define a constant $\alpha$ as
\begin{equation}\label{alpha}
\alpha:=(\sigma-1)\tau.
\end{equation}

Thus, the racetrack economy we consider is 
\begin{equation}\label{racetrack}
\left\{
\begin{aligned}
&G(t, \rho r) = \left[\frac{1}{F}\int_{-\pi}^\pi \lambda(t, \rho s)e^{-\alpha D(\rho r, \rho s)}\rho ds\right]^{\frac{1}{1-\sigma}},\\
&w(t, \rho r)=\frac{\mu}{\sigma F}\int_{-\pi}^\pi \left(\phi(\rho s)+\lambda(t, \rho s)\right)G(t, \rho s)^{\sigma-1}e^{-\alpha D(\rho r, \rho s)}\rho ds,\\ 
&\omega(t, \rho r) = w(t, \rho r)-\mu\ln G(t, \rho r),\\
&\frac{\partial \lambda}{\partial t}(t, \rho r) + 
\frac{a}{\rho^2}\frac{\partial}{\partial r}\left(\lambda(t, \rho r)\frac{\partial \omega}{\partial r}(t, \rho r)\right)
= \frac{d}{\rho^2}\frac{\partial^2 \lambda}{\partial{r}^2}(t, \rho r)
\end{aligned}\right.
\end{equation}
for $(t, r) \in [0,\infty)\times [-\pi, \pi]$ with an initial condition $\lambda(0, \rho r)=\lambda_0(\rho r)$.

\section{Stationary solution}\label{sec:stationarysol}

\subsection{Homogeneous stationary solution}
We consider a homogeneous stationary solution where the mobile and immobile populations are uniformly distributed on $S$. Same as \citet{Ohtake2023cont}, by substituting the homogeneous population density of immobile workers $\ol{\phi}$ and that of mobile workers $\ol{\lambda}$
\begin{align}
&\ol{\phi} = \frac{\Phi}{2\pi \rho},\\
&\ol{\lambda} = \frac{\Lambda}{2\pi \rho}\label{lamb}
\end{align}
into the first and  second equations of \eqref{racetrack}, we have a homogeneous nominal wage $\ol{w}$ and price index $\ol{G}$ as 
\begin{align}
&\ol{w} = \frac{\mu\left(\ol{\phi}+\ol{\lambda}\right)}{\sigma\ol{\lambda}},\label{wb}\\
&\ol{G} = \left[\frac{2\ol{\lambda}\left(1-e^{-\alpha \rho\pi}\right)}{F\alpha}\right]^{\frac{1}{1-\sigma}},\label{Gb}
\end{align}
respectively. Then, from the third equation of \eqref{1}, it is obvious that the real wage is also homogeneous as 
\begin{equation}
\ol{\omega}=\ol{w}-\mu\ln \ol{G}.\label{omb}
\end{equation}

\subsection{Stability of the homogeneous stationary solution}

\subsubsection{Linearized equations}
Let $\varDelta\lambda$, $\varDelta w$, $\varDelta G$, and $\varDelta\omega$ be  small perturbations added to the homogeneous stationary states \eqref{lamb}, \eqref{wb}, \eqref{Gb}, and \eqref{omb}, respectively. Note that 
\begin{equation}\label{intzero}
\int_S\varDelta\lambda(t,x)dx=0,~\forall t\geq 0
\end{equation}
must hold because of \eqref{conlmd}. Substituting $\lambda=\ol{\lambda}+\varDelta\lambda$, $w=\ol{w}+\varDelta w$, $G=\ol{G}+\varDelta G$, and $\omega=\ol{\omega}+\varDelta\omega$ into \eqref{racetrack}, and leaving only the first-order terms concerning the small perturbations, we obtain the following linearized equations.
\begin{equation}\label{linsys}
\left\{
\begin{aligned}
&\varDelta w(t,\rho r)\\
& = \frac{\mu(\sigma -1)(\ol{\phi}+\ol{\lambda})\ol{G}^{\sigma-2}}{F\sigma}\int_{-\pi}^{\pi}\varDelta G(t,\rho s)e^{-\alpha D(\rho r, \rho s)}\rho ds\\
&\hspace{25mm}+ \frac{\mu\ol{G}^{\sigma-1}}{F\sigma}\int_{-\pi}^{\pi}\varDelta \lambda(t, \rho s)e^{-\alpha D(\rho r, \rho s)}\rho ds,\\
&\varDelta G(t, \rho r) = -\frac{\ol{G}^\sigma}{(\sigma-1)F}\int_{-\pi}^{\pi}\varDelta \lambda(t, \rho s)e^{-\alpha D(\rho r, \rho s)}\rho ds,\\
&\varDelta\omega(t, \rho r) = \varDelta w(t, \rho r) -\frac{\mu}{\ol{G}}\varDelta G(t, \rho r),\\
&\frac{\partial \varDelta \lambda}{\partial t}(t, \rho r) + \frac{a\ol{\lambda}}{\rho^2}\frac{\partial^2\varDelta\omega}{{\partial r}^2}(t, \rho r) = \frac{d}{\rho^2}\frac{\partial^2\varDelta\lambda}{{\partial r}^2}(t, \rho r)
\end{aligned}
\right.
\end{equation}
for $[0,\infty)\times [-\pi,\pi]$.

\subsubsection{Eigenvalue analysis}
Let us define the Fourier series of a small perturbation $\varDelta h$ by
\begin{equation}\label{Fourierseries}
\varDelta h(t, \rho r) = \frac{1}{2\pi}\sum_{k=0,\pm1,\pm2,\cdots}\hat{h}_k(t) e^{ikr},
\end{equation}
where the Fourier coefficient is defined by
\[
\hat{h}_k(t) = \int_{-\pi}^\pi \varDelta h(t, \rho r) e^{-ik r} dr.
\]
In \eqref{Fourierseries}, we refer to $k$ and $e^{ikr}$ as the spatial frequency and the $k$-th mode, respectively. By expanding the small perturbations of equations \eqref{linsys} into the Fourier series, we obtain the equations for the Fourier coefficients
\begin{equation}\label{hatsys}
\left\{
\begin{aligned}
&\hat{w}_k = \frac{\mu(\sigma-1)(\ol{\phi}+\ol{\lambda})}{\sigma\ol{\lambda}\ol{G}}Z_k\hat{G}_k + \frac{\mu}{\sigma\ol{\lambda}}Z_k\hat{\lambda}_k,\\
&\hat{G}_k = -\frac{\ol{G}}{(\sigma-1)\ol{\lambda}}Z_k\hat{\lambda}_k,\\
&\hat{\omega}_k = \hat{w}_k - \frac{\mu}{\ol{G}}\hat{G}_k,\\
&\frac{d\hat{\lambda}_k}{dt} -\frac{a\ol{\lambda}k^2}{\rho^2}\hat{\omega}_k = -\frac{dk^2}{\rho^2}\hat{\lambda}_k
\end{aligned}
\right.
\end{equation}
for each $k=\pm1,\pm2,\cdots$. We do not need to consider the case $k=0$ because $\hat{\lambda}_0=0$ is implied by \eqref{intzero}. The variable $Z_k$ in \eqref{hatsys} is defined by \footnote{See Subsection \ref{subsec:Zk} for how the variable $Z_k$ appears. This variable is essentially the same as the variable $Z$, which is referred to as ``a sort of index of trade cost'' in \citet[Eq. (4.41) on p.~57]{FujiKrugVenab} and \citet[Eq. ~(6.14) on p.90]{FujiKrugVenab}. This variable is monotonically increasing in relation to transport costs. It is interesting to note that all the effects of transport costs on the linearized model are summarized in this variable.}
\[
Z_k := \frac{\alpha^2\rho^2\left(1-(-1)^ke^{-\alpha \rho\pi}\right)}{\left(k^2+\alpha^2\rho^2\right)\left(1-e^{-\alpha\rho\pi}\right)}.
\]
It is known that $Z_k$ is monotonically increasing with respect to $\alpha \rho\geq 0$ and $\lim_{\alpha\rho\to0} Z_k=0$ and $\lim_{\alpha\rho\to\infty} Z_k=1$ hold.\footnote{\citet[Theorem 2]{Ohtake2023city}} By solving \eqref{hatsys}, we have
\begin{equation}\label{eigdif}
\frac{d\hat{\lambda}_k}{dt}(t) = \Gamma_k \hat{\lambda}_k(t),
\end{equation}
where the {\it eigenvalue} $\Gamma_k$ for the $k$-th mode is
\begin{equation}
\Gamma_k :=
\frac{k^2}{\rho^2}\left[a\mu Z_k\left(-\frac{\ol{\phi}+\ol{\lambda}}{\sigma\ol{\lambda}}Z_k+\frac{2\sigma-1}{\sigma(\sigma-1)}\right)-d\right].\label{eigenvalue}
\end{equation}
It is easy to verify that $\Gamma_k=\Gamma_{-k}$ from $Z_k=Z_{-k}$. From \eqref{eigdif}, we have
\begin{equation}
\hat{\lambda}_k(t) = e^{t\Gamma_k}\hat{\lambda}_k(0).
\end{equation}
Hence, the time evolution of the $k$-th mode is governed by its corresponding eigenvalue. If $\Gamma_k<0$, then the amplitude of the $k$-th mode decays to $0$ over time, and we call the mode the {\it stable mode}. Meanwhile, if $\Gamma_k>0$, then the amplitude of the $k$-th mode grows with time, and we will refer to this mode as the {\it unstable mode}. As shown in Fig.~\ref{fig:eigenvz}, the eigenvalue $\Gamma_k$ is a quadratic function of $Z_k$, which is upward convex. Therefore, as $Z_k$ increases (i.e., $Z_k$ is monotonically increasing concerning transport costs), the $k$-th mode can be stabilized. If the diffusion coefficient $d$ is zero, then the eigenvalues are also zero when the transport costs are zero. However, a positive diffusion coefficient uniformly shifts the entire eigenvalue downward, as seen in \eqref{eigenvalue}. As a result, the $k$-th mode is stabilized not only where transport costs are high, but also where they are low.

\begin{figure}[H]
\centering
\begin{tikzpicture}
 \draw[name path=xaxis, ->, thin] (0, 0)--(4.5, 0)node[right]{$Z_k$};
 \draw[name path=yaxis, ->, thin] (0, -1.5)--(0, 1)node[above]{};
 \draw (0,0)node[above left]{O};
 \draw[name path=Gammak,red,thin,domain=0:4] plot(\x,{-0.30*pow(\x,2)+\x-0.5})node[right]{$\Gamma_k$};
 \path[name intersections={of=Gammak and xaxis, by={A, B}}];
 \path[name intersections={of=Gammak and yaxis, by ={C}}];
 \fill[black] (A) circle (0.05) node[above right]{};
 \fill[black] (B) circle (0.05) node[above right]{};
 \draw (C) node[left]{$-d$};
\end{tikzpicture}
\caption{Sketch of $\Gamma_k$}
\label{fig:eigenvz}
\end{figure}

\begin{figure}[H]
\begin{subfigure}[H]{0.5\columnwidth}
\includegraphics[width=\columnwidth]{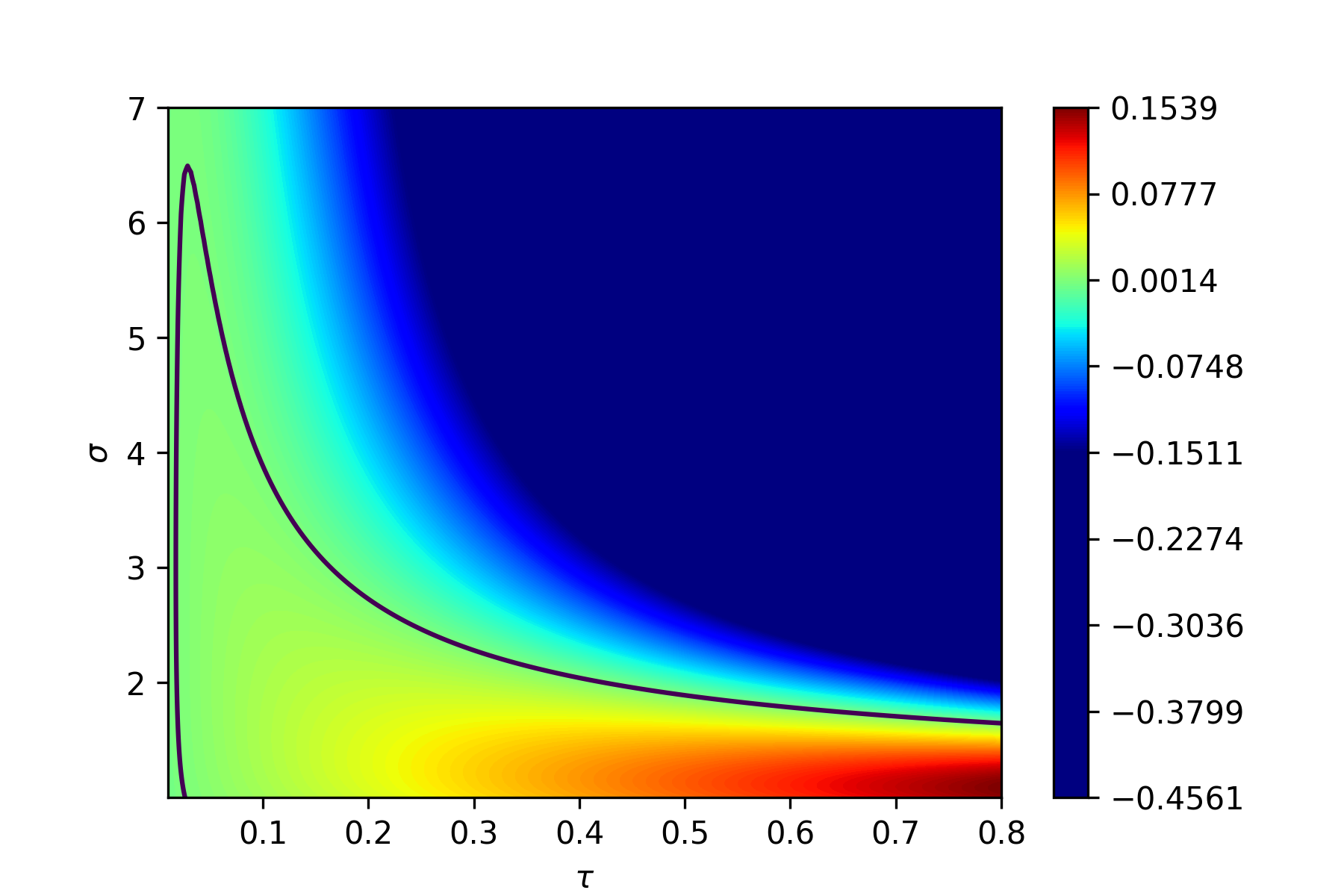}
\caption{$k=1$}\label{hm_1}
\end{subfigure}
\begin{subfigure}[H]{0.5\columnwidth}
\includegraphics[width=\columnwidth]{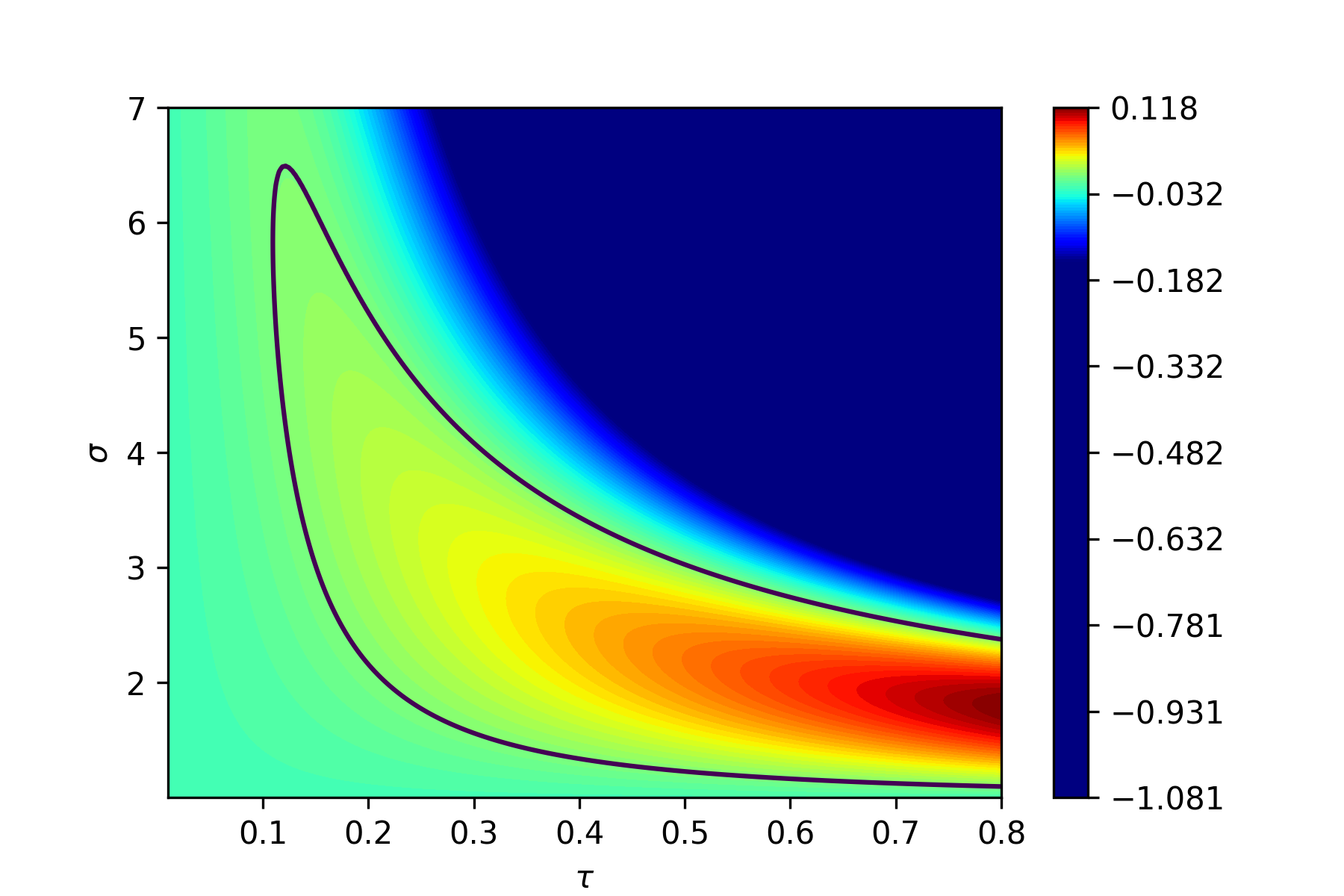}
\caption{$k=2$}\label{hm_2}
\end{subfigure}
\begin{subfigure}[H]{0.5\columnwidth}
\includegraphics[width=\columnwidth]{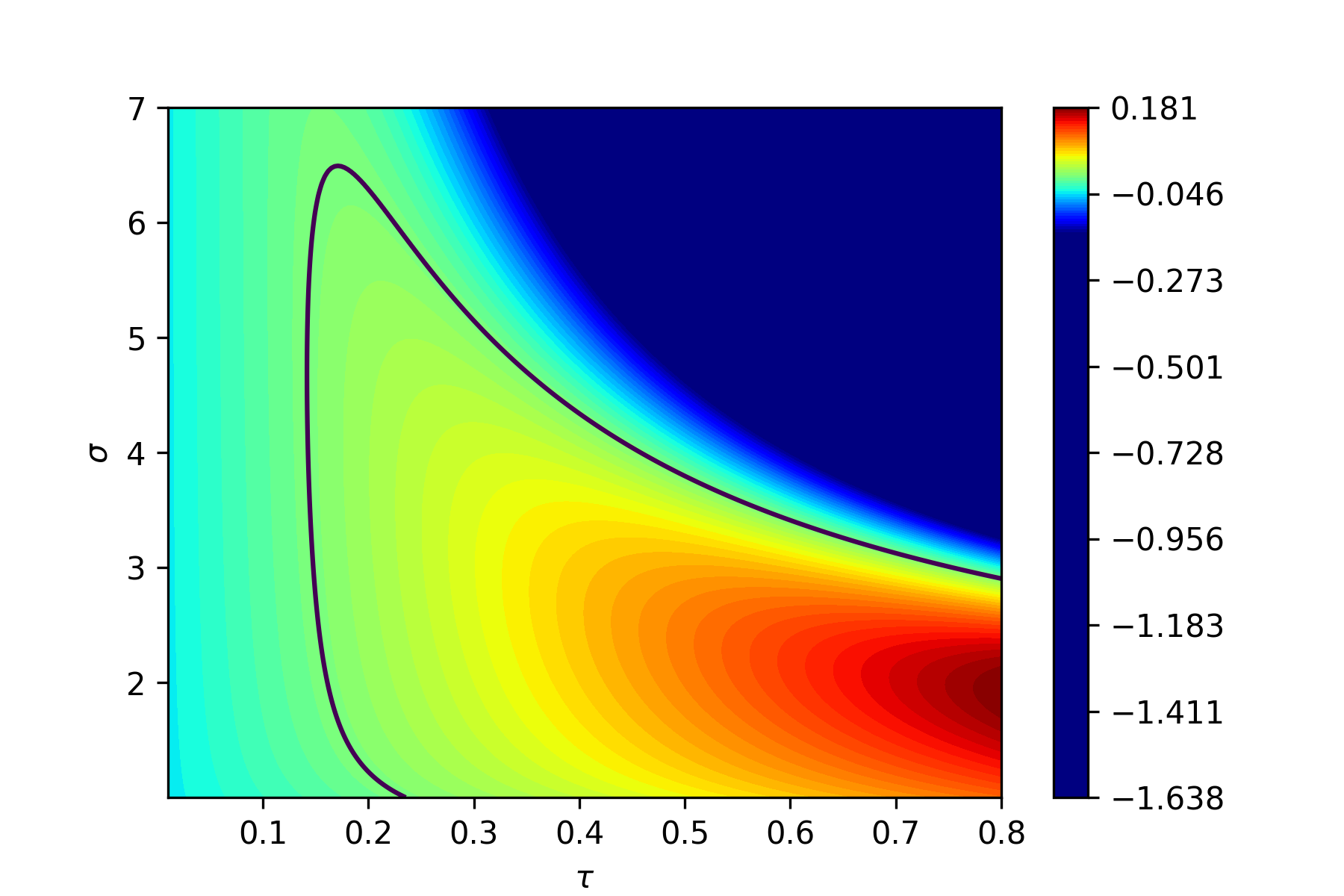}
\caption{$k=3$}\label{hm_3}
\end{subfigure}
\begin{subfigure}[H]{0.5\columnwidth}
\includegraphics[width=\columnwidth]{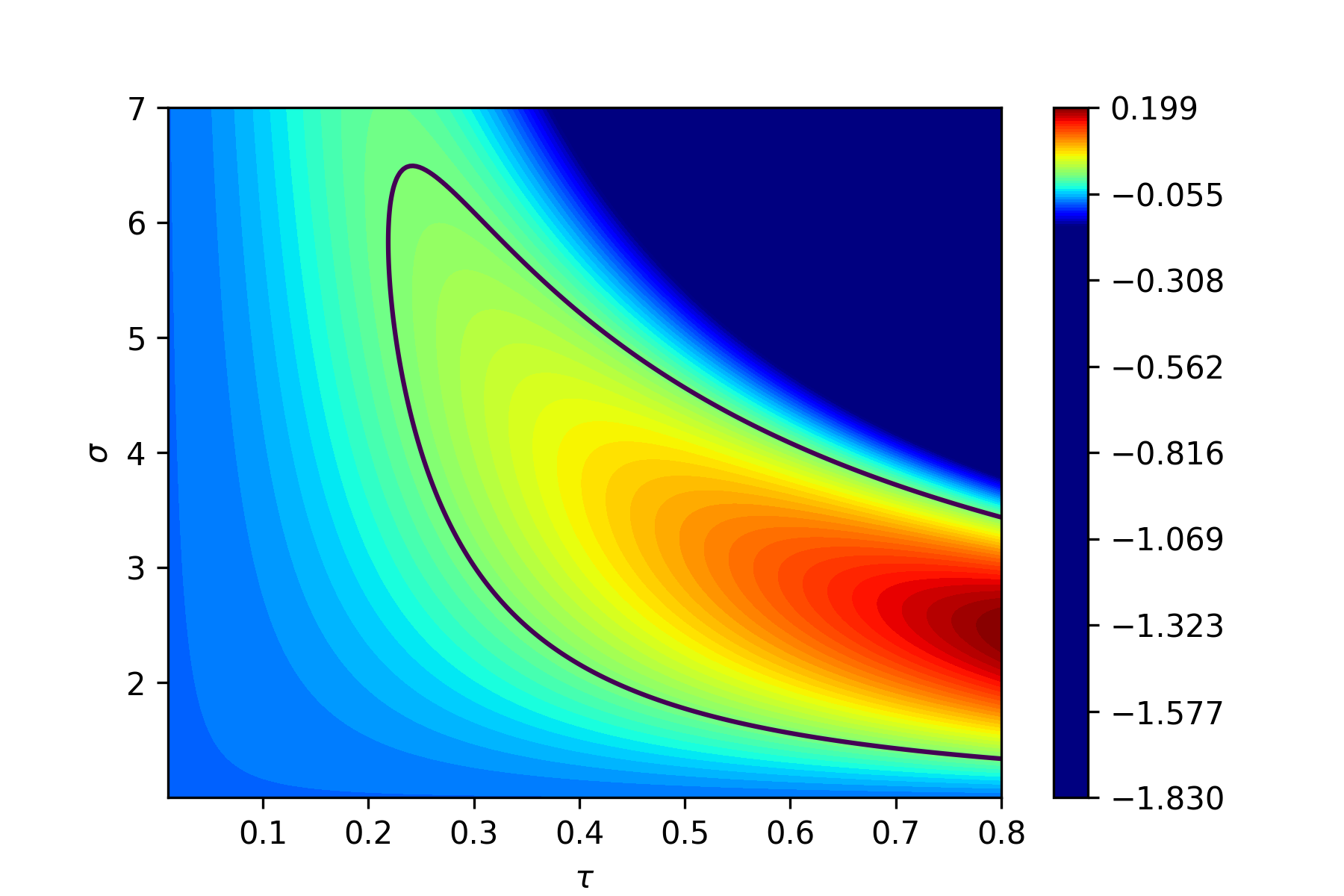}
\caption{$k=4$}\label{hm_4}
\end{subfigure}
\begin{subfigure}[H]{0.5\columnwidth}
\includegraphics[width=\columnwidth]{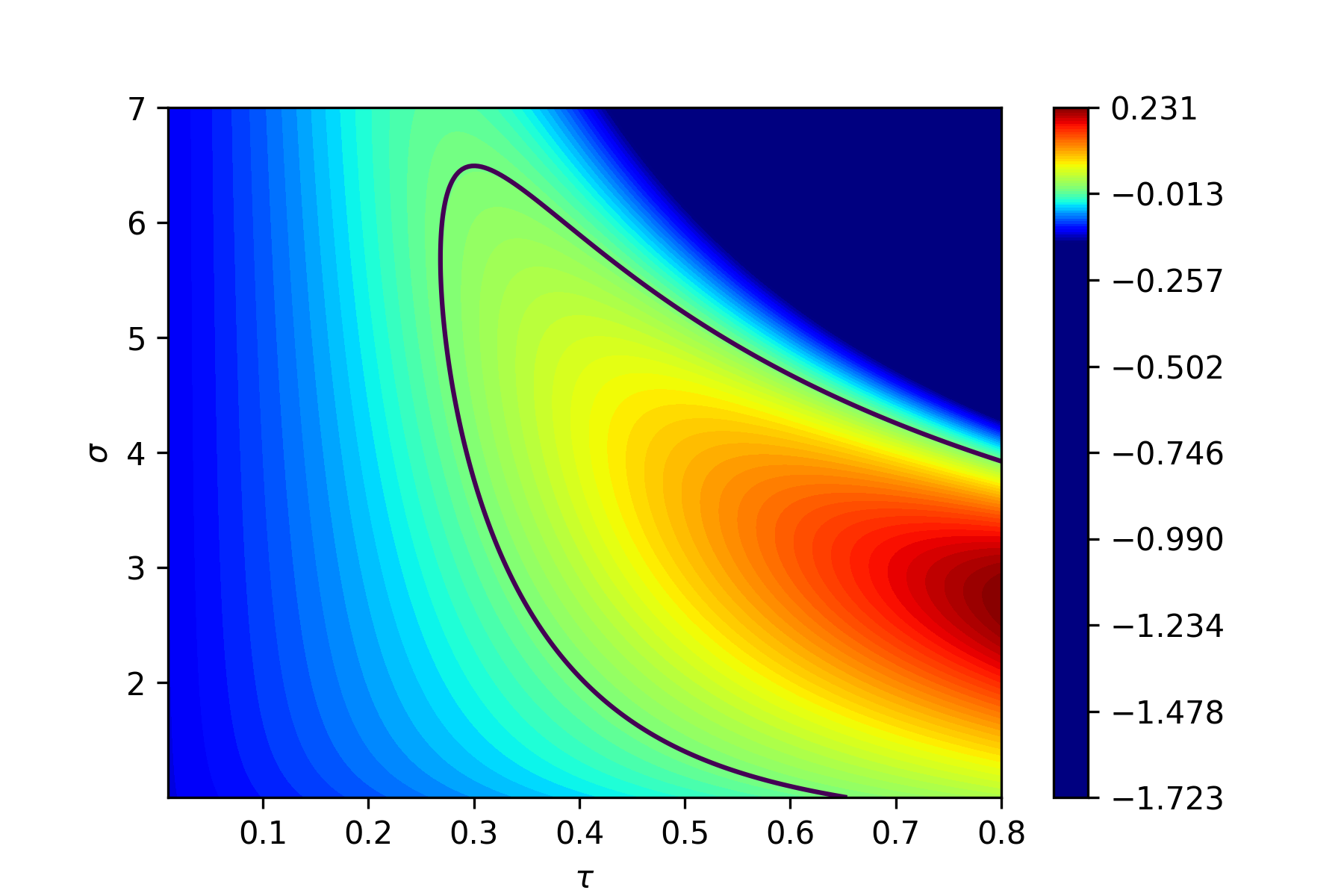}
\caption{$k=5$}\label{hm_5}
\end{subfigure}
\begin{subfigure}[H]{0.5\columnwidth}
\includegraphics[width=\columnwidth]{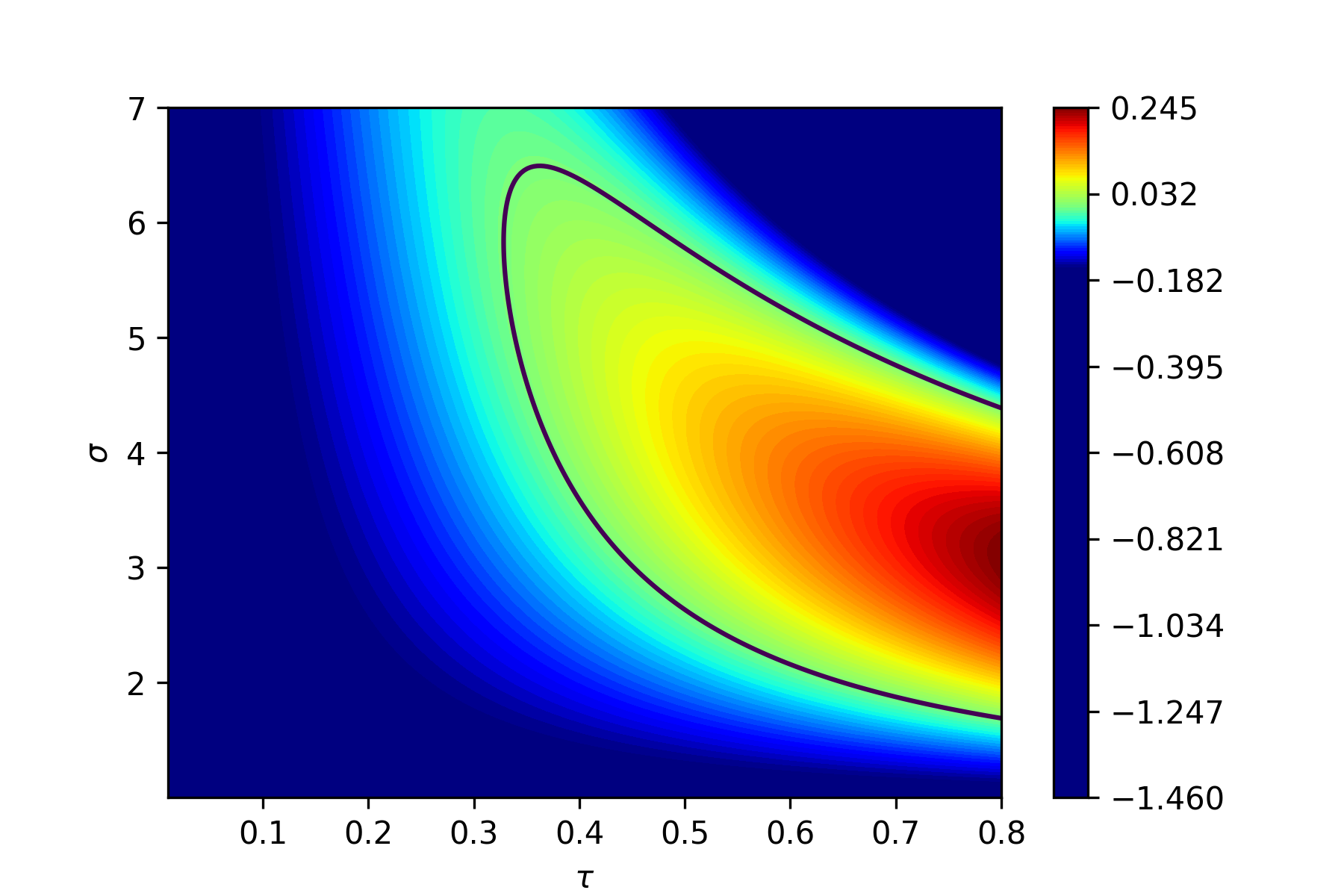}
\caption{$k=6$}\label{hm_6}
\end{subfigure}
\caption{Heatmaps of $\Gamma_k$ in $(\tau, \sigma)$-plane}\label{figs:hm_sigma}
\end{figure}

\begin{figure}[H]
\centering
\includegraphics[width=\columnwidth]{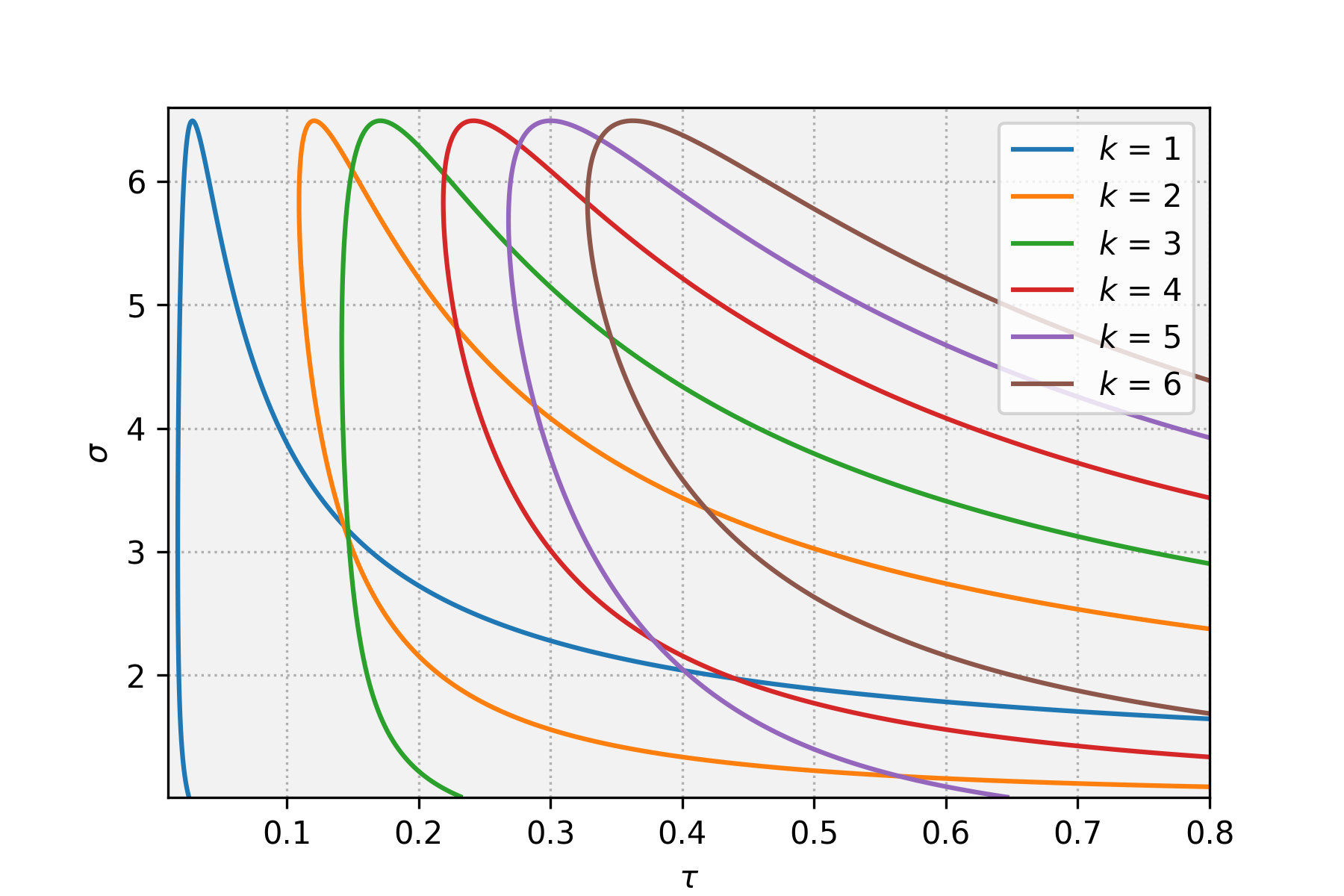}
\caption{Critical curves in $(\tau, \sigma)$-plane}\label{figs:cc_sigma}
\end{figure}

Fig.~\ref{figs:hm_sigma} shows in heatmaps how eigenvalues depend on the control parameters $\tau>0.01$ and $\sigma>1.01$ when the other parameters are fixed to $\mu=0.6$,  $a=0.5$, $d=0.005$, $F=1$, $\Lambda=1$, $\Phi=10$, and $\rho=1$.\footnote{Numpy \citep{numpy} and Matplotlib \citep{matplotlib} are used to compute eigenvalues. The source code is available at \url{https://github.com/k-ohtake/advection-diffusion-neg}.} Each sub-figure corresponds to a specific frequency mode, $k=1$, $2$, $\cdots$, $6$. For each $\sigma$, when $\tau$ is moved from larger to smaller values, $\Gamma_k$ is initially negative. However, after passing a certain critical point at which $\Gamma_k=0$, it becomes positive. If $\tau$ is made even smaller, it passes another critical point and becomes negative again. The set of critical points forms a {\it critical curve} as shown by the black line in the figure. Fig.~\ref{figs:cc_sigma} shows only the critical curves for each frequency.\footnote{Numpy \citep{numpy} and Matplotlib \citep{matplotlib} are used to compute eigenvalues. The source code is available at \url{https://github.com/k-ohtake/advection-diffusion-neg}.}

The fact that there are two critical points for each $\sigma$, i.e., that the mode stabilizes if the transport costs are sufficiently small, is a direct consequence of diffusion. Actually, in \eqref{eigenvalue}, if $\tau\to 0$ ($Z_k\to 0$), we see that $\Gamma_k\to -\frac{k^2}{\rho^2}d < 0$. This is in contrast to the most basic CP model, where a decrease in transport costs promotes agglomeration.\footnote{This property of the stability, i.e., the redispersion associated with lower transport costs, is known to occur in models that incorporate agricultural transport costs. See for example, \citet{PicaZeng05}, \citet[Chapter 7]{FujiKrugVenab}, and \citet{Ohtake2025agriculture}.} This is because, when transport costs fall sufficiently, firms can enjoy the benefits of agglomeration through increasing returns without needing to agglomerate, and consumers can enjoy diverse goods without needing to agglomerate; thus, the diffusion overcomes agglomeration forces.

Diffusion causes even more interesting phenomena. Around $\sigma=6.4$ in Fig.~\ref{figs:cc_sigma}, there is not always a destabilized mode,  an interesting cascade of unstable and stable states occurs as $\tau$ decreases; a mode with $k=6$ is unstable → all modes are stable → a mode with $k=5$ is unstable → all modes are stable → ... → a mode with $k=1$ is unstable → all modes are stable. This indicates that agglomeration forces, driven by a preference for variety, and diffusion forces, resulting from congestion costs or idiosyncratic tastes, are in a delicate balance. As can be seen from Fig.~\ref{figs:cc_sigma}, when the preference for variety is too weak (i.e., when $\sigma$ is large), diffusion forces prevail altogether. Conversely, when preference for variety is quite strong (when $\sigma$ is close to $1$), there is no cascade of unstable and stable states. Cascades appear only when the preference for variety is moderate. The mechanism of this cascade can probably be understood as follows. First, assume that transport costs have decreased from a state where a mode with a frequency $|k|+1$ is unstable. Consumers can now easily obtain a wide variety of manufactured goods without having to agglomerate in urban areas, enabling them to avoid congestion costs and follow their idiosyncratic tastes. Given this tendency toward population dispersion, it is advantageous for firms to disperse their locations so as not to lose demand. Therefore, the homogeneous distribution can be stable as a result of the alignment of interests between consumers and firms. Next, assume that transport costs have fallen further from this point. In this case, firms do not lose demand from immobile consumers even if they agglomerate, and consumers now choose to enjoy the variety offered by agglomeration rather than avoid congestion costs or follow their idiosyncratic tastes (their preference for variety is strong to that extent). This alignment of interests creates instability of a mode with a frequency $|k|$.

Thus, unlike the standard CP model, agglomeration does not progress monotonically as transport costs decrease. Even so, we can see that the absolute value of the frequencies of unstable modes becomes smaller as the transport costs decrease. This means that the number of agglomerated areas (assuming they are formed) tends to decrease with lower transport costs. This can be interpreted in a manner consistent with the basic theory of NEG. In other words, by reducing transport costs, firms can enjoy the benefits of agglomeration arising from increasing returns without losing demand from immobile consumers.

Observations on the role of preference for variety show that if $\tau$ is not too small, the smaller $\sigma$ is, the smaller $|k|$ of the unstable mode becomes. This means that the stronger the preference for variety, the more agglomeration is facilitated. This can also be interpreted in line with the basic theory of NEG. That is, as the preference for variety increases, consumers are attracted to areas with many firms in agglomeration. However, the condition that the transport costs are not too small is essential here. For example, for $\tau=0.1$, it can be seen in Fig.~\ref{figs:cc_sigma} that only the mode with $|k|=1$ is destabilized, regardless of how small $\sigma$ becomes. This is because, when transport costs are sufficiently low, consumers can enjoy a wide variety of manufactured goods without agglomeration, and dispersion forces, stemming from idiosyncratic preferences or congestion costs, exceed agglomeration forces. Therefore, consumers do not bother to agglomerate, even if their preference for variety strengthens slightly.

\section{Numerical simulations}\label{sec:numerical}

\subsection{Numerical scheme}
The time variable $t\in[0,\infty)$ is discretized by a small fragment $\varDelta t$ as $t_n=n\varDelta t$ for $n=0,1,2,\cdots$. The space variable $r\in [-\pi,\pi)$ is descretized with $N$ equally spaced nodes denoted by $r_i=-\pi+(i-1)\varDelta r$ for $i=1,2,\cdots,N$ where $\varDelta r= 2\pi/N$. A function $h$ on $[0,\infty)\times S$ is approximated by a vector $h^n=[h_1^n, h_2^n, \cdots, h_{N}^n]\in\mathbb{R}^{N}$ each of which element corresponds to an approximated value of $h(t_n, \rho r_i)$. We often have to mention elements such as $h_{N+1}^n$ and $h_{0}^n$. In that case, the periodicity naturally leads to $h_{N+1}^n:=h^n_1$ and $h_{0}^n:=h_{N}^n$. The integration of $h$ on $S$ is approximated as
\[
\int_{-\pi}^\pi h(t_n, \rho r)\rho dr \fallingdotseq \sum_{i=1}^{N} h_i^n\rho \varDelta r,
\]
which is equivalent to the approximation using the trapezoidal rule under the periodic condition.

By defining the flux $f$ as
\begin{equation}\label{flux}
f(t, \rho r) := \lambda(t, \rho r)\frac{\partial \omega}{\partial r}(t, \rho r),
\end{equation}
we see that \eqref{PDErtheta} becomes
\[
\frac{\partial \lambda}{\partial t}(t,\rho r) + 
\frac{a}{\rho^2}\frac{\partial f}{\partial r}(t,\rho r)
= \frac{d}{\rho^2}\frac{\partial^2 \lambda}{\partial r^2}(t,\rho r).
\]
We approximate the time derivative by the forward-difference method as
\begin{equation}\label{timediff}
\frac{\partial \lambda}{\partial t}(t^n, \rho r_j) \fallingdotseq \frac{\lambda^{n+1}_j-\lambda^n_j}{\varDelta t}.
\end{equation}
The diffusion term is approximated by the well-known formula \footnote{\citet[p. 15]{strang2007}}
\begin{equation}\label{diffusiondiff}
\frac{\partial^2 \lambda}{\partial r^2}(t,\rho r) \fallingdotseq \frac{\lambda^n_{j+1}-2\lambda^n_j+\lambda^n_{j-1}}{\varDelta r^2}.
\end{equation}
To approximate the second term on the left-hand side, we follow the finite-volume method used in numerical fluid dynamics.\footnote{\citet[pp. 526-527]{strang2007}} First, let us introduce the virtual nodes $r_{j-\frac{1}{2}}\in(r_{j-1},r_{j})$ and $r_{j+\frac{1}{2}}\in(r_{j},r_{j+1})$, and consider the values $\tilde{f}^n_{j-\frac{1}{2}}$ and $\tilde{f}^n_{j+\frac{1}{2}}$ of a numerical flux there. The method for assigning these values is discussed below. Anyway, using the numerical flux, we approximate
\begin{equation}\label{dfdrdiff}
\frac{\partial f}{\partial r}(t_n, r_j) \fallingdotseq \frac{\tilde{f}_{j+\frac{1}{2}}^n - \tilde{f}_{j-\frac{1}{2}}^n}{\varDelta r}.
\end{equation}
By using \eqref{timediff}, \eqref{diffusiondiff}, and \eqref{dfdrdiff}, we see that \eqref{PDErtheta} is approximated as
\begin{equation}
\frac{\lambda^{n+1}_j-\lambda^n_j}{\varDelta t}
+ \frac{a}{\rho^2}\frac{\tilde{f}_{j+\frac{1}{2}}^n - \tilde{f}_{j-\frac{1}{2}}^n}{\varDelta r}
= \frac{d}{\rho^2}\frac{\lambda^n_{j+1}-2\lambda^n_j+\lambda^n_{j-1}}{\varDelta r^2},
\end{equation}
which is equivalent to the explicit form
\begin{equation}\label{exm}
\lambda^{n+1}_j = \lambda^n_j -\frac{a}{\rho^2}\frac{\varDelta t}{\varDelta r}\left(\tilde{f}_{j+\frac{1}{2}}^n - \tilde{f}_{j-\frac{1}{2}}^n\right)
+\frac{d}{\rho^2}\frac{\varDelta t}{\varDelta r^2}\left(\lambda^n_{j+1}-2\lambda^n_j+\lambda^n_{j-1}\right).
\end{equation}

Let us consider the numerical flux. If mobile workers flow into the virtual interval $[r_{j-\frac{1}{2}}, r_{j+\frac{1}{2}}]$ through the boundary $j-\frac{1}{2}$, i.e., $\omega_{j-1}<\omega_j$, then we approximate the flux \eqref{flux} by \footnote{The use of $\lambda^n_{j-1}$ here is an application of the {\it upwind} concept in numerical fluid dynamics, in which the value on the side from which information comes is used. The same concept applies to the following other cases. See \citet[pp. 526-528]{strang2007} for the upwind method.}
\[
\tilde{f}_{j-\frac{1}{2}}^n = \lambda^n_{j-1}\frac{\omega^n_j-\omega^n_{j-1}}{\varDelta r} > 0.
\]
Conversely, if mobile workers flow out of the interval through the boundary at $j-\frac{1}{2}$, i.e., $\omega_{j-1}>\omega_j$, then we approximate the flux by
\[
\tilde{f}_{j-\frac{1}{2}}^n = \lambda^n_{j}\frac{\omega^n_j-\omega^n_{j-1}}{\varDelta r} < 0.
\]
A similar consideration applies to the inflow and outflow of mobile workers at the boundary $j+\frac{1}{2}$. These ideas can be summarized as follows.
\[\left\{
\begin{aligned}
&\text{if}~~\omega^n_j - \omega^n_{j-1}>0 \Rightarrow \tilde{f}_{j-\frac{1}{2}}^n = \lambda^n_{j-1}\frac{\omega^n_j-\omega^n_{j-1}}{\varDelta r}, \\
&\text{if}~~\omega^n_j - \omega^n_{j-1}<0 \Rightarrow \tilde{f}_{j-\frac{1}{2}}^n = \lambda^n_{j}\frac{\omega^n_j-\omega^n_{j-1}}{\varDelta r}.
\end{aligned}
\right.
\]
Similarly, 
\[\left\{
\begin{aligned}
&\text{if}~~\omega^n_{j+1} - \omega^n_{j}>0 \Rightarrow \tilde{f}_{j+\frac{1}{2}}^n = \lambda^n_{j}\frac{\omega^n_{j+1}-\omega^n_{j}}{\varDelta r}, \\
&\text{if}~~\omega^n_{j+1} - \omega^n_{j}<0 \Rightarrow \tilde{f}_{j+\frac{1}{2}}^n = \lambda^n_{j+1}\frac{\omega^n_{j+1}-\omega^n_{j}}{\varDelta r}.
\end{aligned}
\right.
\]

\subsection{Numerical results \footnotemark}
\begin{figure}[H]
 \begin{subfigure}{0.5\columnwidth}
  \centering
  \includegraphics[width=\columnwidth]{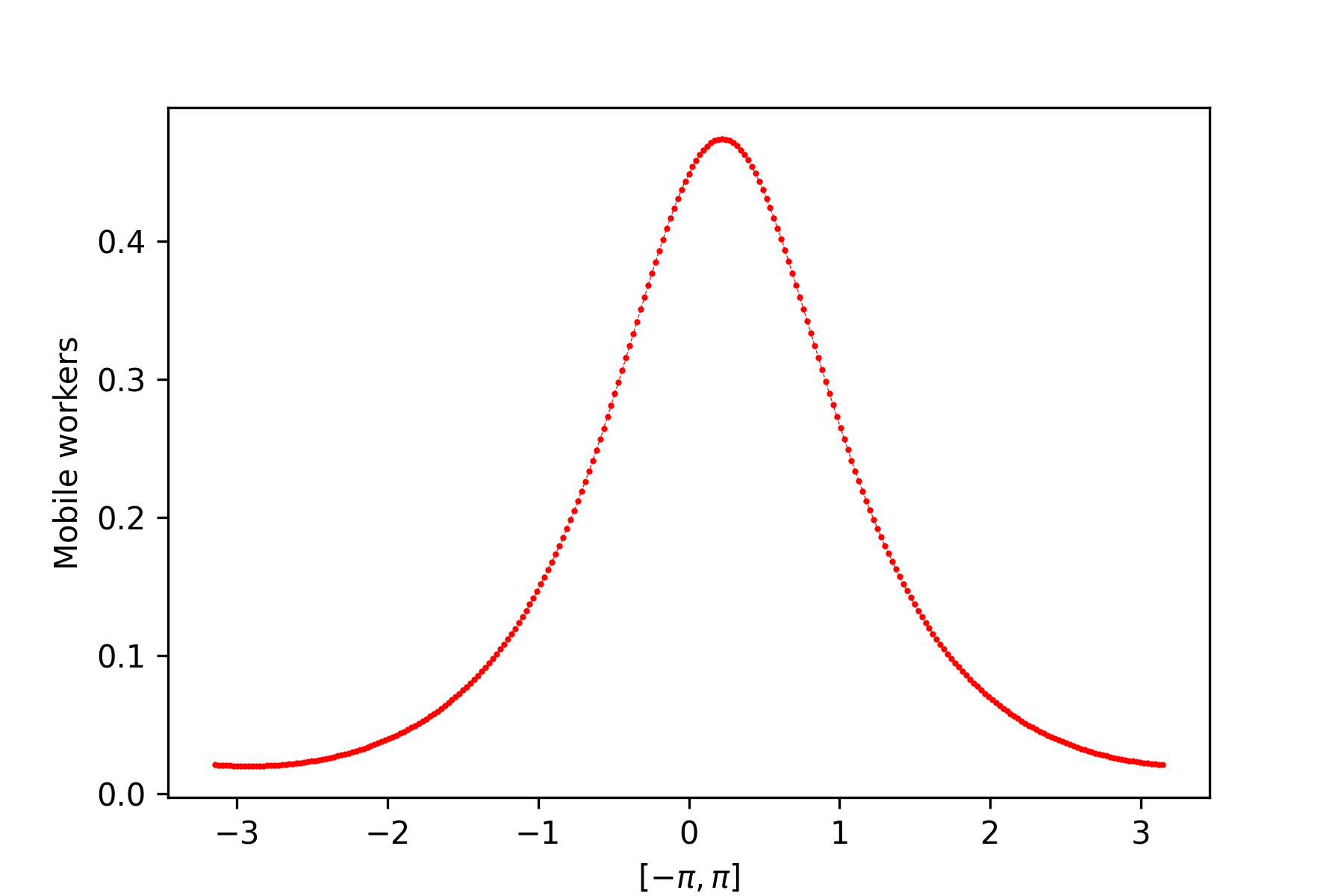}
  \caption{Mobile population density $\lambda^*$}
 \end{subfigure}
 \begin{subfigure}{0.5\columnwidth}
  \centering
  \includegraphics[width=\columnwidth]{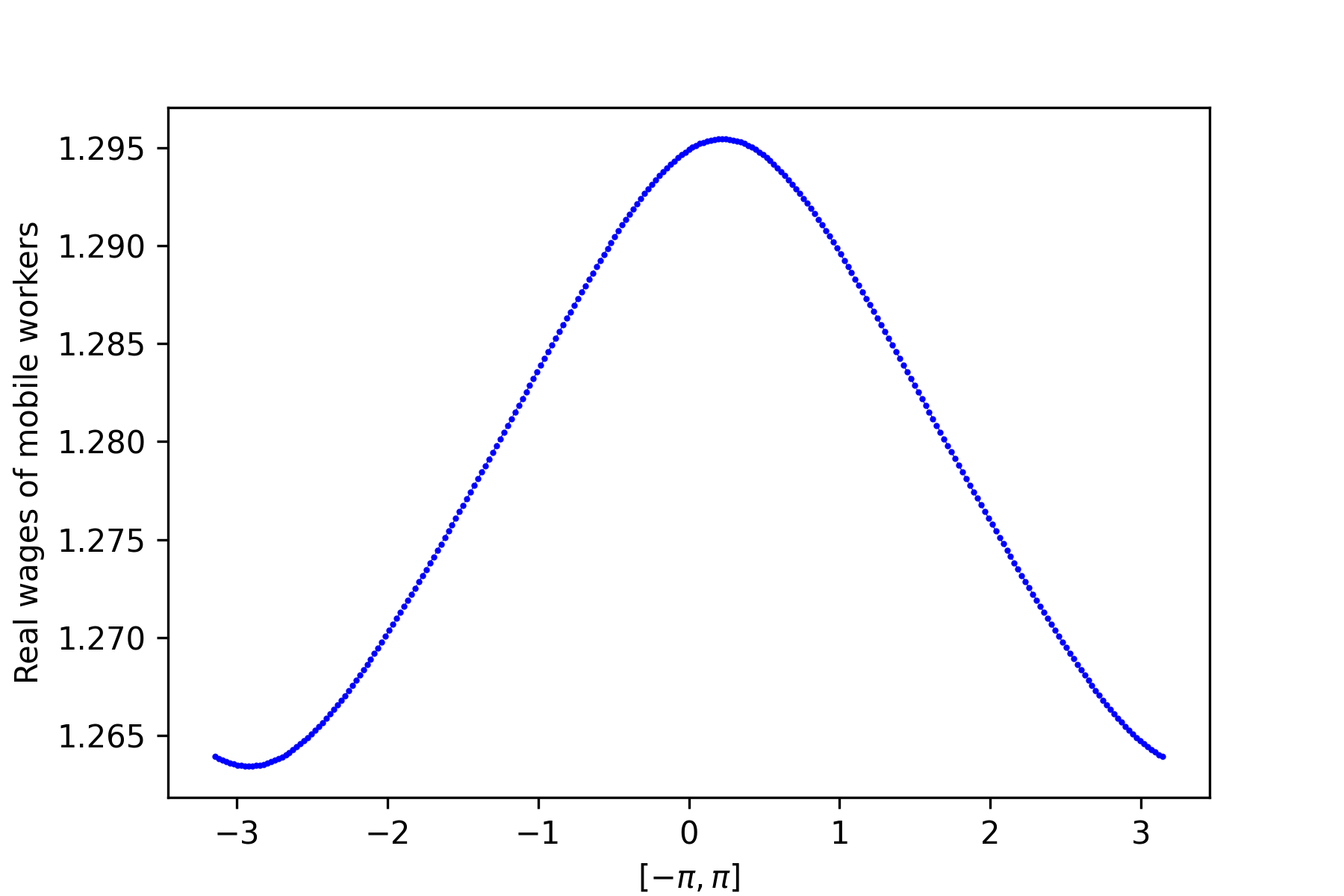}
  \caption{Real wage $\omega^*$}
 \end{subfigure}\\
 \caption{Numerical stationary solution for $(\sigma,\tau)=(5.0, 0.05)$}
 \label{fig:one-city-tau}
\end{figure}

\begin{figure}[H]
 \begin{subfigure}{0.5\columnwidth}
  \centering
  \includegraphics[width=\columnwidth]{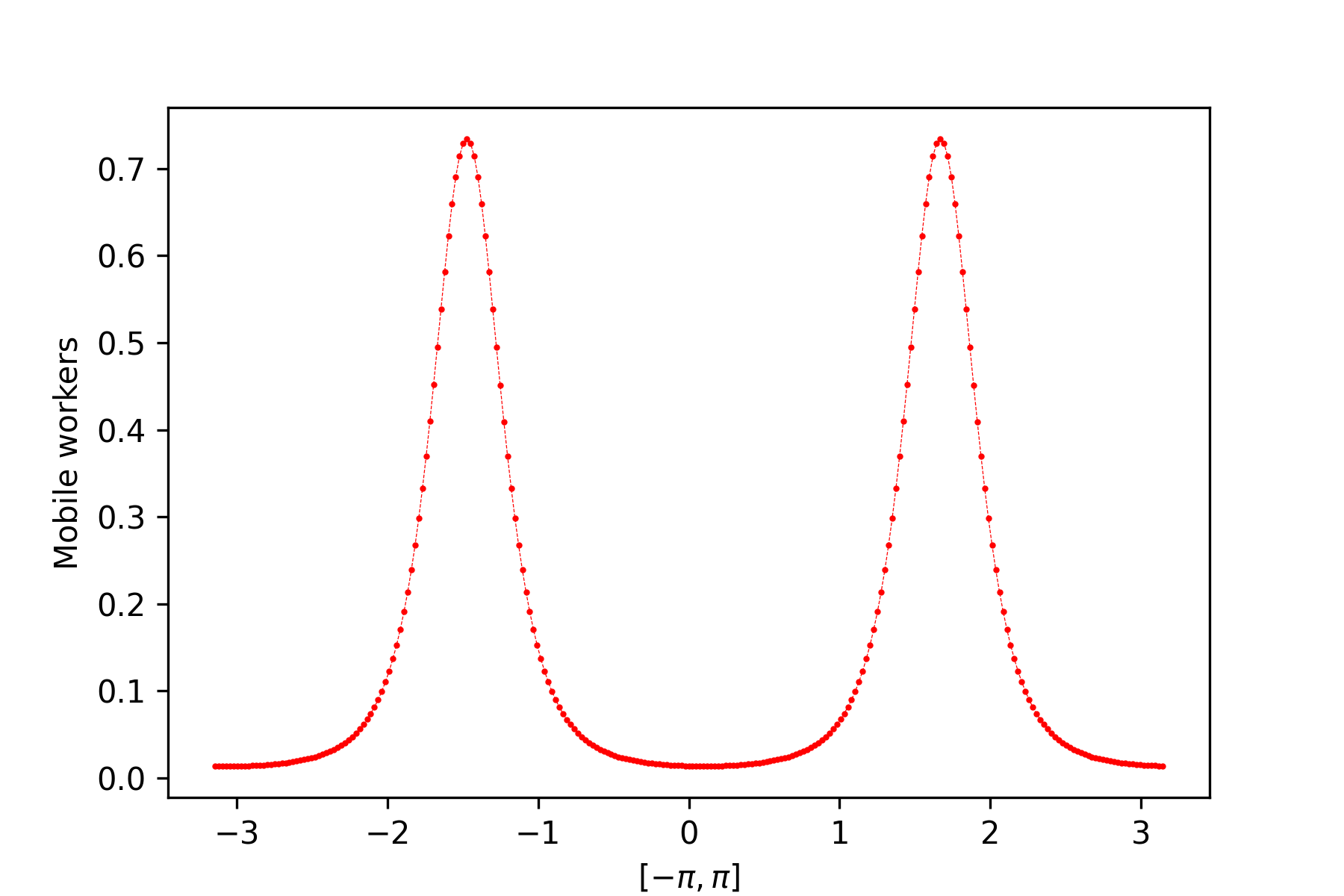}
  \caption{Mobile population density $\lambda^*$}
 \end{subfigure}
 \begin{subfigure}{0.5\columnwidth}
  \centering
  \includegraphics[width=\columnwidth]{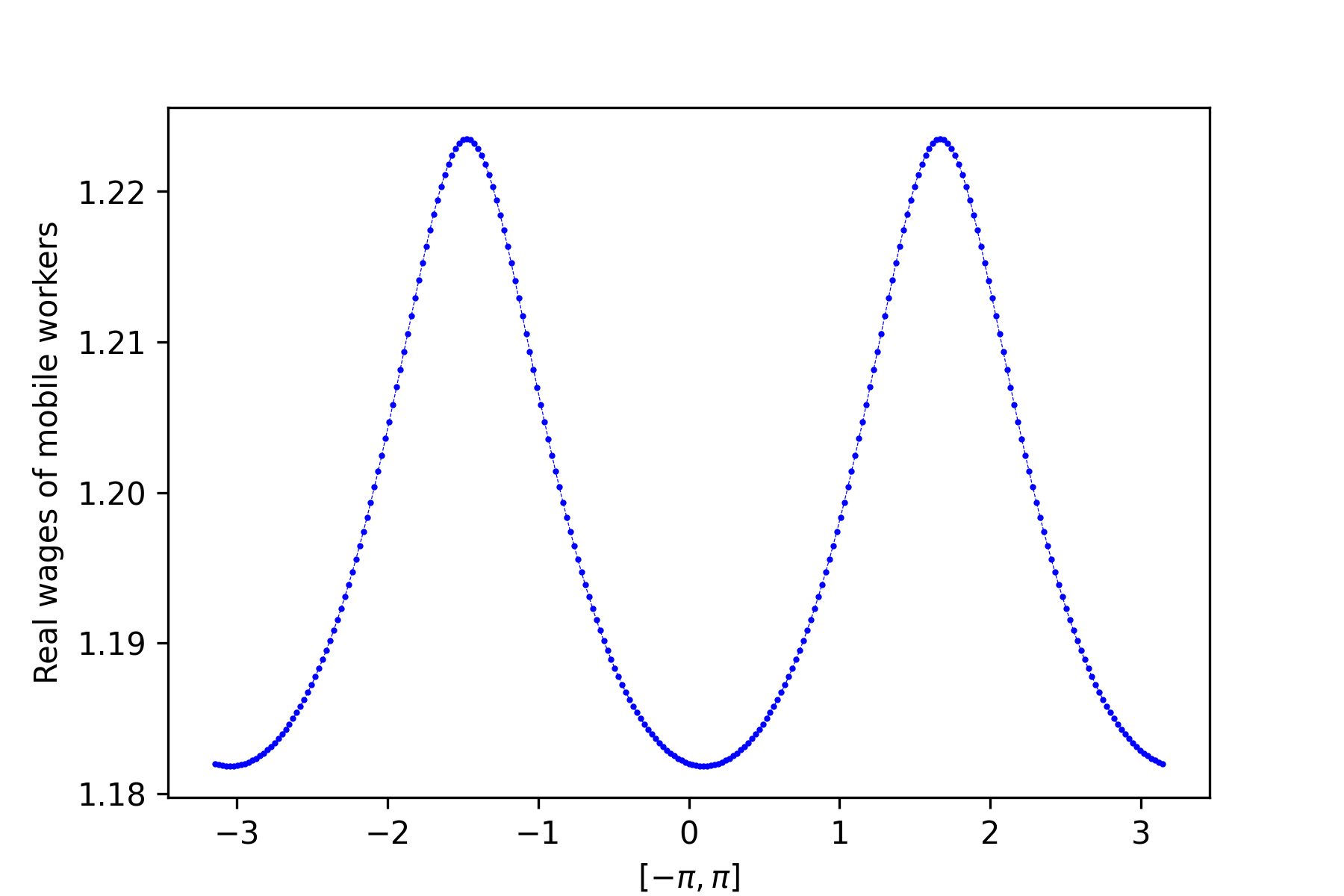}
  \caption{Real wage $\omega^*$}
 \end{subfigure}\\
 \caption{Numerical stationary solution for $(\sigma,\tau)=(5.0, 0.15)$}
 \label{fig:two-city-tau}
\end{figure}

\begin{figure}[H]
 \begin{subfigure}{0.5\columnwidth}
  \centering
  \includegraphics[width=\columnwidth]{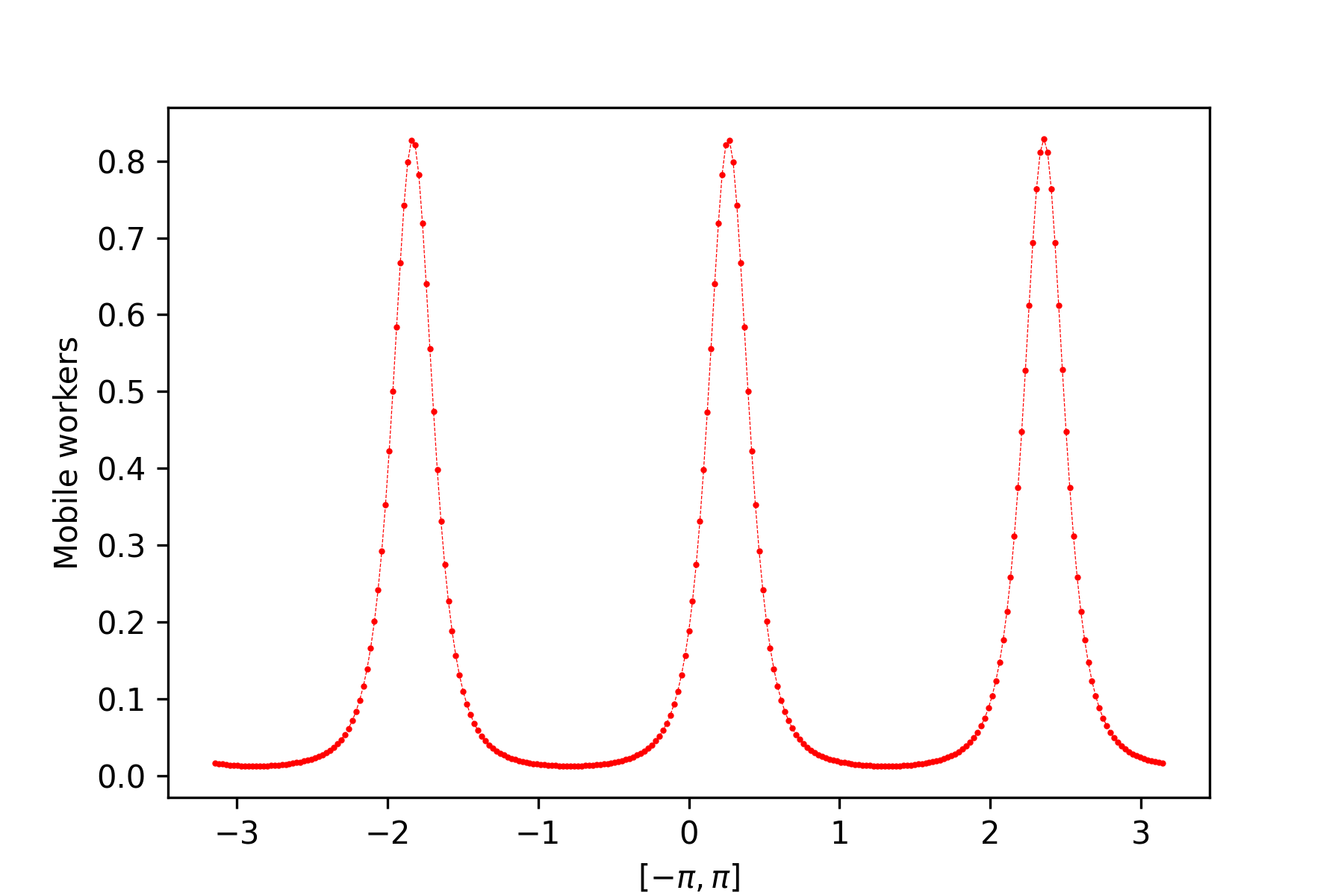}
  \caption{Mobile population density $\lambda^*$}
 \end{subfigure}
 \begin{subfigure}{0.5\columnwidth}
  \centering
  \includegraphics[width=\columnwidth]{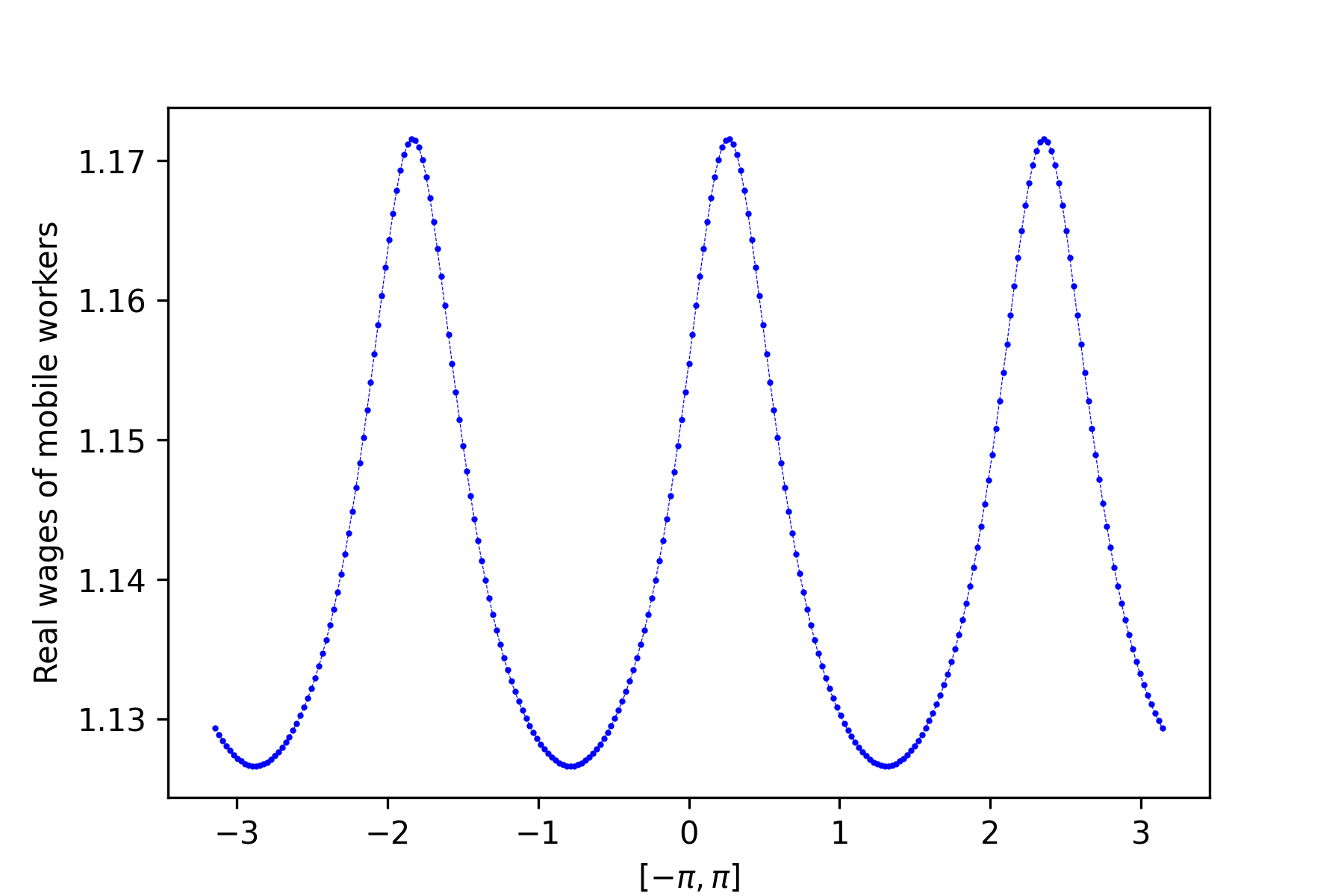}
  \caption{Real wage $\omega^*$}
 \end{subfigure}\\
 \caption{Numerical stationary solution for $(\sigma,\tau)=(5.0, 0.25)$}
 \label{fig:three-city-tau}
\end{figure}

\begin{figure}[H]
 \begin{subfigure}{0.5\columnwidth}
  \centering
  \includegraphics[width=\columnwidth]{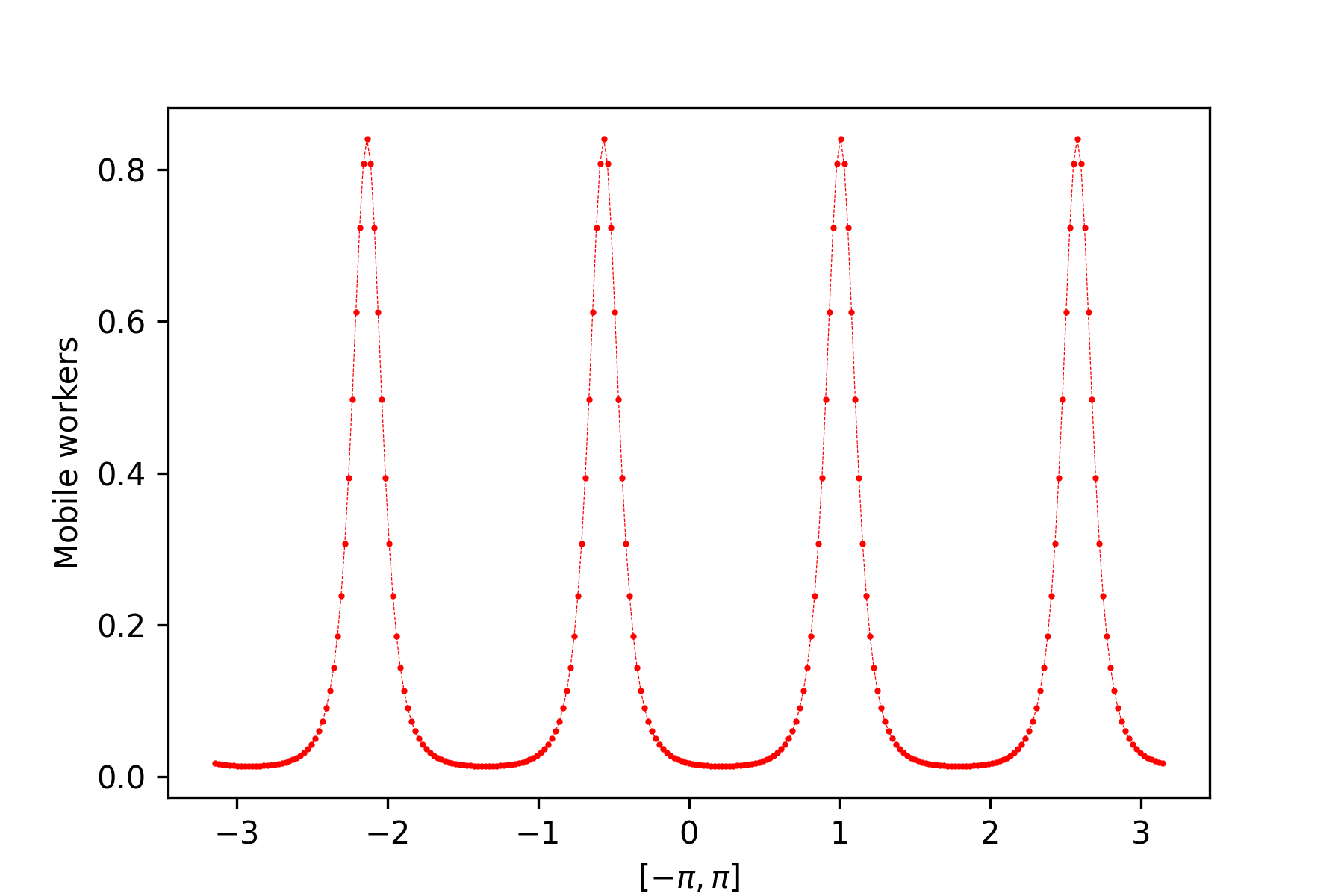}
  \caption{Mobile population density $\lambda^*$}
 \end{subfigure}
 \begin{subfigure}{0.5\columnwidth}
  \centering
  \includegraphics[width=\columnwidth]{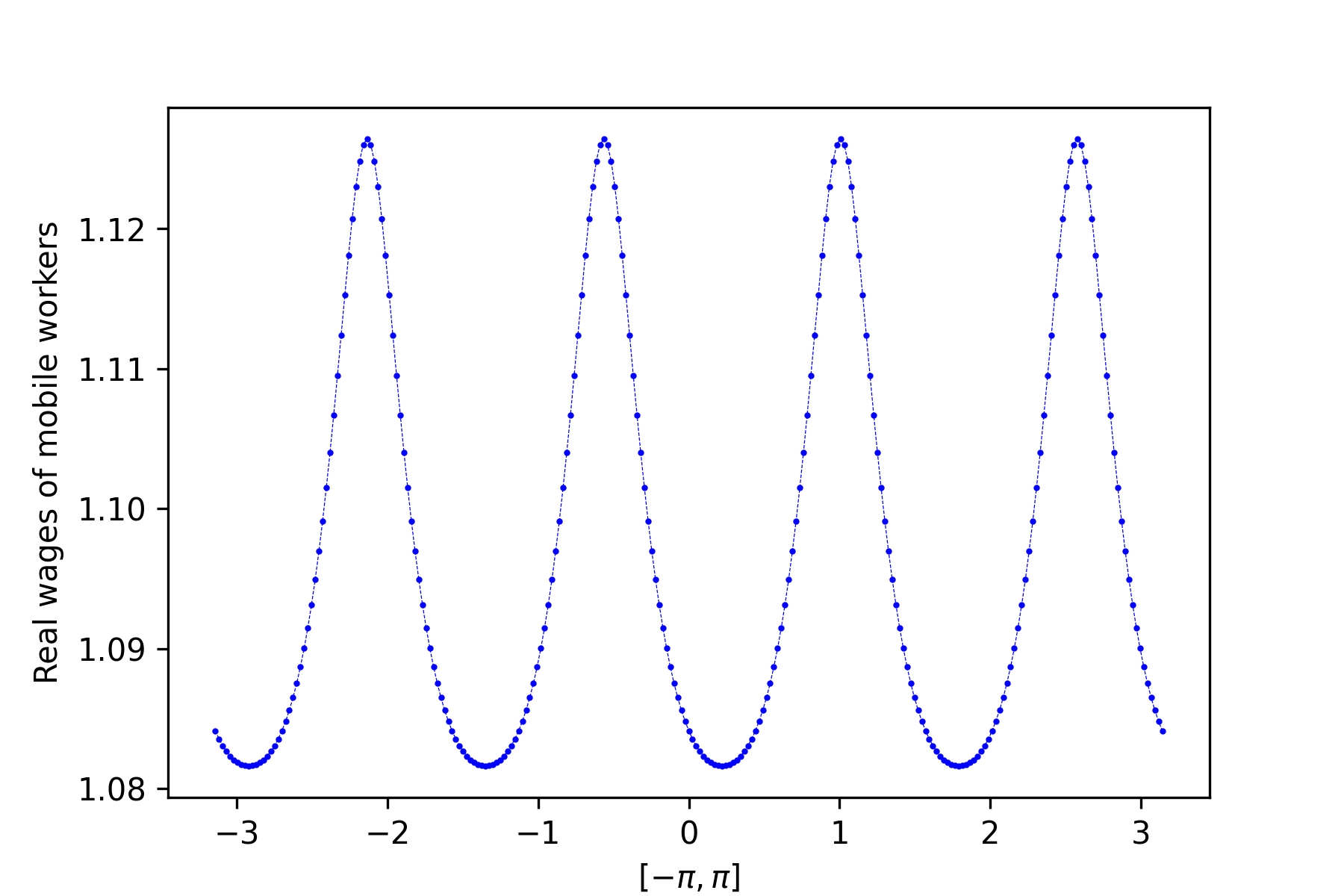}
  \caption{Real wage $\omega^*$}
 \end{subfigure}\\
 \caption{Numerical stationary solution for $(\sigma,\tau)=(5.0, 0.35)$}
 \label{fig:four-city-tau}
\end{figure}

\begin{figure}[H]
 \begin{subfigure}{0.5\columnwidth}
  \centering
  \includegraphics[width=\columnwidth]{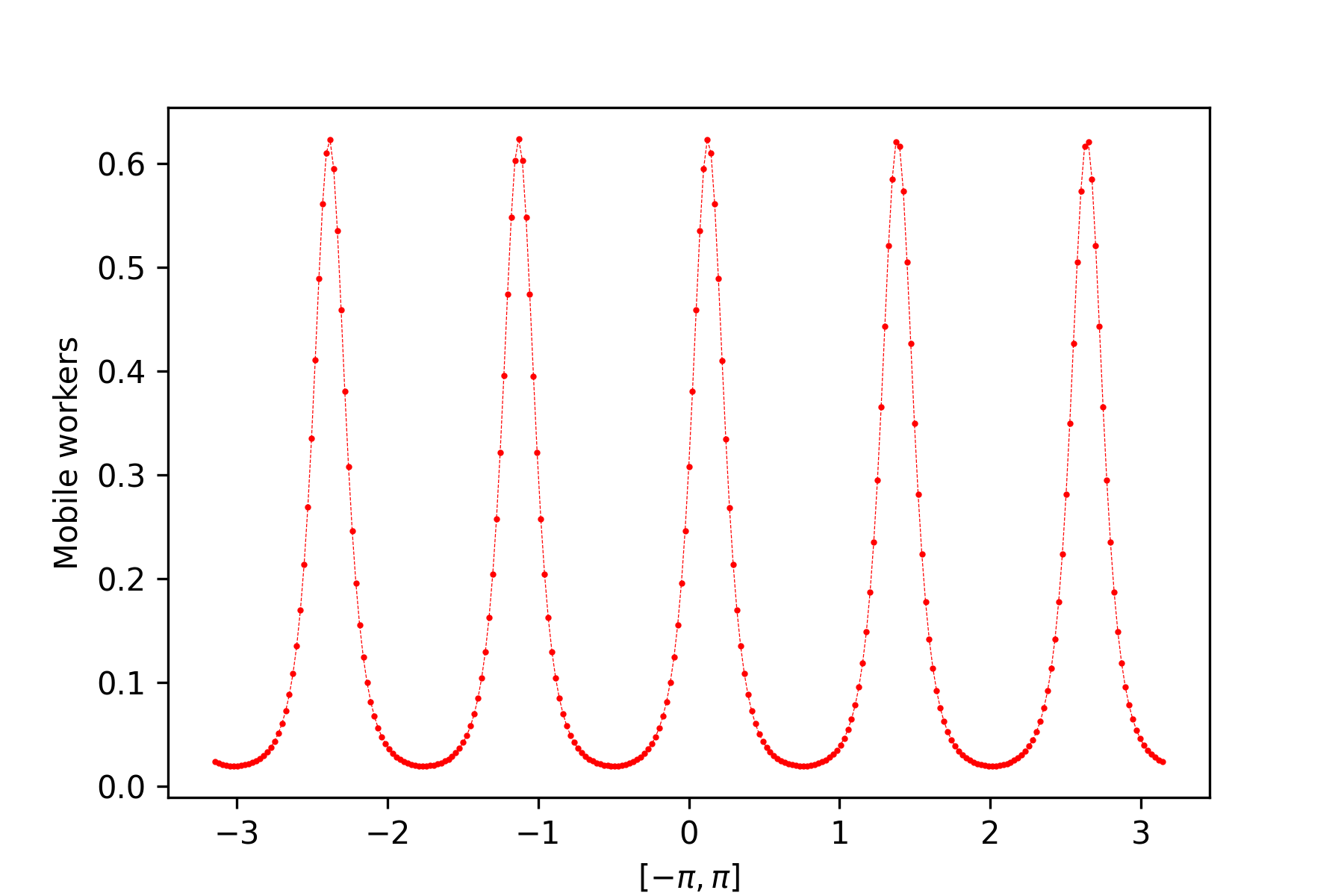}
  \caption{Mobile population density $\lambda^*$}
 \end{subfigure}
 \begin{subfigure}{0.5\columnwidth}
  \centering
  \includegraphics[width=\columnwidth]{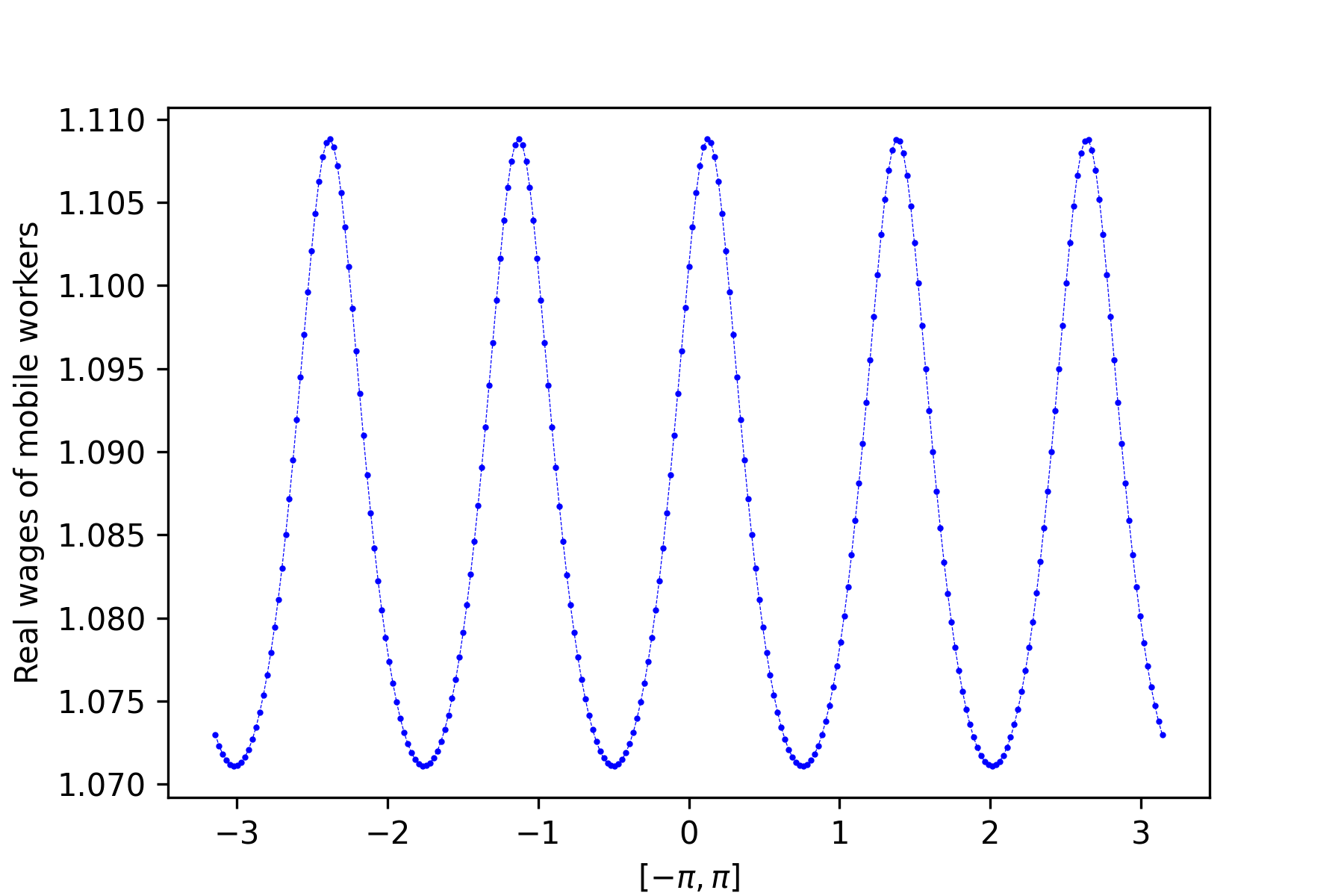}
  \caption{Real wage $\omega^*$}
 \end{subfigure}\\
 \caption{Numerical stationary solution for $(\sigma,\tau)=(5.0, 0.37)$}
 \label{fig:five-city-tau}
\end{figure}

\begin{figure}[H]
 \begin{subfigure}{0.5\columnwidth}
  \centering
  \includegraphics[width=\columnwidth]{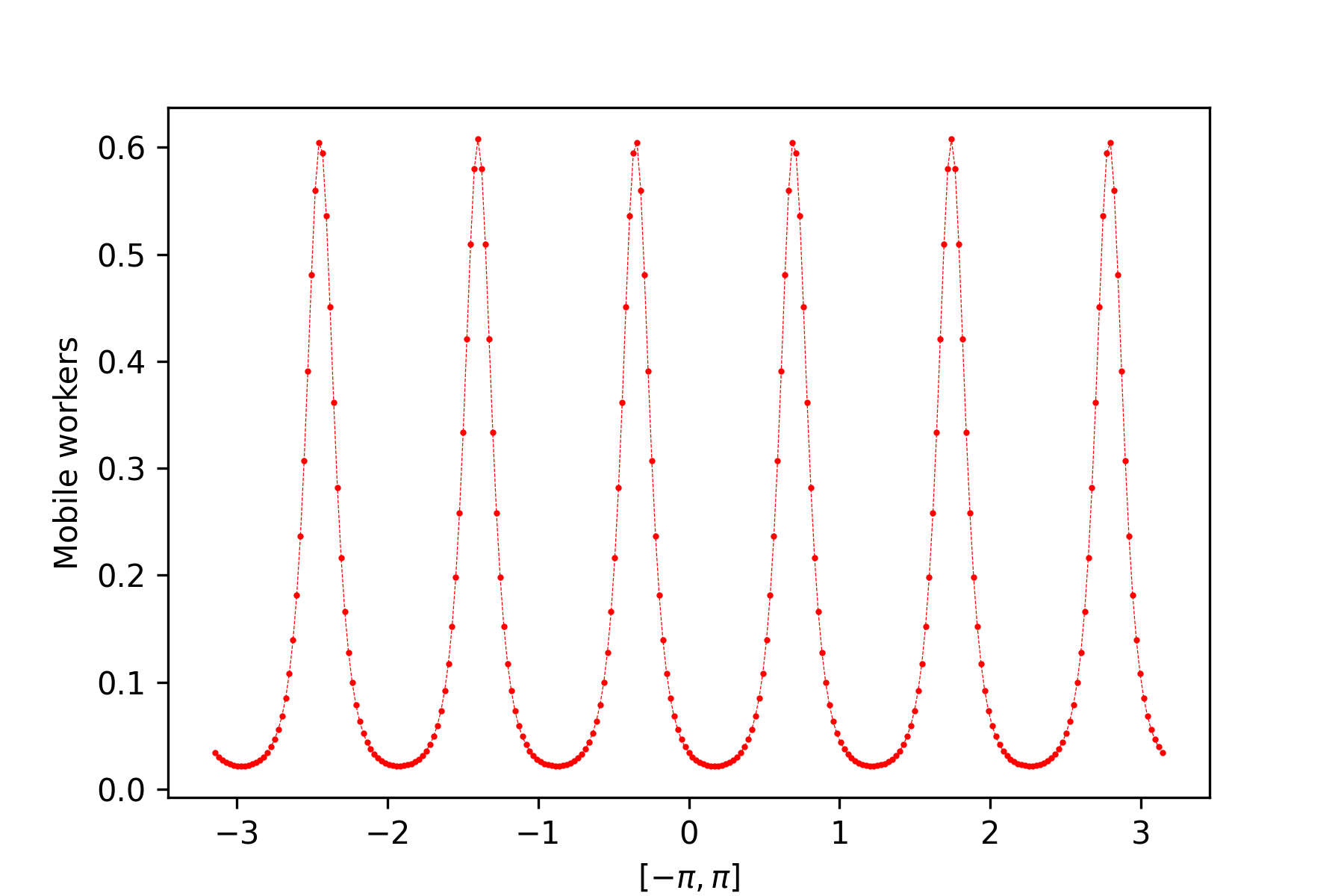}
  \caption{Mobile population density $\lambda^*$}
 \end{subfigure}
 \begin{subfigure}{0.5\columnwidth}
  \centering
  \includegraphics[width=\columnwidth]{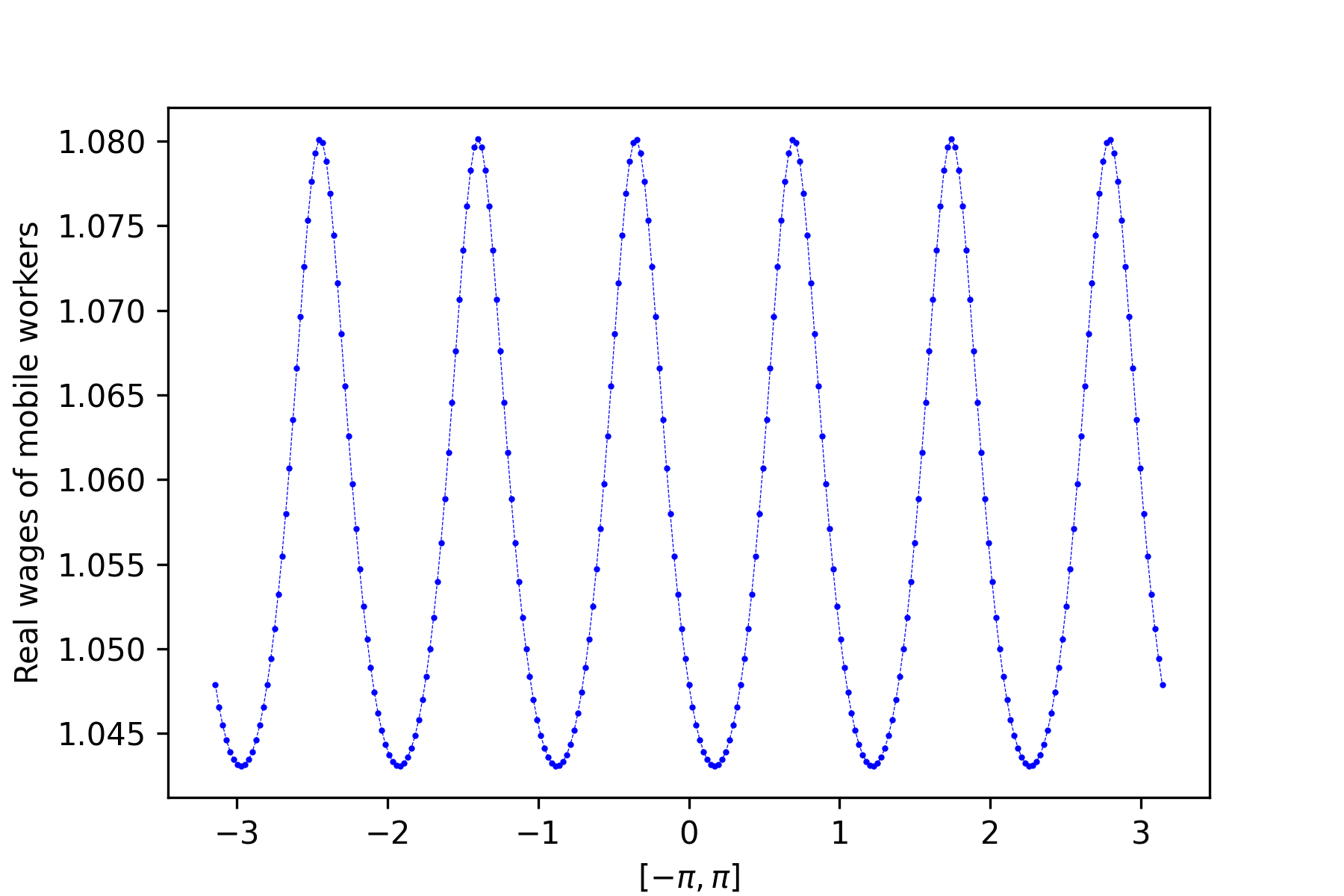}
  \caption{Real wage $\omega^*$}
 \end{subfigure}\\
 \caption{Numerical stationary solution for $(\sigma,\tau)=(5.0, 0.45)$}
 \label{fig:six-city-tau}
\end{figure}

\begin{figure}[H]
 \begin{subfigure}{0.5\columnwidth}
  \centering
  \includegraphics[width=\columnwidth]{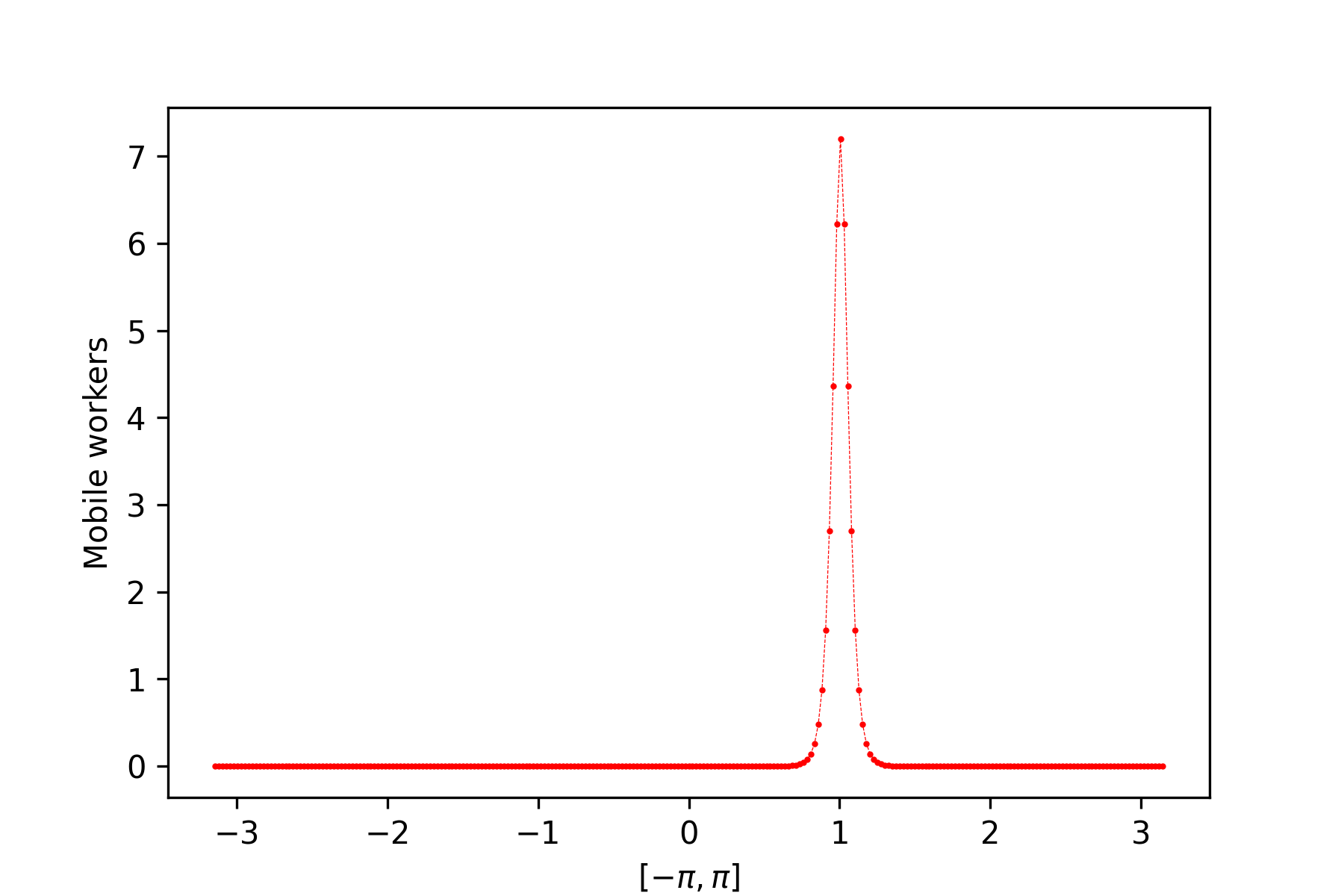}
  \caption{Mobile population density $\lambda^*$}
 \end{subfigure}
 \begin{subfigure}{0.5\columnwidth}
  \centering
  \includegraphics[width=\columnwidth]{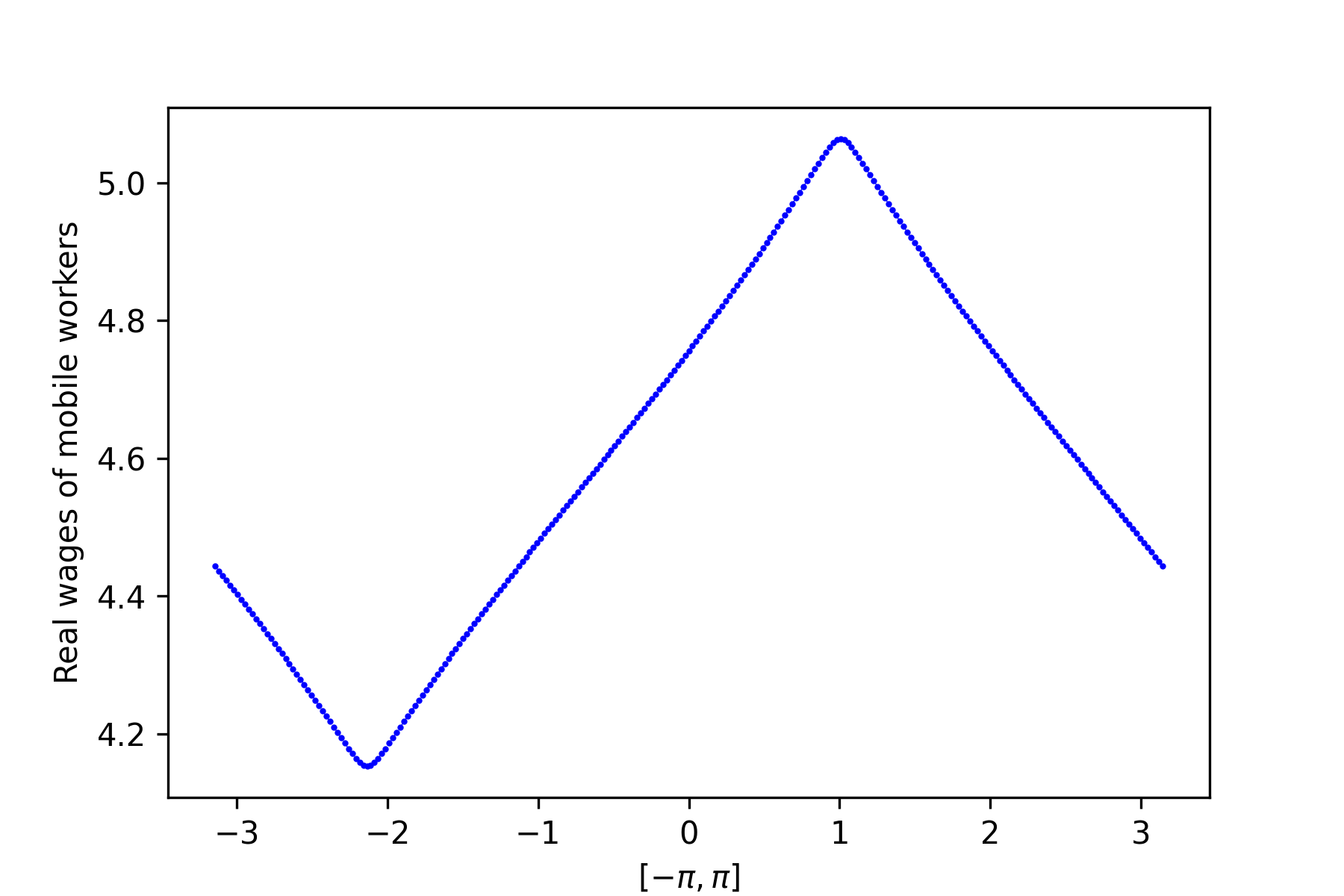}
  \caption{Real wage $\omega^*$}
 \end{subfigure}\\ 
 \caption{Numerical stationary solution for $(\sigma,\tau)=(1.3, 0.5)$}
 \label{fig:one-city-sigma}
\end{figure}

\begin{figure}[H]
 \begin{subfigure}{0.5\columnwidth}
  \centering
  \includegraphics[width=\columnwidth]{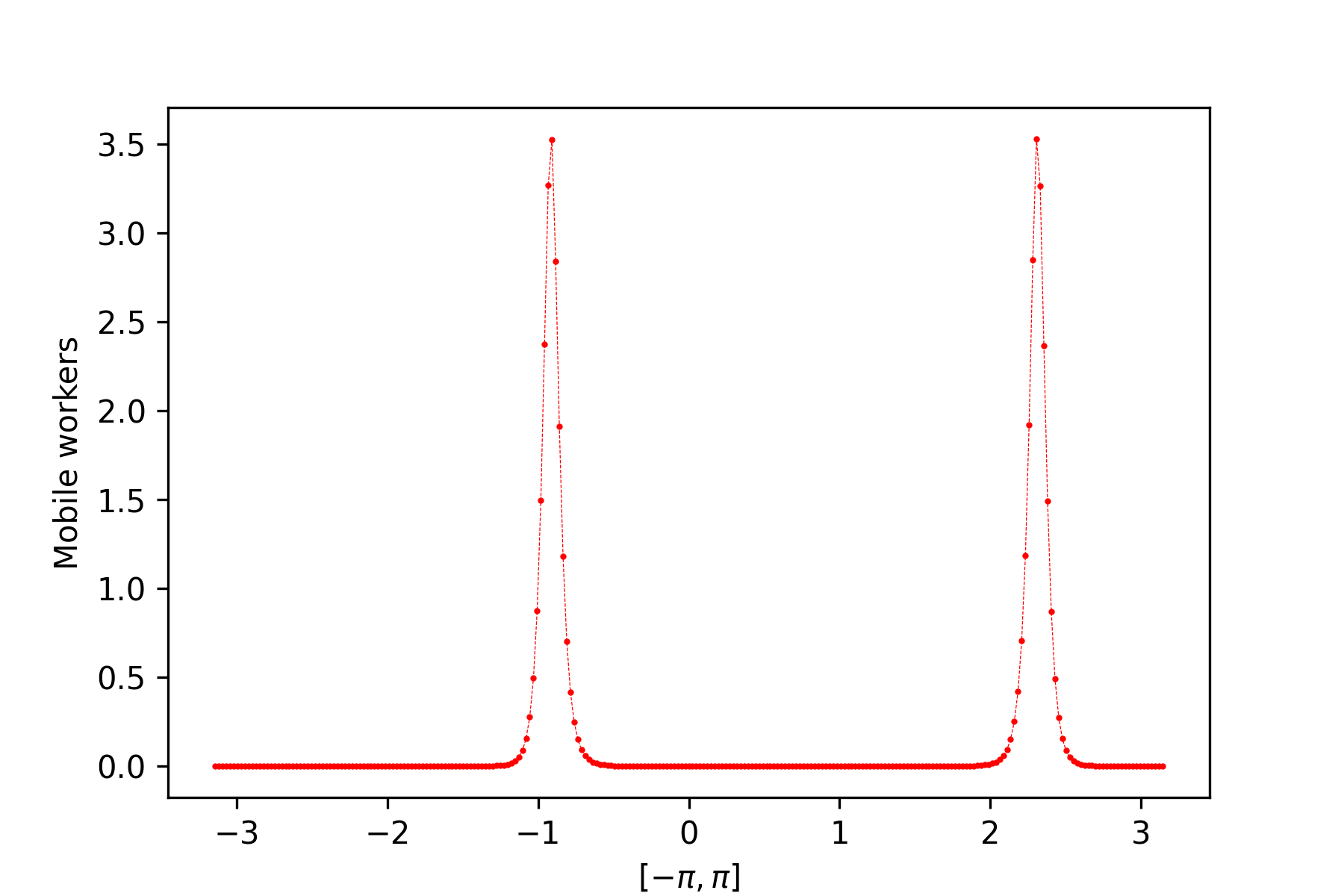}
  \caption{Mobile population density $\lambda^*$}
 \end{subfigure}
 \begin{subfigure}{0.5\columnwidth}
  \centering
  \includegraphics[width=\columnwidth]{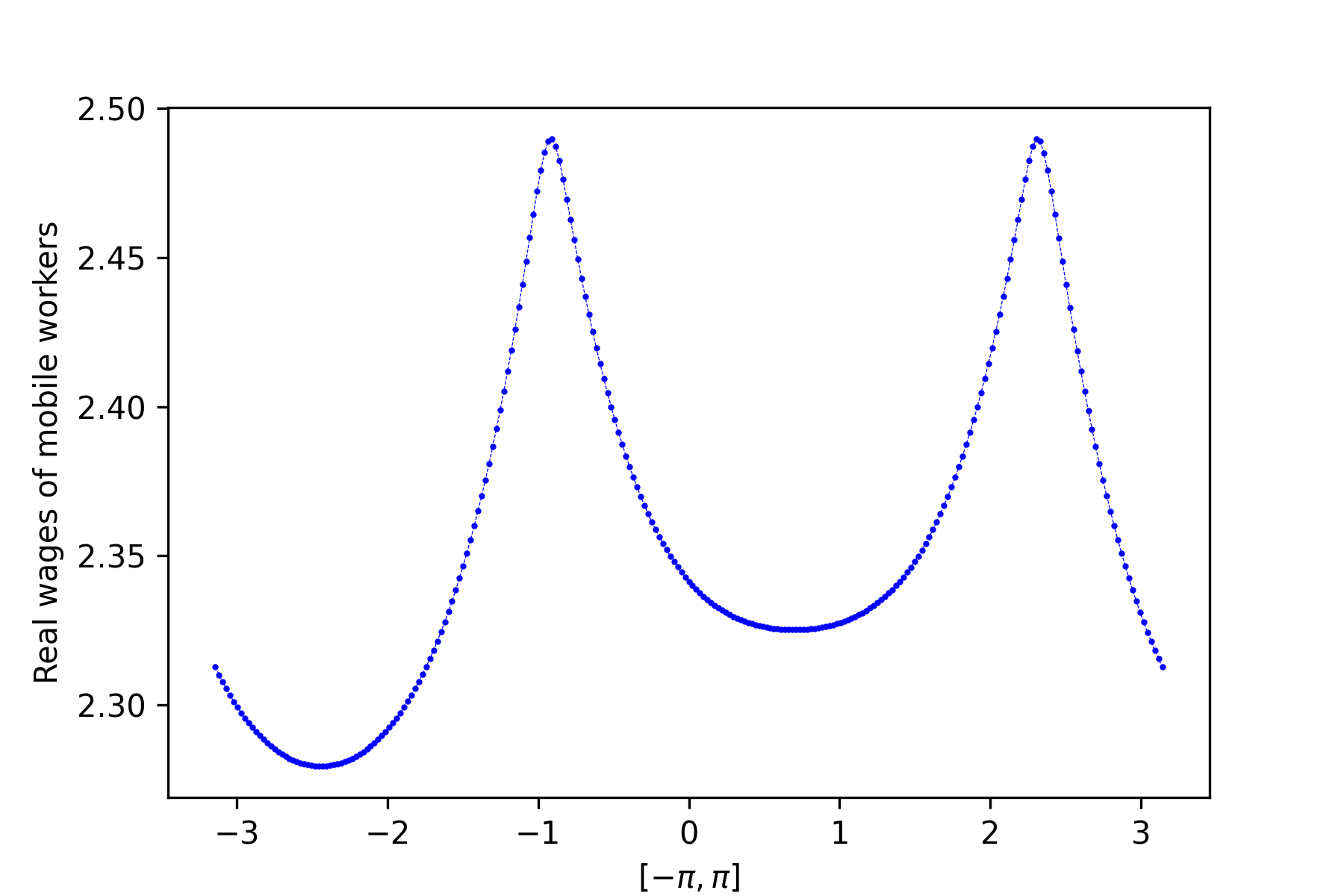}
  \caption{Real wage $\omega^*$}
 \end{subfigure}\\ 
 \caption{Numerical stationary solution for $(\sigma,\tau)=(2.4, 0.5)$}
 \label{fig:two-city-sigma}
\end{figure}

\begin{figure}[H]
 \begin{subfigure}{0.5\columnwidth}
  \centering
  \includegraphics[width=\columnwidth]{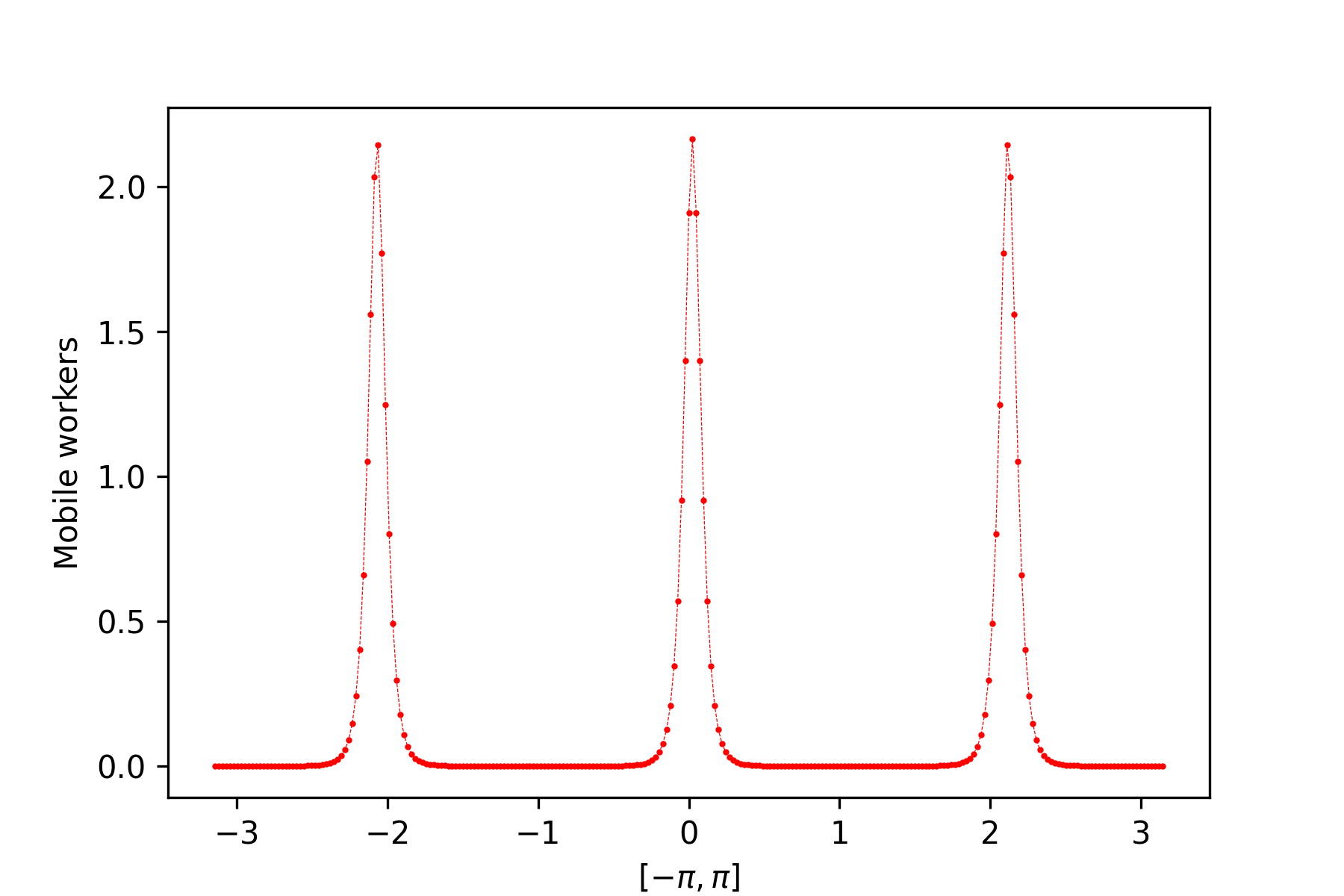}
  \caption{Mobile population density $\lambda^*$}
 \end{subfigure}
 \begin{subfigure}{0.5\columnwidth}
  \centering
  \includegraphics[width=\columnwidth]{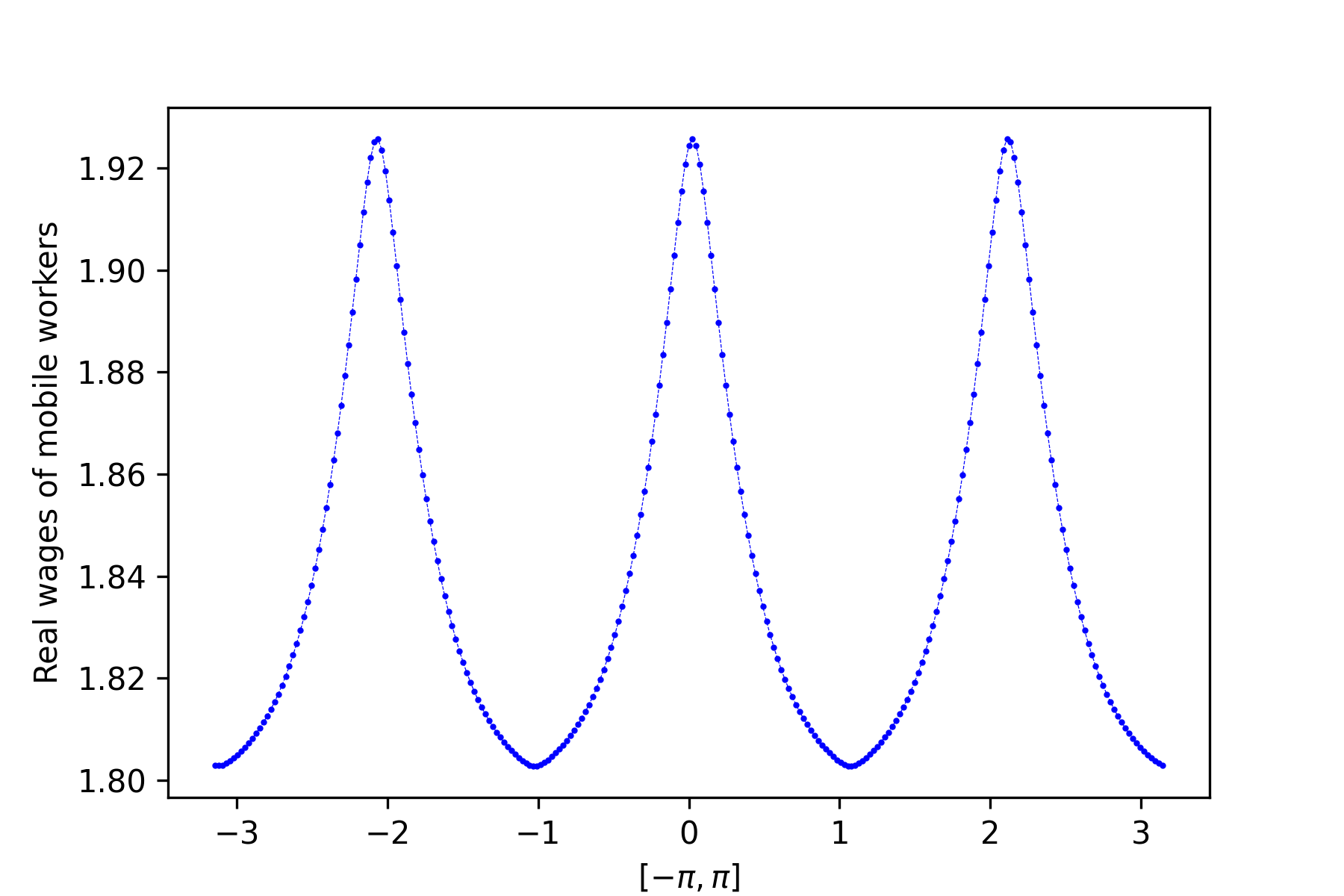}
  \caption{Real wage $\omega^*$}
 \end{subfigure}\\ 
 \caption{Numerical stationary solution for $(\sigma,\tau)=(3.0, 0.5)$}
 \label{fig:three-city-sigma}
\end{figure}

\begin{figure}[H]
 \begin{subfigure}{0.5\columnwidth}
  \centering
  \includegraphics[width=\columnwidth]{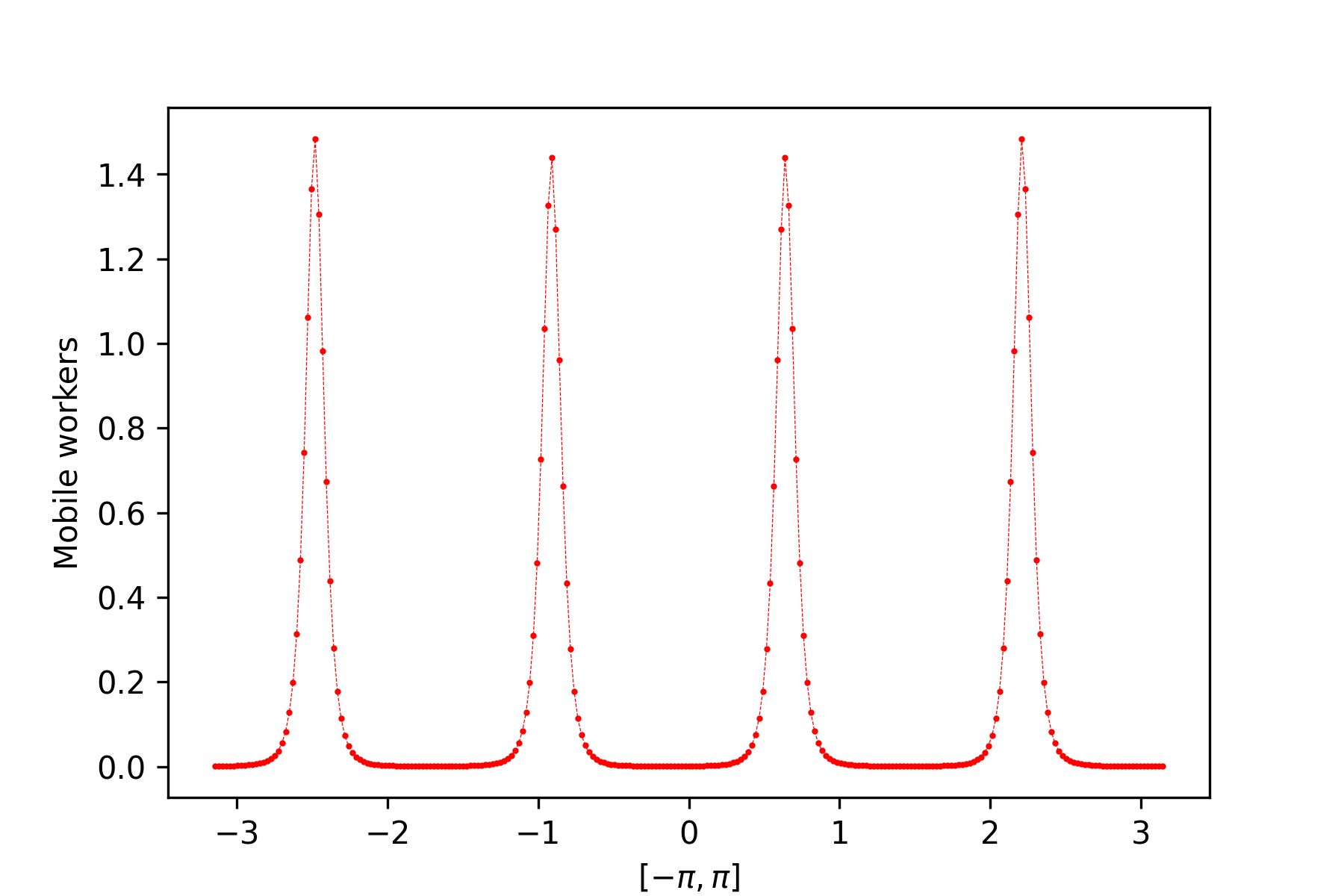}
  \caption{Mobile population density $\lambda^*$}
 \end{subfigure}
 \begin{subfigure}{0.5\columnwidth}
  \centering
  \includegraphics[width=\columnwidth]{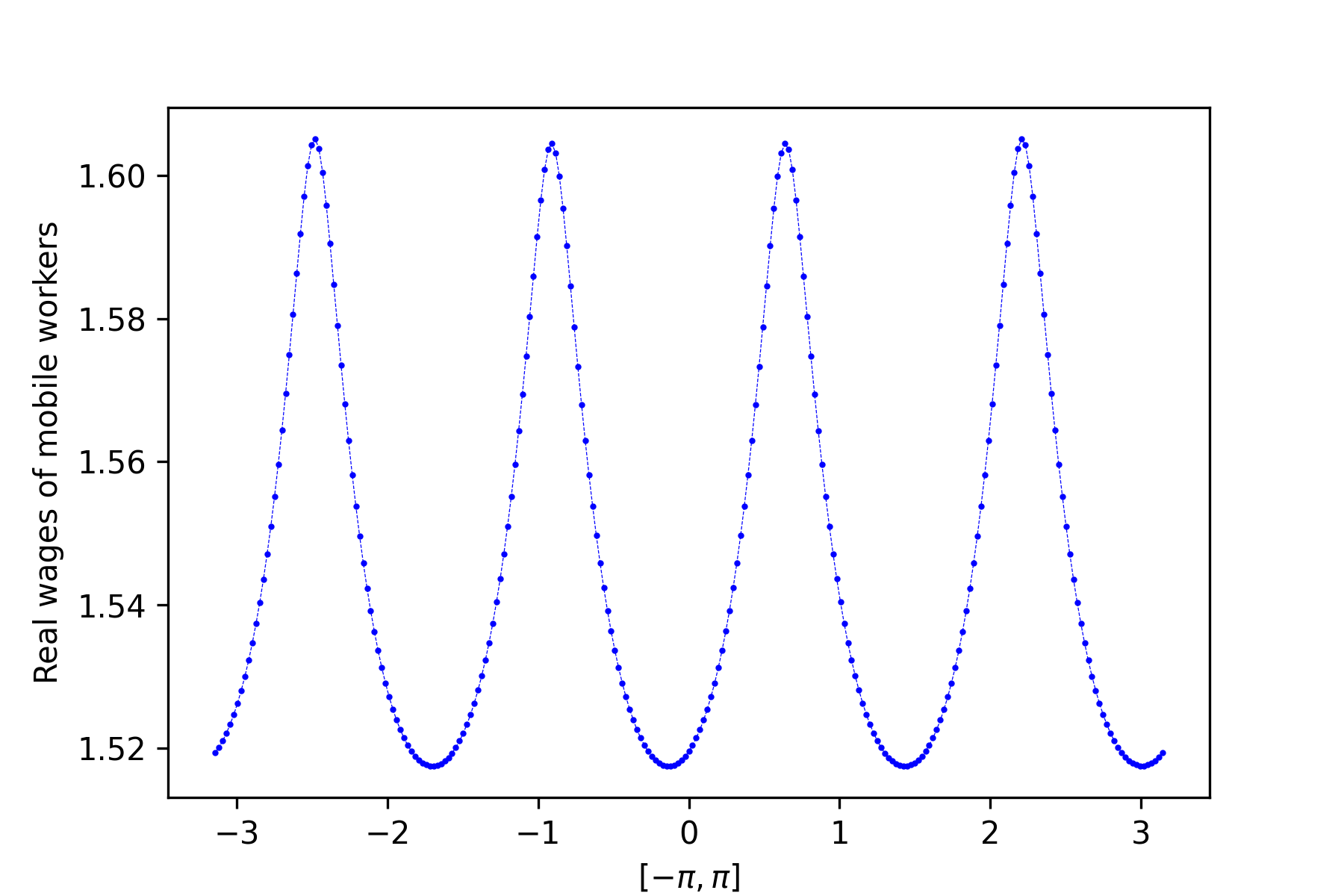}
  \caption{Real wage $\omega^*$}
 \end{subfigure}\\ 
 \caption{Numerical stationary solution for $(\sigma,\tau)=(3.5, 0.5)$}
 \label{fig:four-city-sigma}
\end{figure}

\begin{figure}[H]
 \begin{subfigure}{0.5\columnwidth}
  \centering
  \includegraphics[width=\columnwidth]{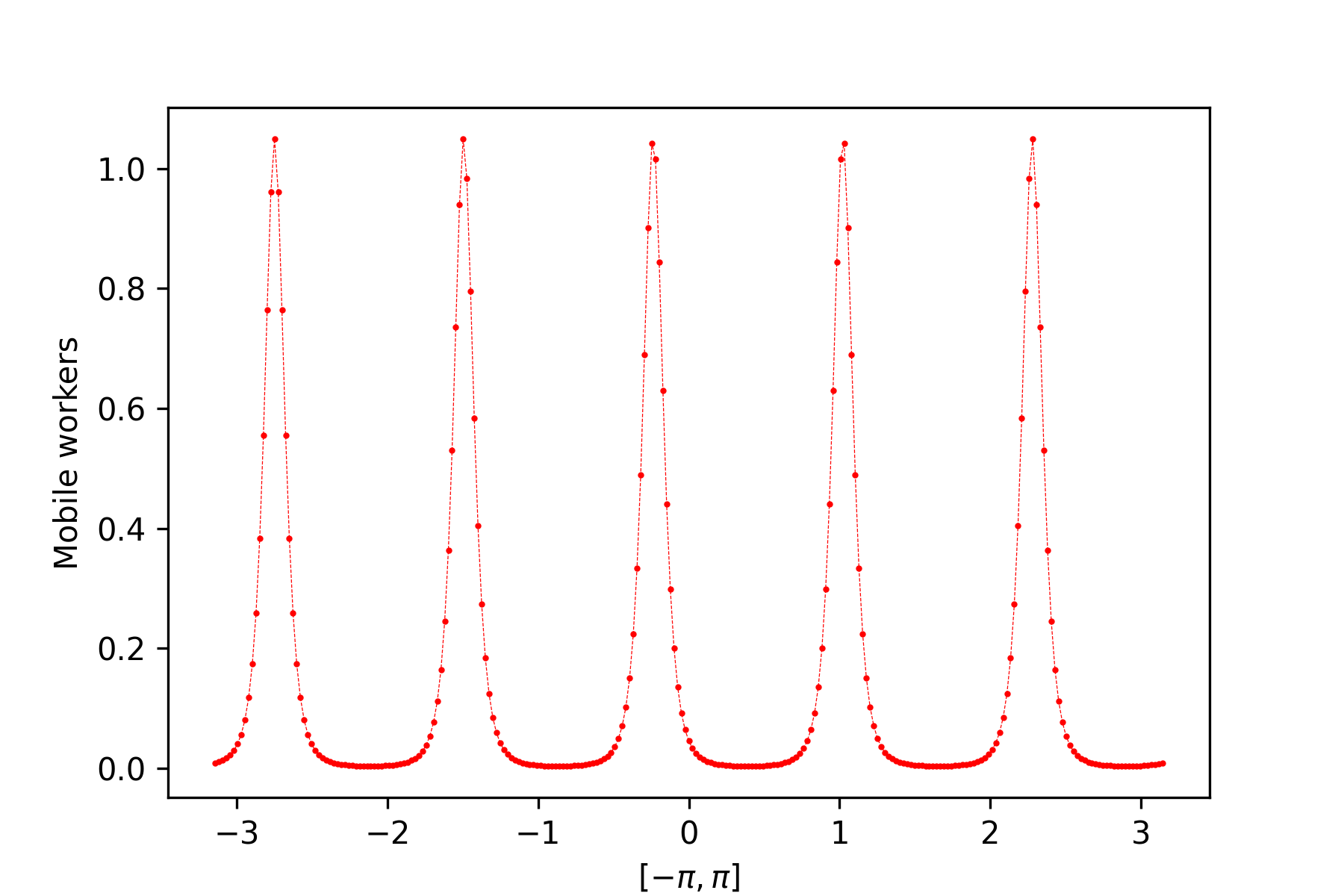}
  \caption{Mobile population density $\lambda^*$}
 \end{subfigure}
 \begin{subfigure}{0.5\columnwidth}
  \centering
  \includegraphics[width=\columnwidth]{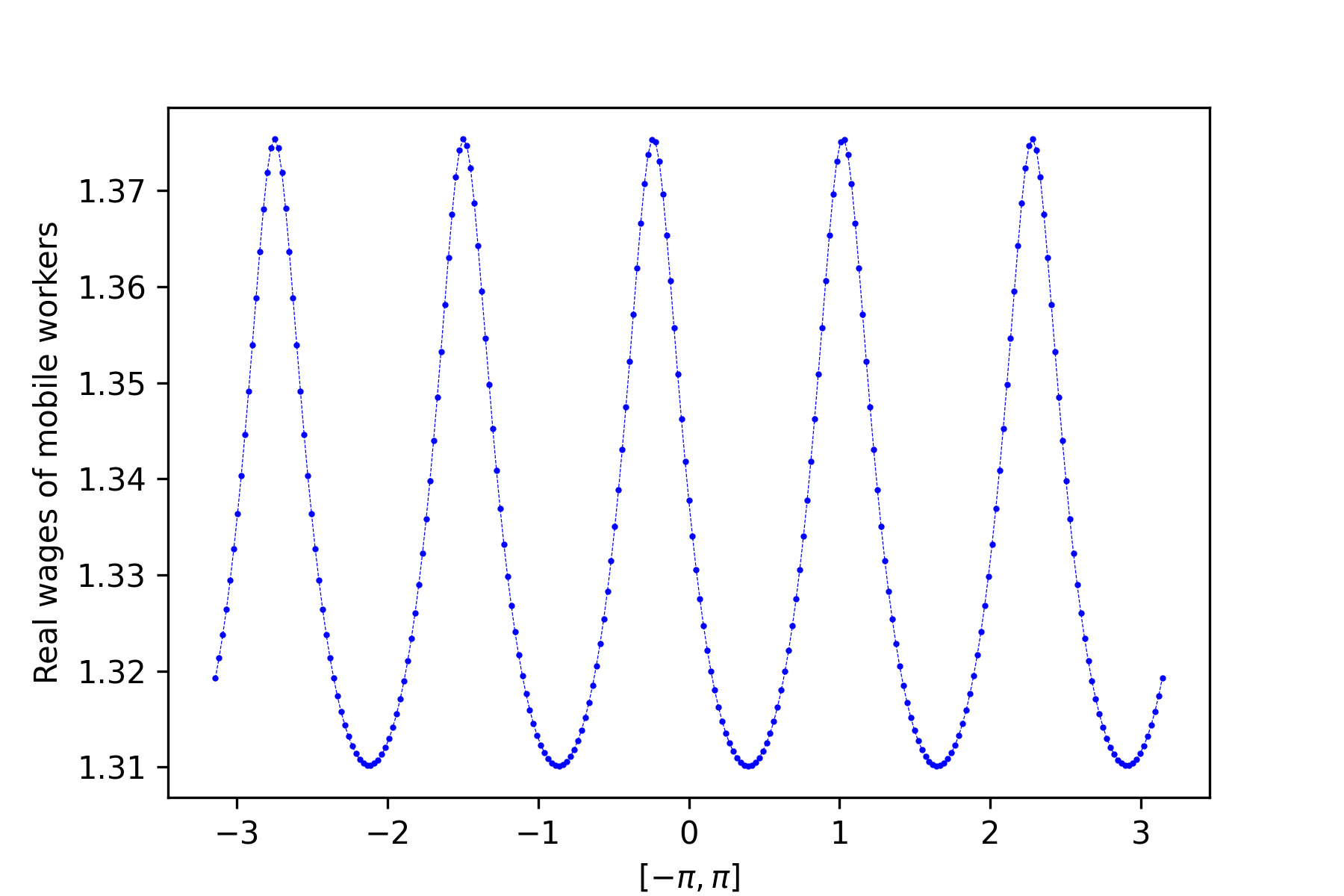}
  \caption{Real wage $\omega^*$}
 \end{subfigure}\\ 
 \caption{Numerical stationary solution for $(\sigma,\tau)=(4.0, 0.5)$}
 \label{fig:five-city-sigma}
\end{figure}

\footnotetext{The numerical simulations in this section were performed by using Julia (\citet{be2017julia}) ver. 1.10.2. The simulation code is available at \url{https://github.com/k-ohtake/advection-diffusion-neg}.}
The basic settings of the simulation are as follows. The parameters are set to $\mu=0.6$,  $a=0.5$, $d=0.005$, $F=1$, $\Lambda=1$, $\Phi=10$, and $\rho=1$. \footnote{The value of $\Phi$ must be sufficiently large for the demand of any consumer for agricultural goods to be positive. See \citet[pp. 911-912]{Ohtake2023cont} for more details. The diffusion coefficient $d$ must be sufficiently small for the numerical scheme of the diffusion equation to be stable.  Specifically, $d\leq0.5\frac{\Delta r^2}{\Delta t}$ is known to be a stability condition for explicit numerical schemes for linear one-dimensional diffusion equations (See \citet[p.~56]{TaAnPl}). In the case of the settings $\Delta r=2\pi/256$ and $\Delta t=0.01$ in this paper, it must be at least $d\leq 0.03$. In the case of this nonlinear model, $d$ must be smaller than this condition for the calculation to converge. The author confirmed that numerical computations could be performed stably even at $d = 0.01$ and that qualitatively similar results to those obtained at $d = 0.005$ could be obtained.} An initial distribution of the population $\lambda^0$ is given by adding a randomly generated small perturbation $\varDelta \lambda^0$ to the homogeneous state as
\[
\lambda^0 = \ol{\lambda} + \varDelta \lambda^0.
\]
Since \eqref{intzero}, the initial perturbation must satisfy
\[
\int_S \varDelta\lambda(0, x)dx \fallingdotseq\sum_{j=1}^N \varDelta\lambda^0_j\rho\varDelta r \fallingdotseq 0.
\]
Starting from an initial value $\lambda^0$, an approximated population density $\lambda^n$ for $n=1,2,3,\cdots$ is computed one after the other according to the explicit numerical scheme \eqref{exm}. At each iteration, for the difference
\[
\delta\lambda^n :=\lambda^n-\lambda^{n-1} \in \mathbb{R}^N
\]
the maximum norm 
\[
\left\|\delta\lambda^n\right\|_\infty := \max\left\{|\delta\lambda^n_1|,|\delta\lambda^n_2|,\cdots,|\delta\lambda^n_N|\right\}
\]
is computed. When $\left\|\delta\lambda^{n+1}\right\|_\infty < 10^{-11}$ is satisfied, the numerical solution is regarded as reaching a sort of stationary solution consiststs of $\lambda^*=\lambda^{n+1}$ and corresponding real wage $\omega^*=\omega^{n+1}$, and the computation starting from the initial value is stopped.\footnote{In rare cases, the iterations do not converge for any length of time. This may be due to the diffusion working against the convergence of numerical solutions to stationary states. Figs.~\ref{fig:one-city-tau}-\ref{fig:five-city-sigma} show only the convergence results for the condition $\left\|\delta\lambda^{n+1}\right\|_\infty < 10^{-11}$.} We regard the state $(\lambda^*, \omega^*)$ thus obtained as a stationary solution to the model. Of course, this is an approximate solution, so it is referred to hereafter as the ``numerical stationary solution'' to distinguish it from an analytical one.

Figs.~\ref{fig:one-city-tau}-\ref{fig:five-city-sigma} that follow show the numerical stationary solutions obtained under various values of the control parameters $\sigma$ and $\tau$. Figs.~\ref{fig:one-city-tau} - \ref{fig:five-city-tau} show stationary states when $\sigma=5$ is fixed and $\tau$ is varied from $0.05$, $0.15$, $0.25$, $0.35$, and $0.37$. Similarly, Figs.~\ref{fig:one-city-sigma} - \ref{fig:five-city-sigma} show stationary states when $\tau=0.5$ is fixed and $\sigma$ is varied from $1.3$, $2.4$, $3.0$, $3.5$, and $4.0$. In each figure, we observe that the mobile population forms several clusters, which might be referred to as {\it urban areas} that enjoy higher real wages than non-urban areas.\footnote{Under $d = 0.01$, the parameter values for the number of urban areas being one, two, three, four, five, and six are $\tau=0.06$, $0.3$, $0.4$, $0.6$, $0.7$, and $0.8$, respectively, when $\sigma$ is fixed to $3.5$, and $\sigma=1.1$, $1.9$, $2.25$, $2.3$, $2.6$, and $3.0$, respectibely, when $\tau$ is fixed to $1.0$.} To investigate the relation between the control parameters and the number of urban areas, multiple simulations were performed while changing the control parameters. Under the same value of the control parameter, the number of urban areas formed in the numerical stationary solution varies slightly depending on the initial values. Then, to ensure the robustness of the results, five simulations were performed for a single pair of parameters $(\sigma, \tau)$, each with initial values generated by independent random numbers. As a result, it has been found that the maximum number of urban areas obtained in five simulations has a very robust relationship with the value of the control parameters. Figs.~\ref{fig:one-city-tau}-\ref{fig:five-city-sigma} show the numerical stationary solution when the maximum number of cities is obtained under each pair of the control parameters. As shown in these figures, the maximum number of urban areas decreases with decreasing transport costs and a strengthening preference for variety.\footnote{It should be emphasized that this is only applicable in cases where a destabilized mode exists.} In general, the number of urban areas does not necessarily decrease monotonically as $\tau$ or $\sigma$ decreases. For example, in Fig.~\ref{figs:cc_sigma}, there is a range of transport costs near $\tau=0.08$ where all modes are stabilized. This is consistent with the results of the analysis of the linearized equations, in which the absolute value of the frequency of the unstable mode is reduced due to lower transport costs and enhanced preference for variety. This means that the same NEG mechanism is at work in the large-time behavior of the solution as in the early stages of destabilization; 1) low transport costs allow firms to agglomerate and enjoy increasing returns without losing demand from immobile consumers, and 2) strong preference for variety leads to greater incentive for consumers to migrate to regions having large agglomerations of firms.

\begin{figure}[H]
 \begin{subfigure}{0.5\columnwidth}
  \centering
  \includegraphics[width=\columnwidth]{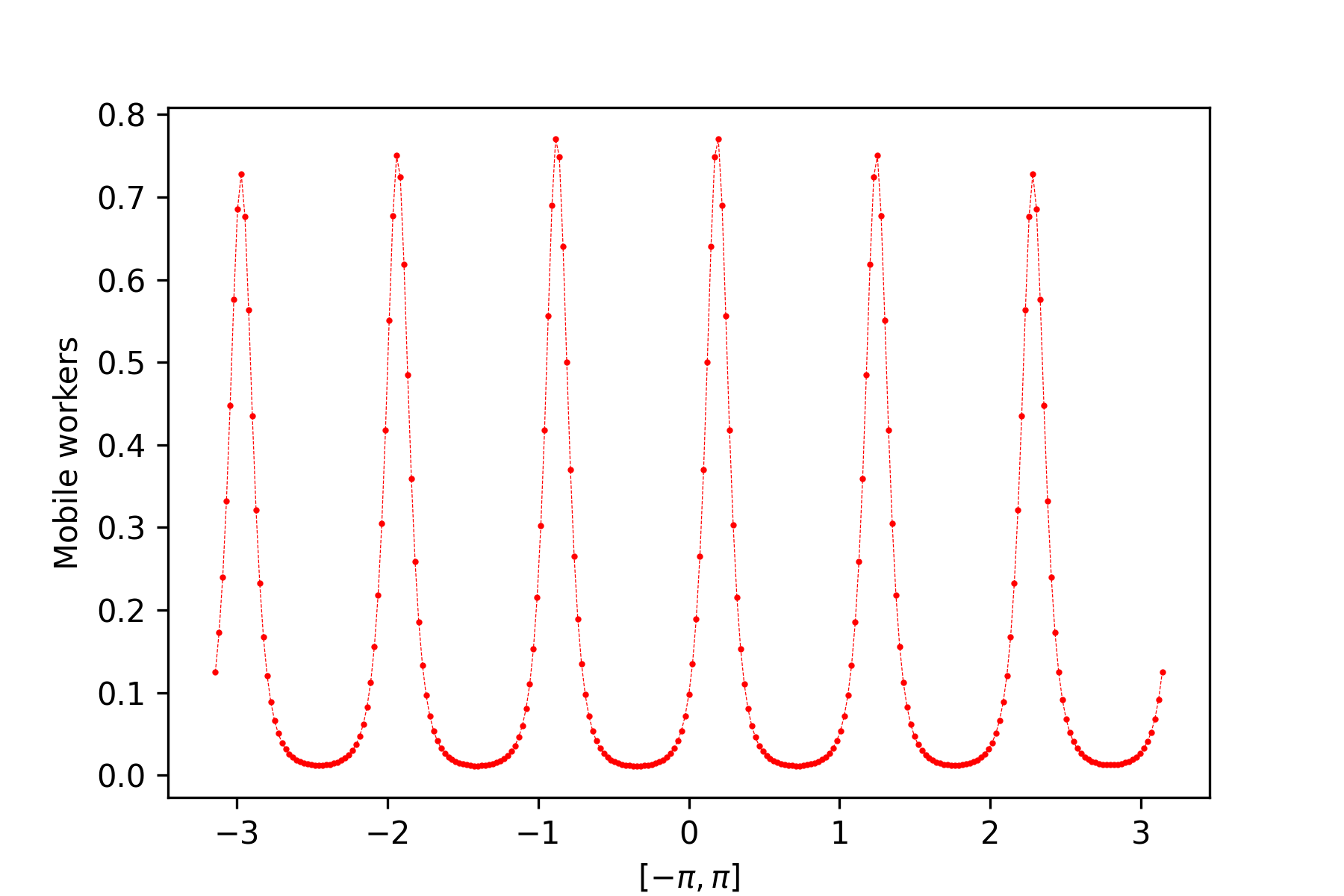}
  \caption{Mobile population density $\lambda^*$}
 \end{subfigure}
 \begin{subfigure}{0.5\columnwidth}
  \centering
  \includegraphics[width=\columnwidth]{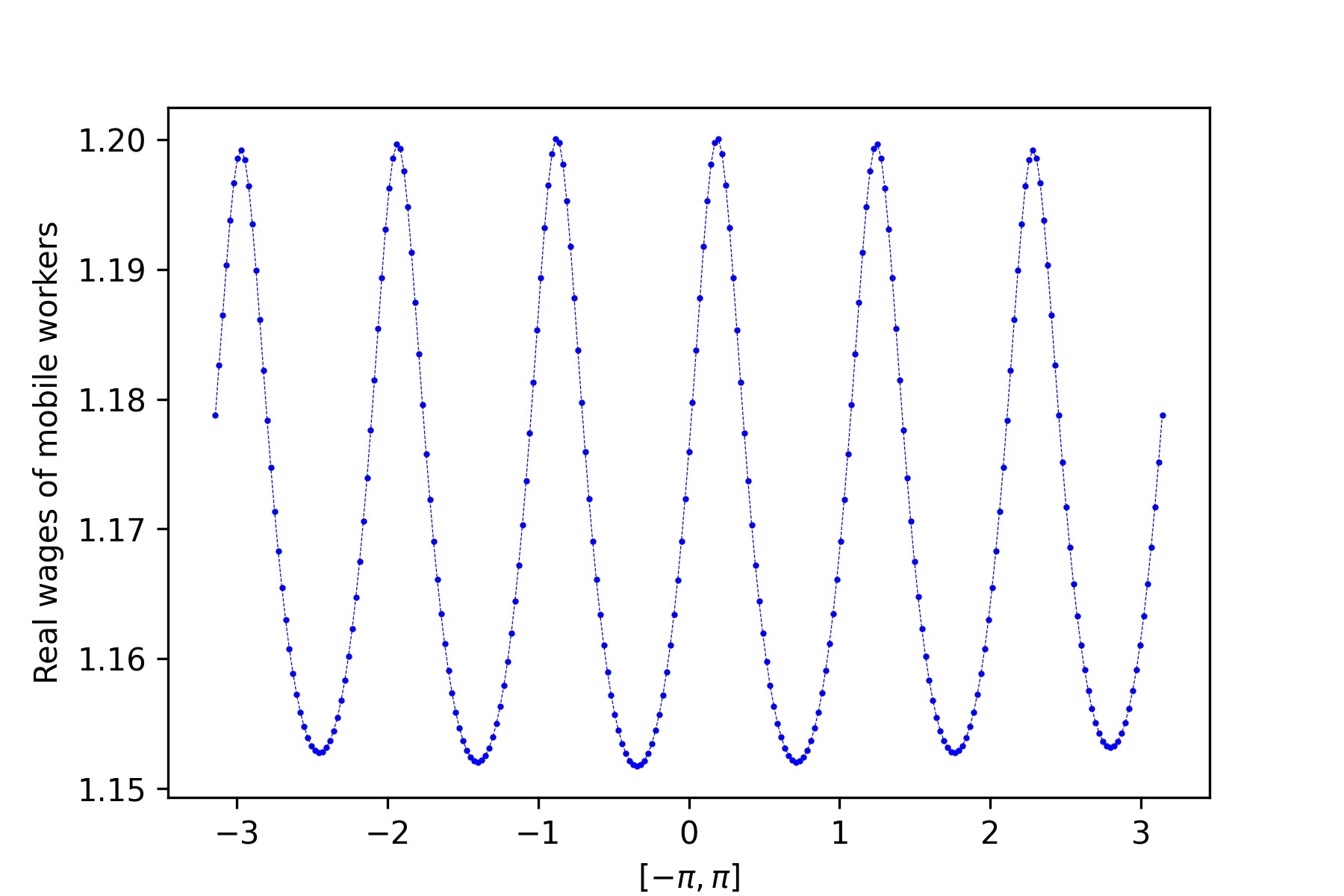}
  \caption{Real wage $\omega^*$}
 \end{subfigure}\\ 
 \caption{Numerical stationary solution for $(\sigma,\tau)=(4.5, 0.5)$}
 \label{fig:six-city-sigma}
\end{figure}

\section{Comparison with similar models}\label{sec:comparison}
Let us compare the results with those obtained from similar models. Each has either the CP or QLLU for its market equilibrium equation, and either the replicator or advection-diffusion for its dynamics, resulting in a total of four models. In the following, the abbreviations shown in the table below will be used for each combination. Hence, the model proposed in this study is written as QLLU-AD.\footnote{CP-R is considered in \citet{OhtakeYagi_Asym}, QLLU-R is considered in \citet{Ohtake2023cont}, and CP-AD is considered in \citet{Moss} and \citet{MossMar2004}. In what follows, the replicator equation refers to \eqref{repl} and the advection-diffusion equation refers to \eqref{PDErtheta} even in these similar models.}
\begin{table}[H]
\centering
  \begin{tabular}{lcc}
  \multicolumn{1}{l|}{} & Replicator & \multicolumn{1}{c}{Advection-Diffusion} \\ \cline{1-3}
  \multicolumn{1}{l|}{CP model} & CP-R & \multicolumn{1}{c}{CP-AD} \\ 
  \multicolumn{1}{l|}{QLLU model} & QLLU-R & \multicolumn{1}{c}{QLLU-AD}
  \end{tabular}
  \caption{Four models}
  \label{tab:similarmodels}
\end{table}

\subsection{Critical curves}
The eigenvalues of the Fourier modes around the homogeneous stationary solution can be computed for other models using the same method as for QLLU-AD. See Appendix for specific eigenvalue equations. In Figs.~\ref{figs:cpr_hm_sigma}-\ref{figs:qllur_hm_sigma}, the eigenvalues of each model are visualized for $\tau>0.01$ and $\sigma>1.01$ by heatmaps.\footnote{In CP-R and QLLU-R, the parameters other than $\tau$ and $\sigma$ are fixed to $\mu=0.6$, $\rho=1$, and $v=1$. In CP-AD, these parameters are fixed at $\mu=0.6$, $\rho=1$, $a=0.5$, and $d=0.005$. Numpy \citep{numpy} and Matplotlib \citep{matplotlib} are used to compute eigenvalues. The source code is available at \url{https://github.com/k-ohtake/advection-diffusion-neg}.}

\begin{figure}[H]
\begin{subfigure}[H]{0.5\columnwidth}
\includegraphics[width=\columnwidth]{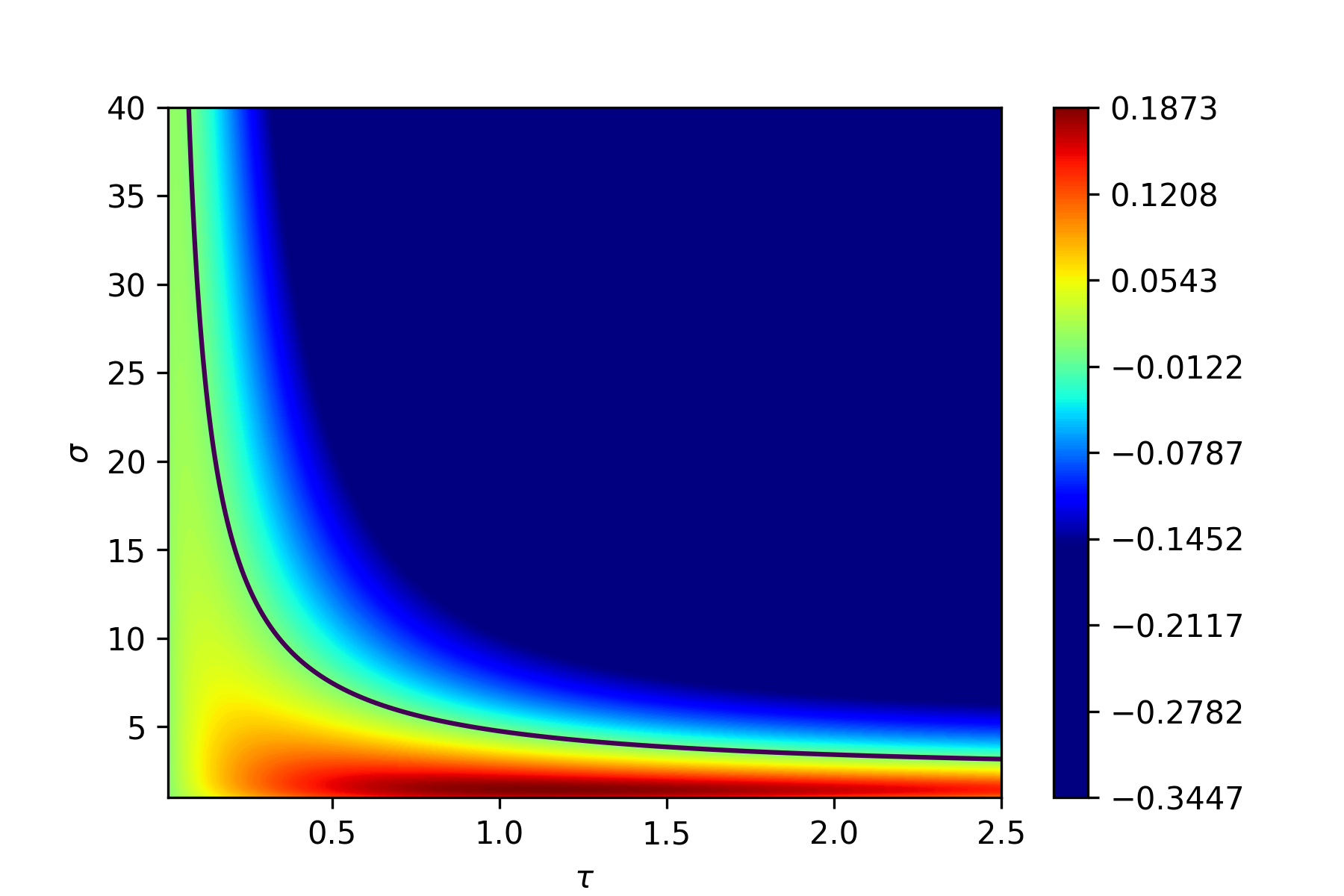}
\caption{$k=1$}\label{hm_1}
\end{subfigure}
\begin{subfigure}[H]{0.5\columnwidth}
\includegraphics[width=\columnwidth]{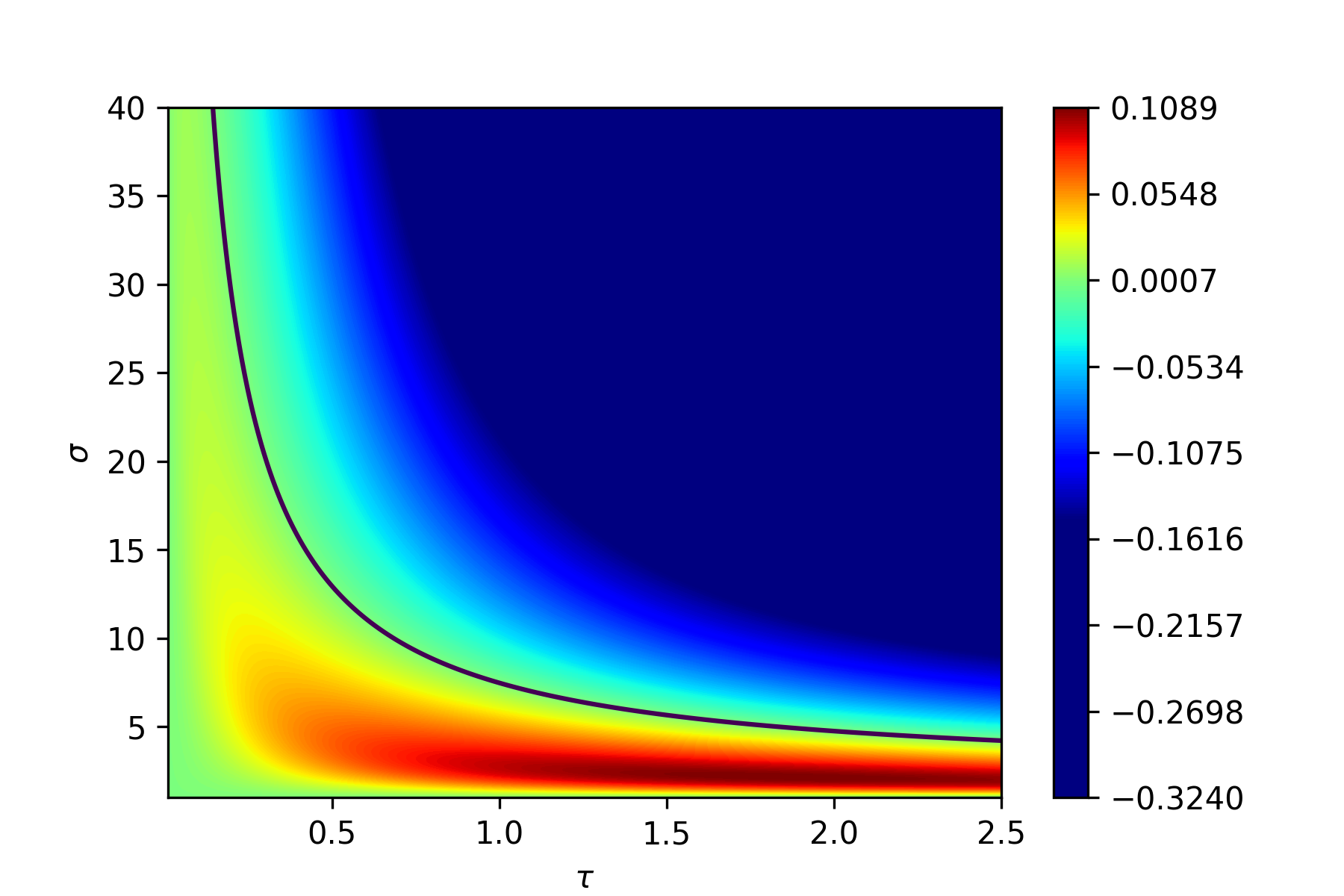}
\caption{$k=2$}\label{hm_2}
\end{subfigure}
\begin{subfigure}[H]{0.5\columnwidth}
\includegraphics[width=\columnwidth]{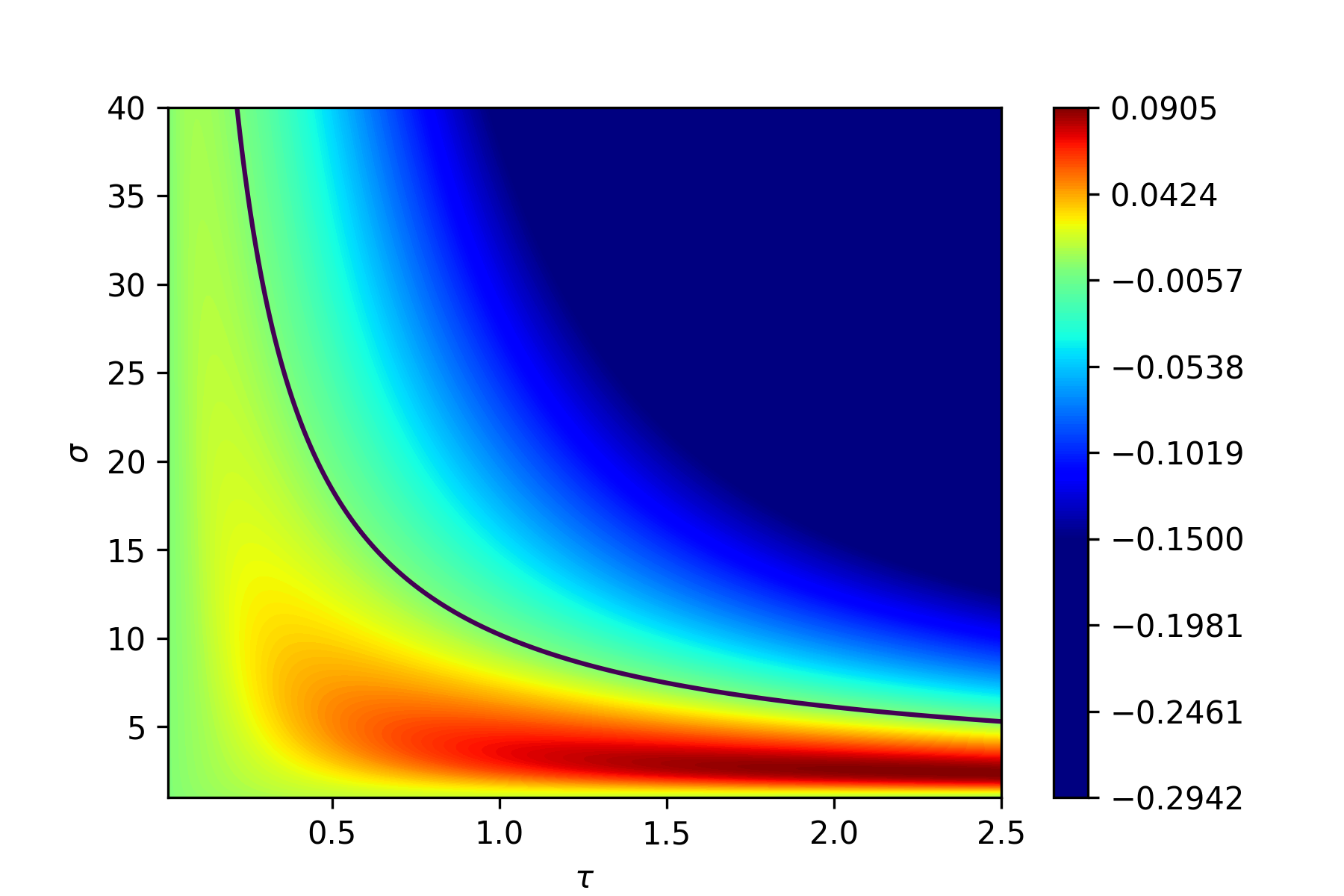}
\caption{$k=3$}\label{hm_3}
\end{subfigure}
\begin{subfigure}[H]{0.5\columnwidth}
\includegraphics[width=\columnwidth]{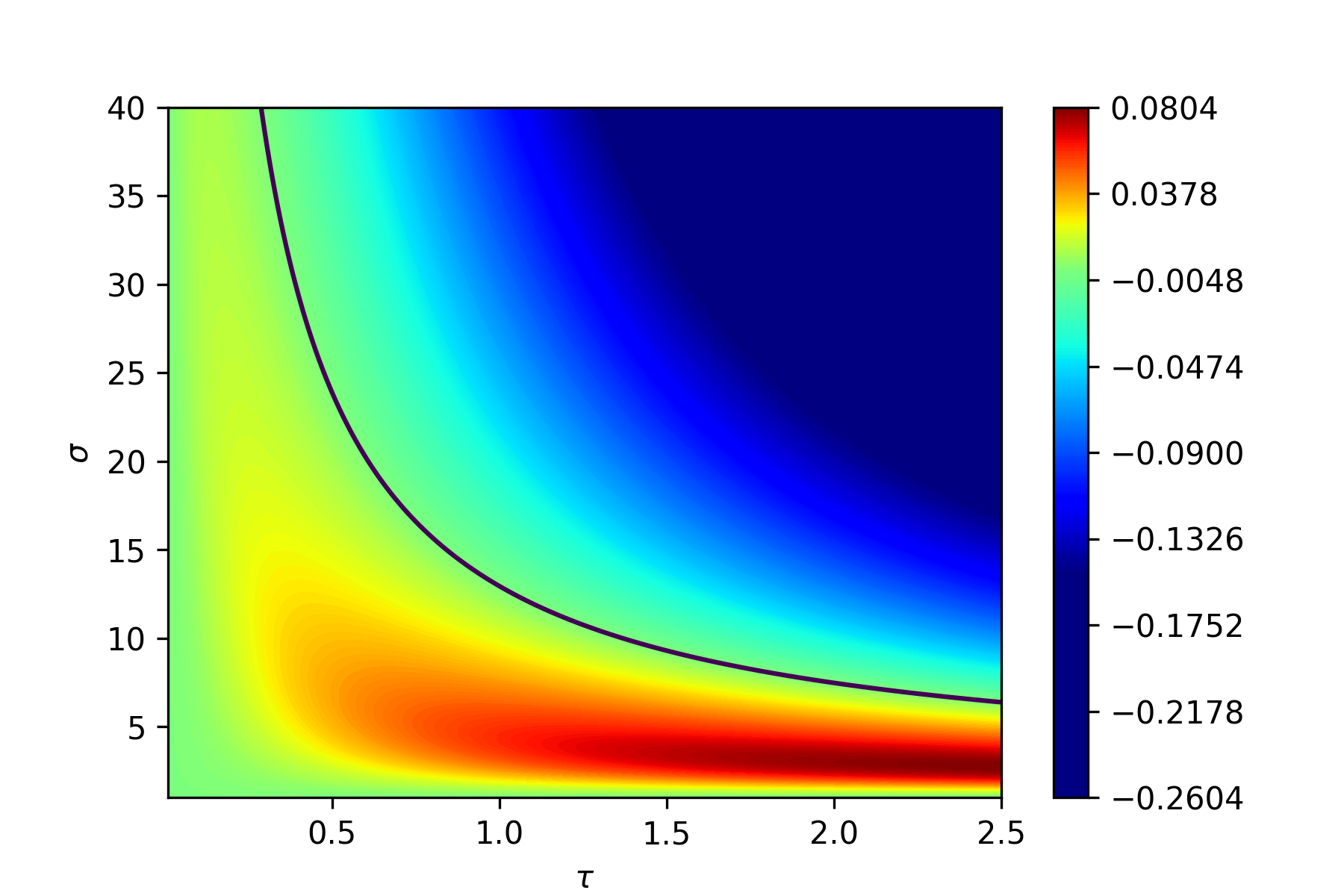}
\caption{$k=4$}\label{hm_4}
\end{subfigure}
\begin{subfigure}[H]{0.5\columnwidth}
\includegraphics[width=\columnwidth]{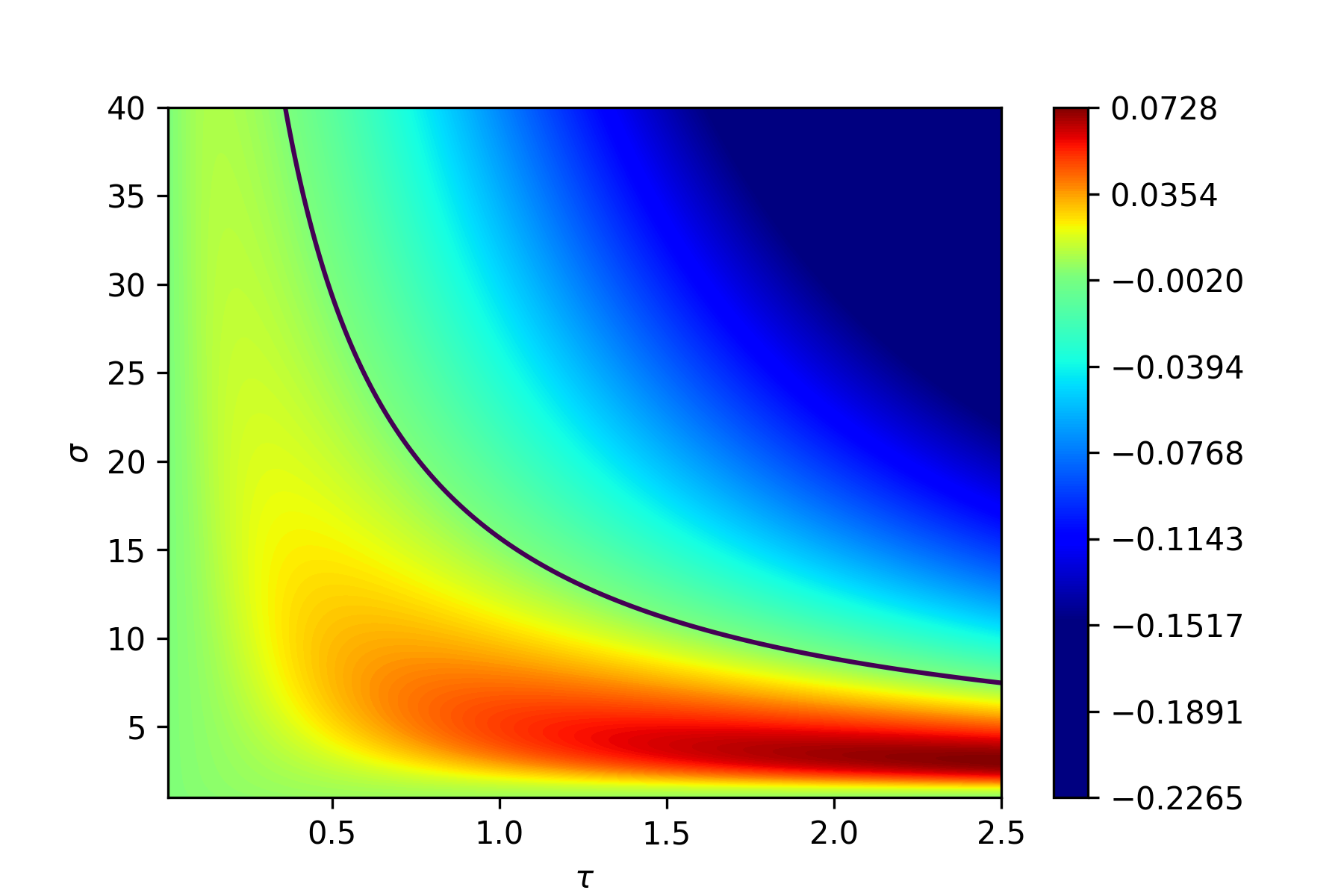}
\caption{$k=5$}\label{hm_5}
\end{subfigure}
\begin{subfigure}[H]{0.5\columnwidth}
\includegraphics[width=\columnwidth]{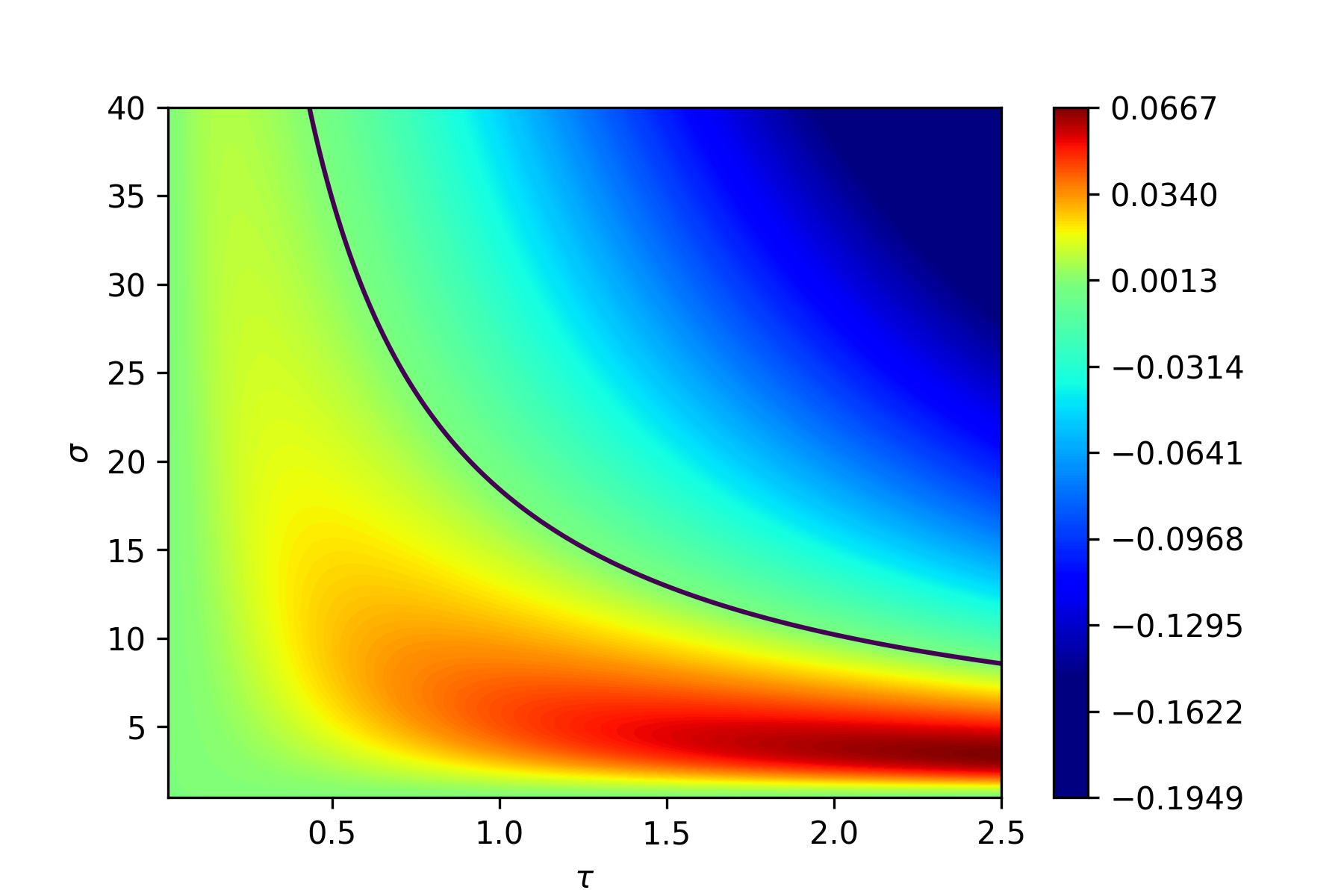}
\caption{$k=6$}\label{hm_6}
\end{subfigure}
\caption{Heatmaps of $\Gamma_k$ of CP-R in $(\tau, \sigma)$-plane}\label{figs:cpr_hm_sigma}
\end{figure}

\begin{figure}[H]
\begin{subfigure}[H]{0.5\columnwidth}
\includegraphics[width=\columnwidth]{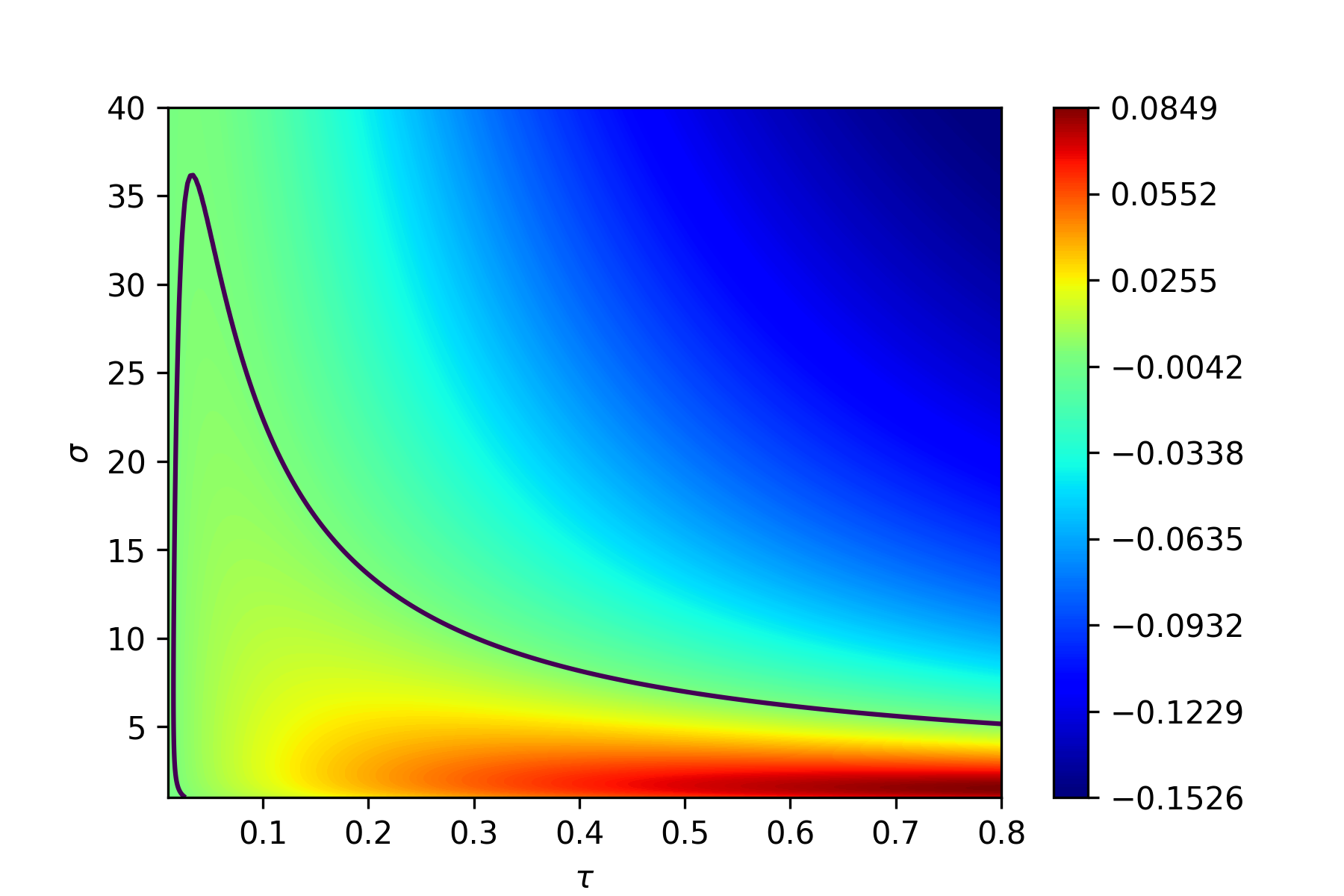}
\caption{$k=1$}\label{hm_1}
\end{subfigure}
\begin{subfigure}[H]{0.5\columnwidth}
\includegraphics[width=\columnwidth]{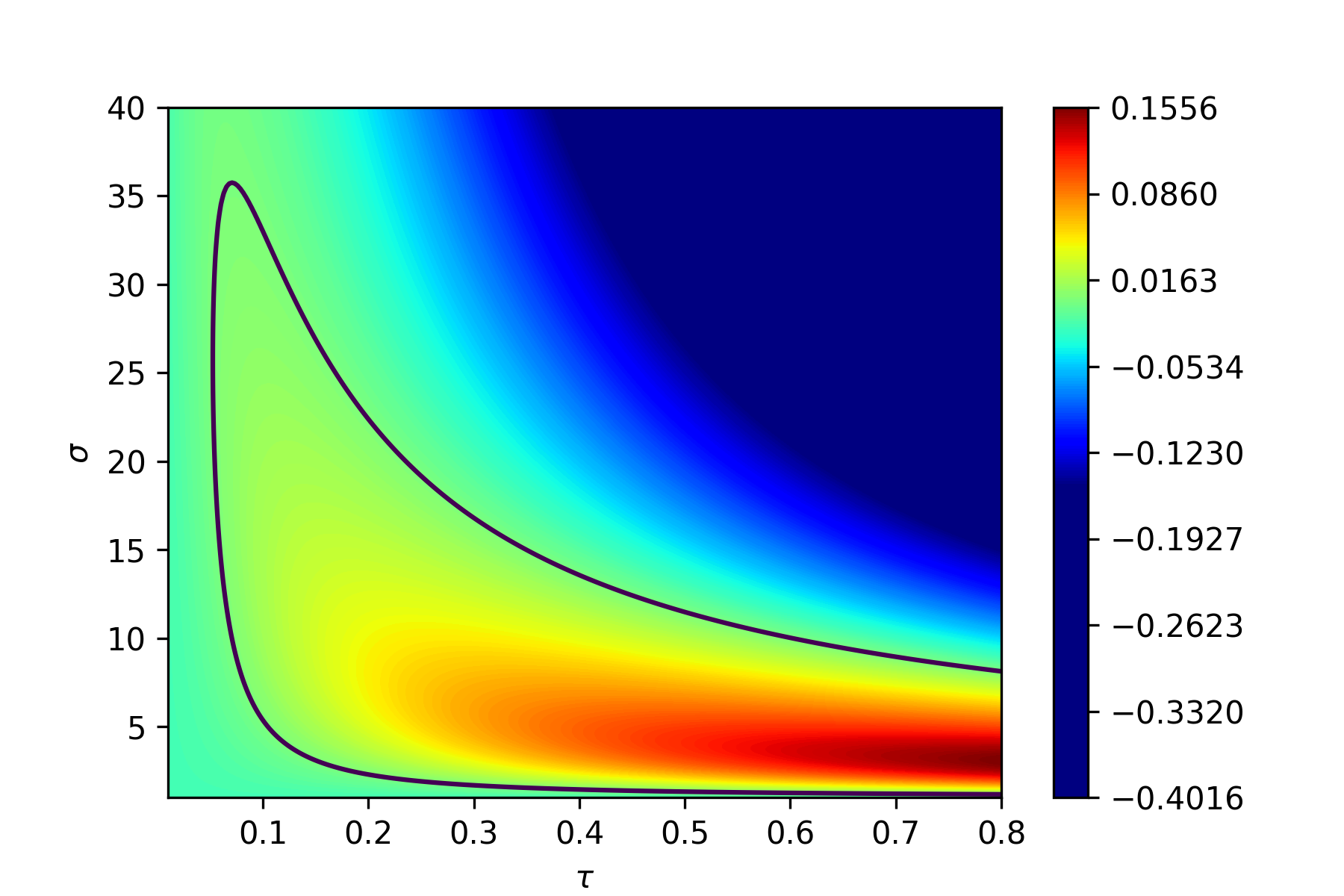}
\caption{$k=2$}\label{hm_2}
\end{subfigure}
\begin{subfigure}[H]{0.5\columnwidth}
\includegraphics[width=\columnwidth]{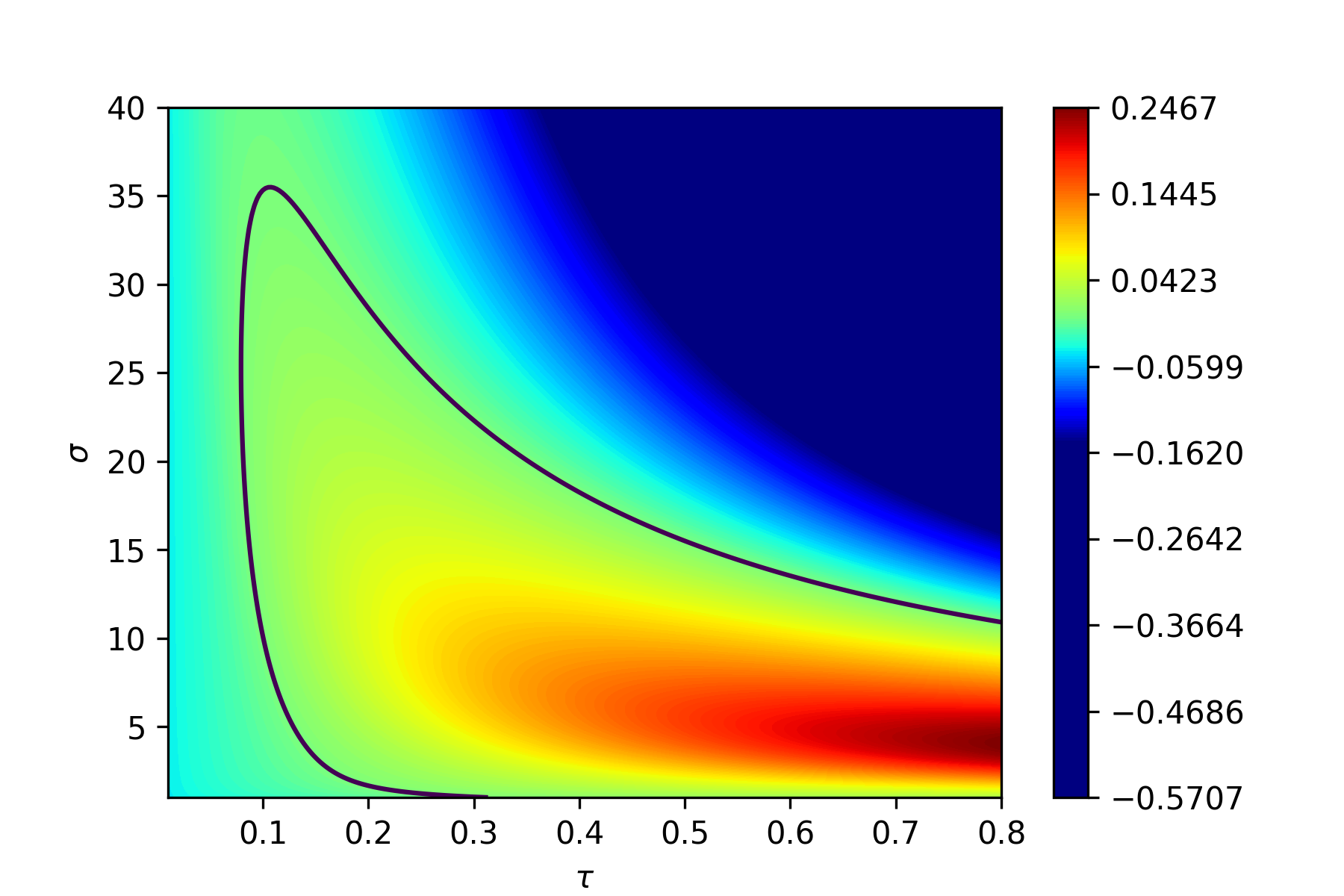}
\caption{$k=3$}\label{hm_3}
\end{subfigure}
\begin{subfigure}[H]{0.5\columnwidth}
\includegraphics[width=\columnwidth]{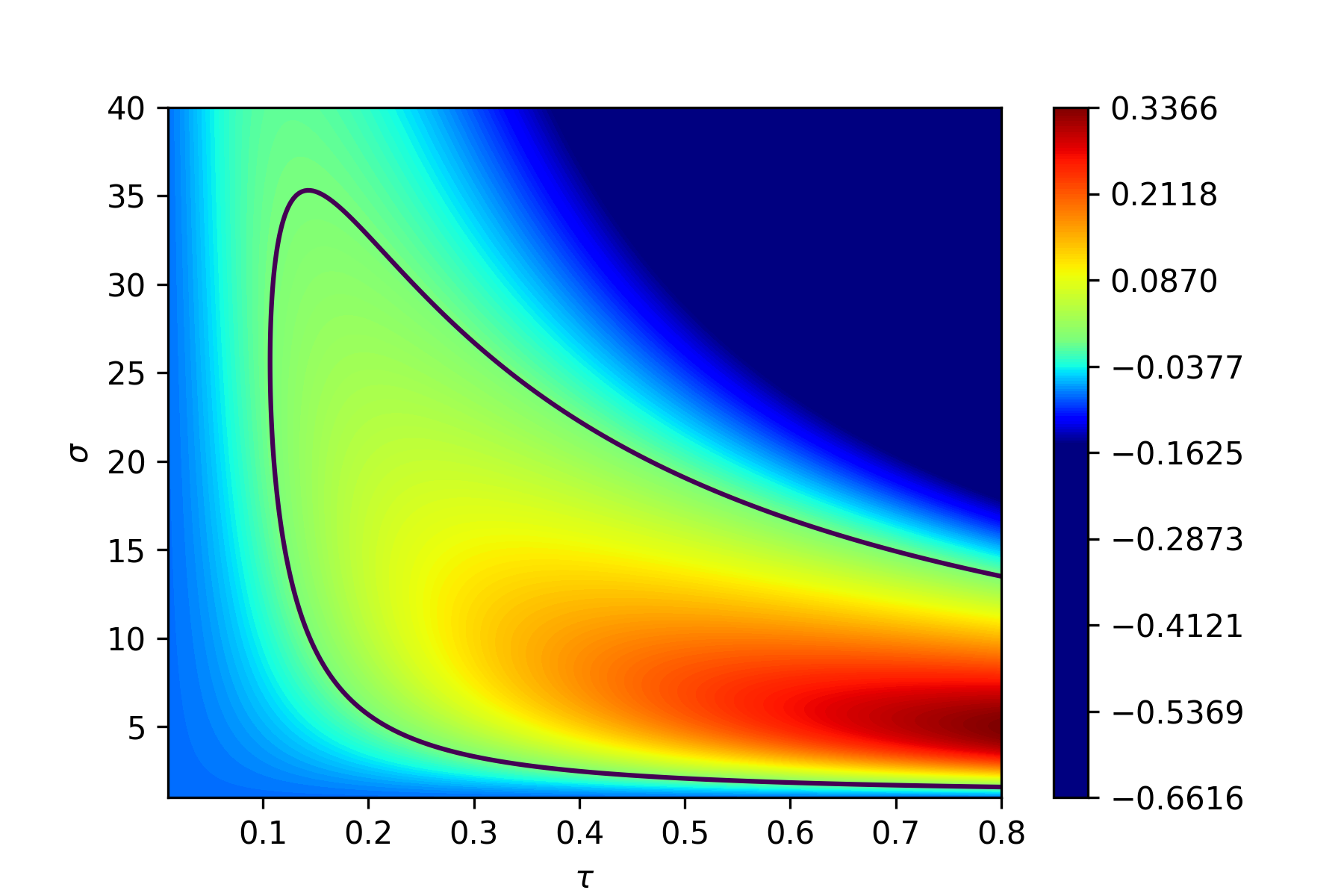}
\caption{$k=4$}\label{hm_4}
\end{subfigure}
\begin{subfigure}[H]{0.5\columnwidth}
\includegraphics[width=\columnwidth]{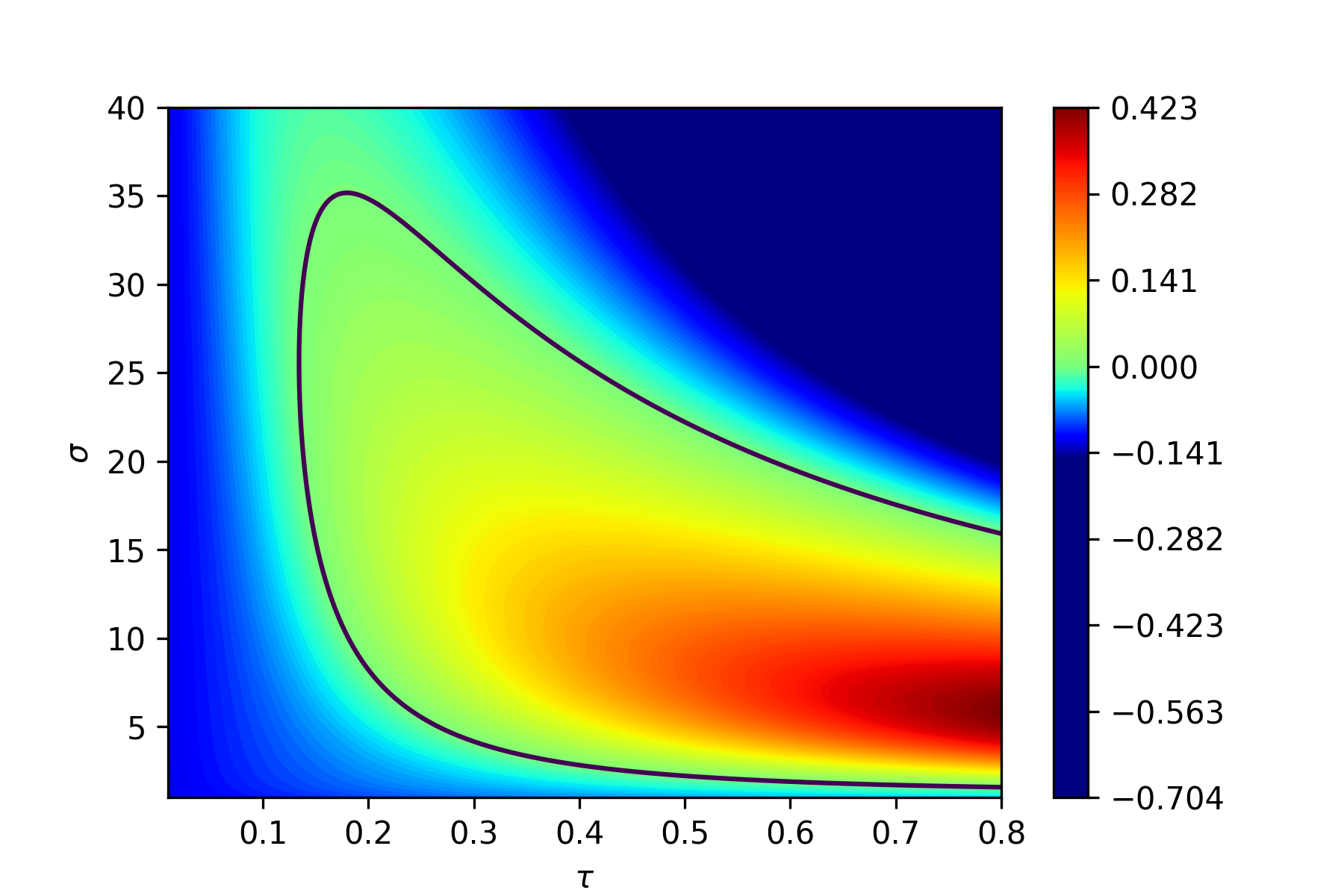}
\caption{$k=5$}\label{hm_5}
\end{subfigure}
\begin{subfigure}[H]{0.5\columnwidth}
\includegraphics[width=\columnwidth]{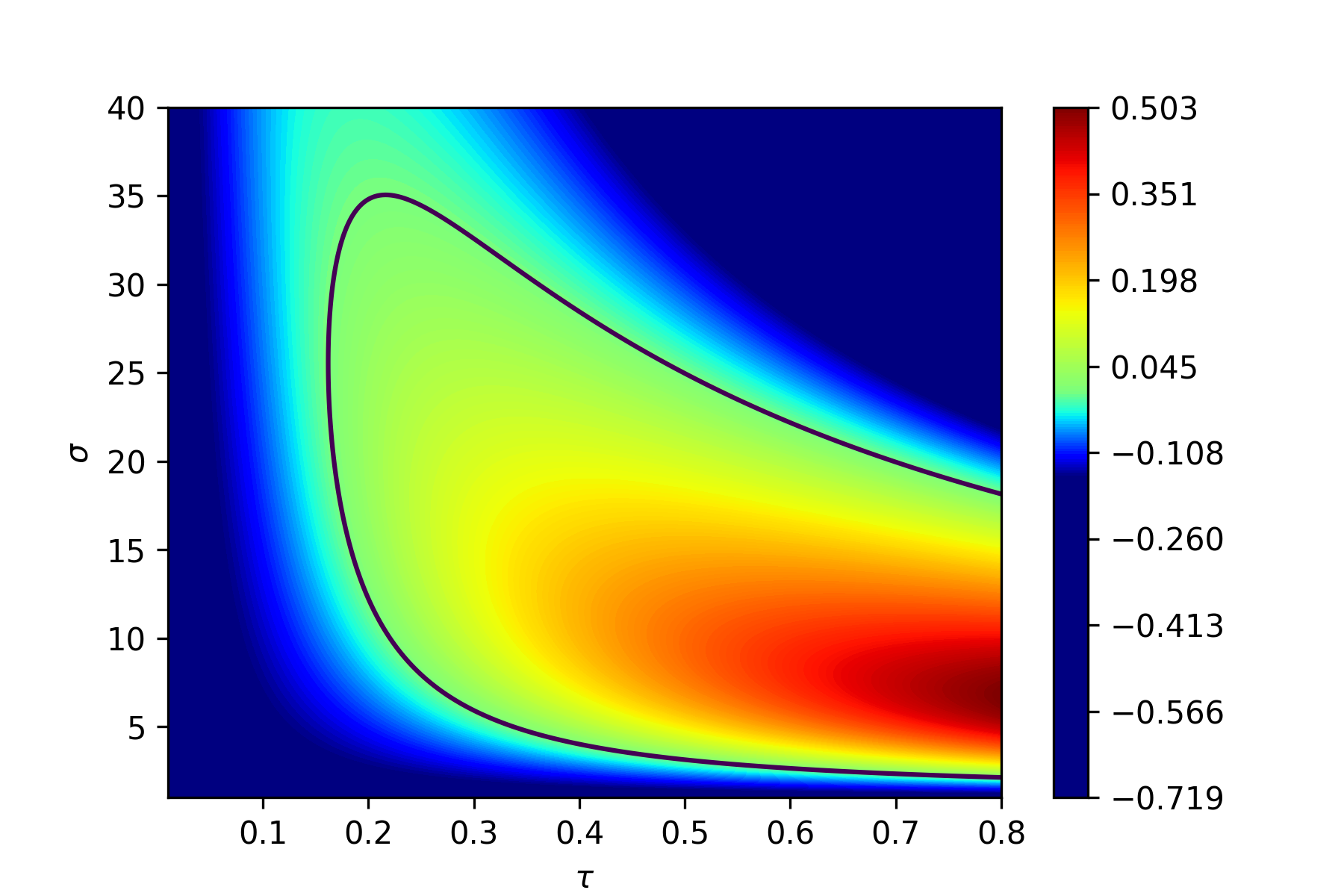}
\caption{$k=6$}\label{hm_6}
\end{subfigure}
\caption{Heatmaps of $\Gamma_k$ of CP-AD in $(\tau, \sigma)$-plane}\label{figs:cpad_hm_sigma}
\end{figure}

\begin{figure}[H]
\begin{subfigure}[H]{0.5\columnwidth}
\includegraphics[width=\columnwidth]{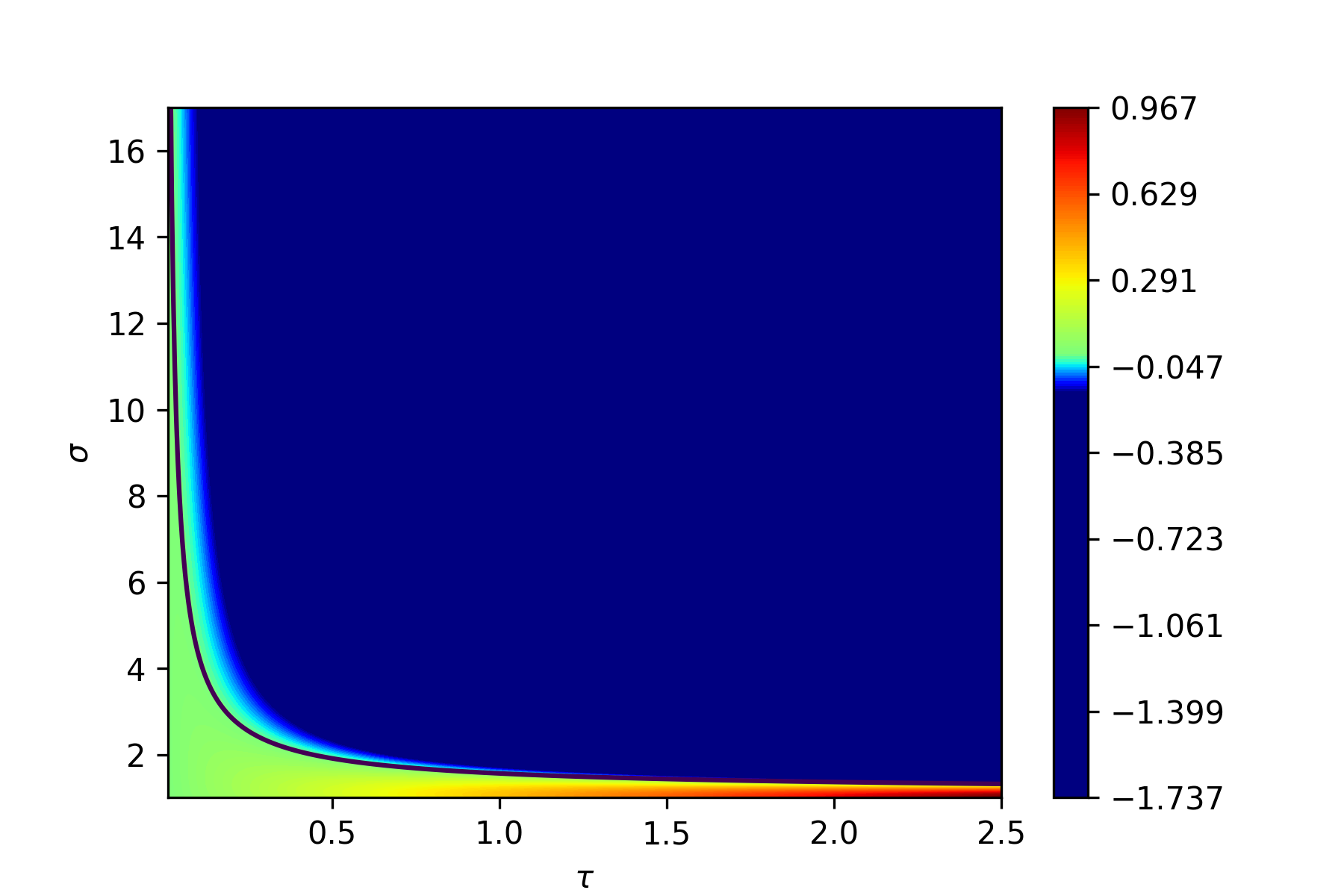}
\caption{$k=1$}\label{hm_1}
\end{subfigure}
\begin{subfigure}[H]{0.5\columnwidth}
\includegraphics[width=\columnwidth]{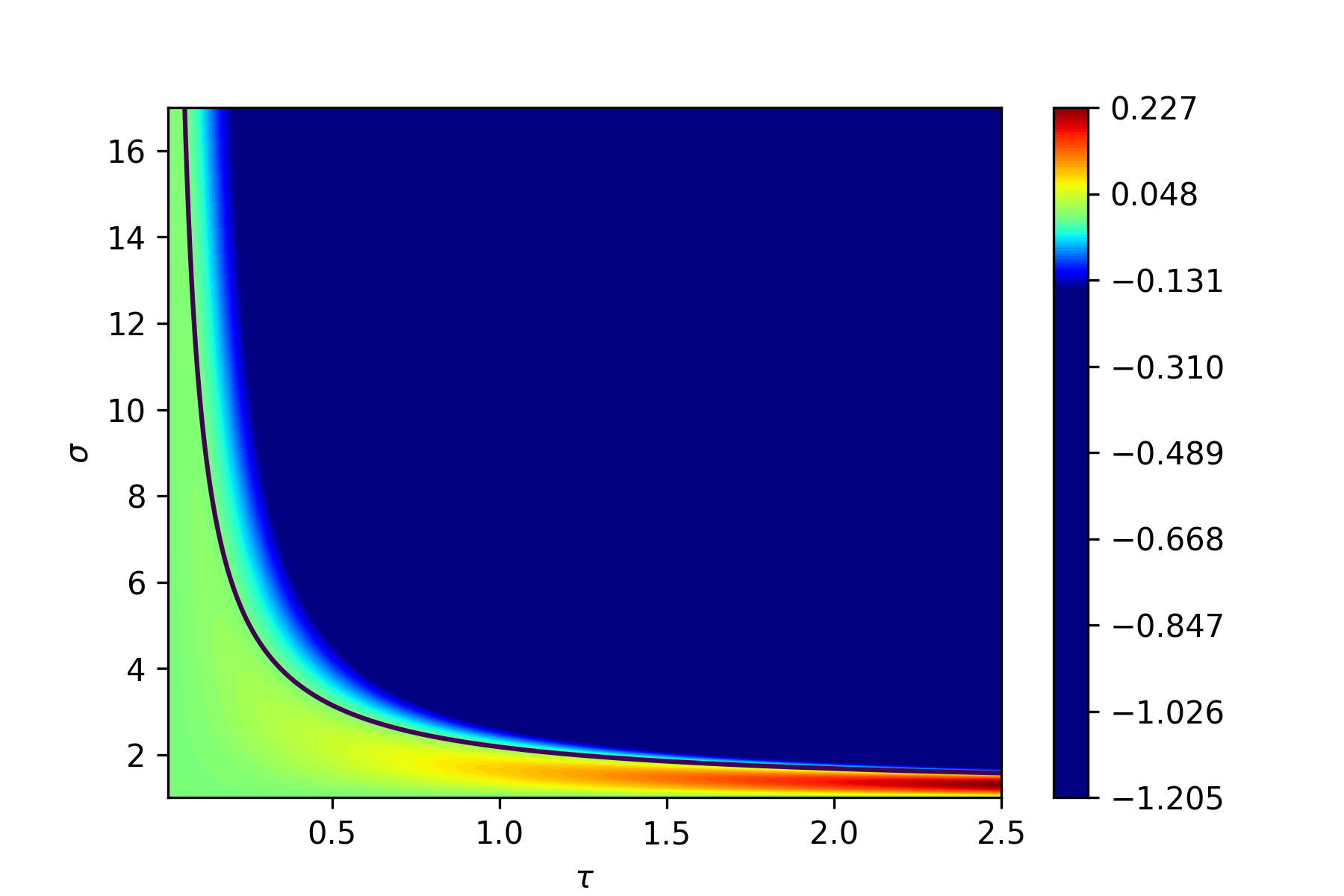}
\caption{$k=2$}\label{hm_2}
\end{subfigure}
\begin{subfigure}[H]{0.5\columnwidth}
\includegraphics[width=\columnwidth]{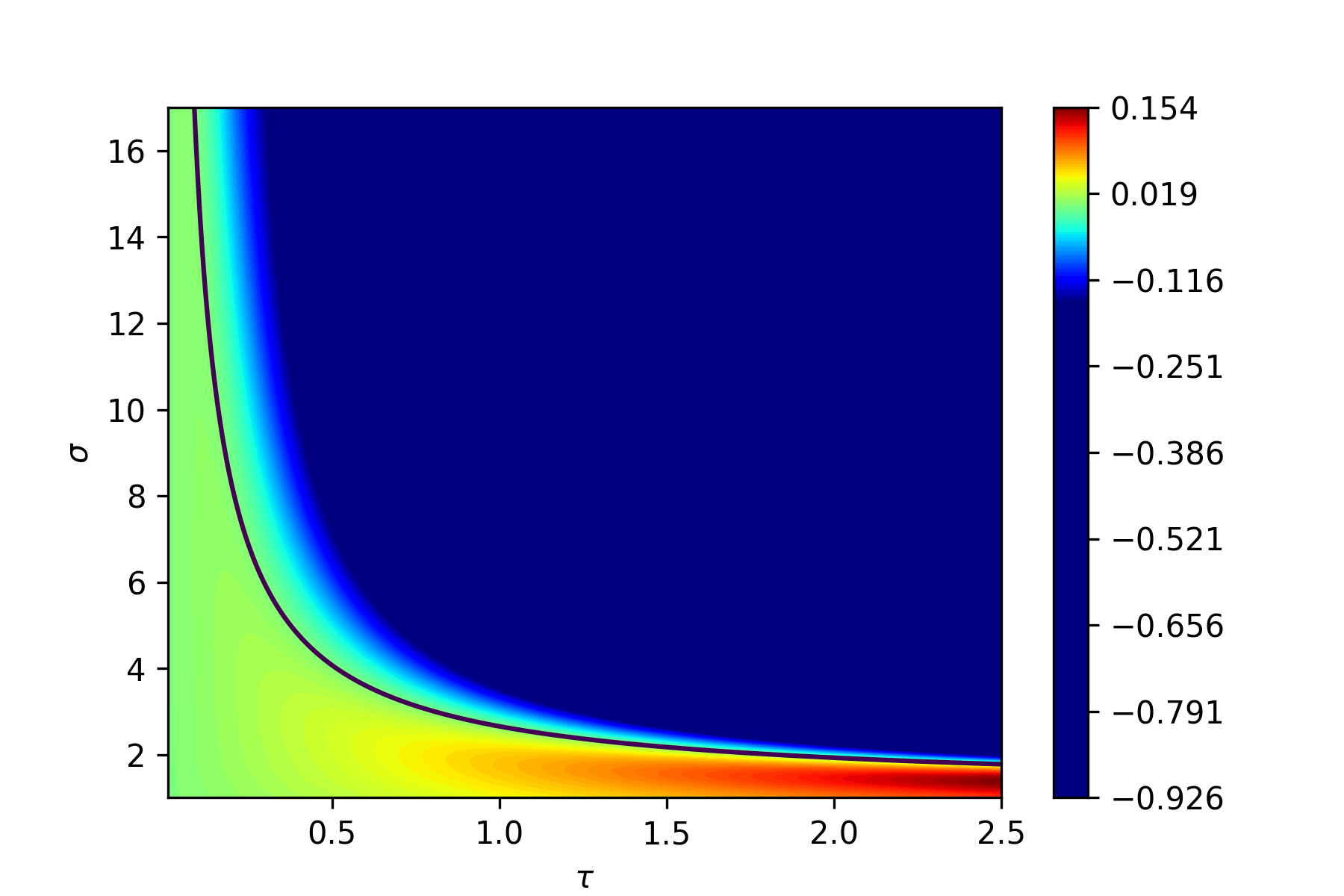}
\caption{$k=3$}\label{hm_3}
\end{subfigure}
\begin{subfigure}[H]{0.5\columnwidth}
\includegraphics[width=\columnwidth]{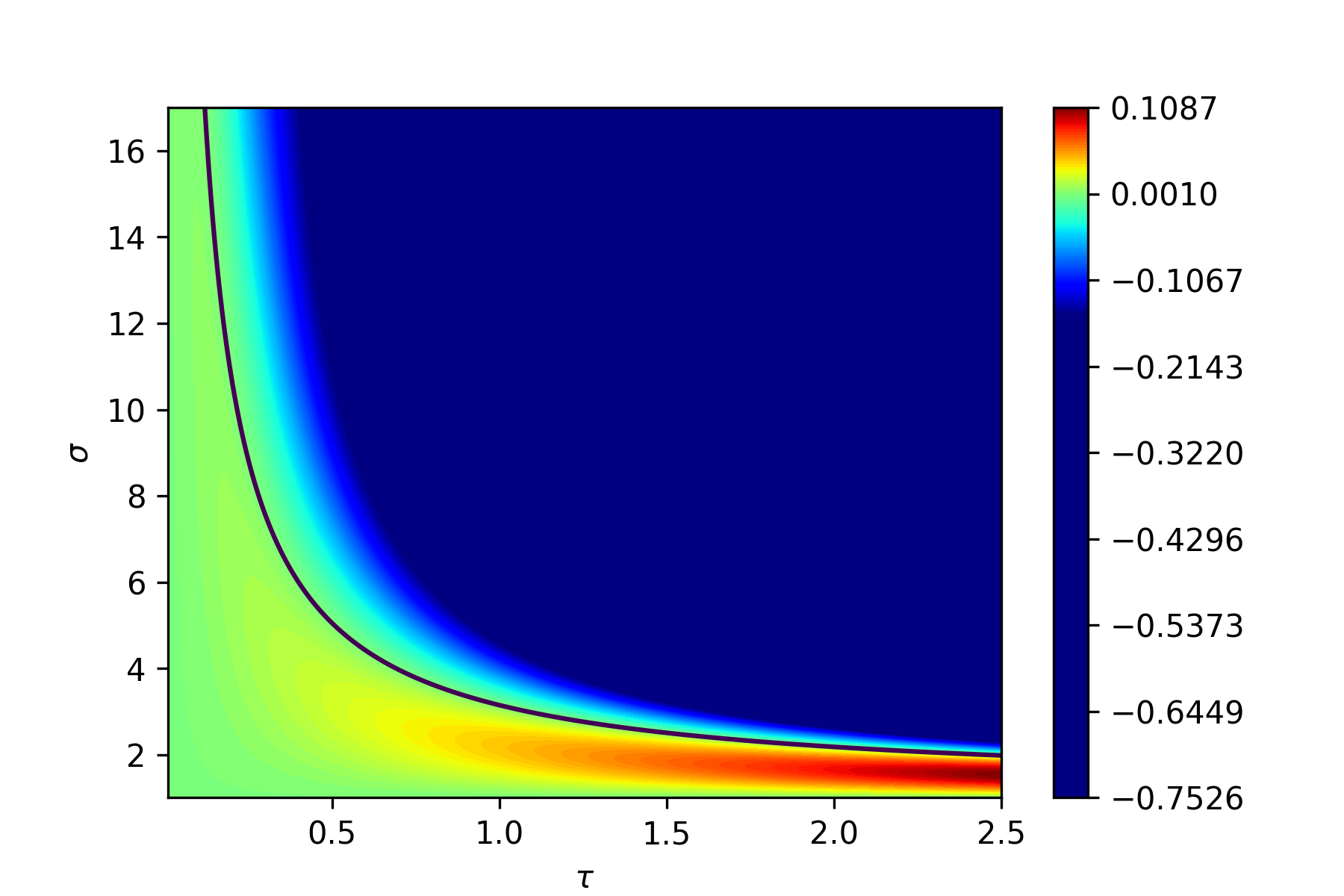}
\caption{$k=4$}\label{hm_4}
\end{subfigure}
\begin{subfigure}[H]{0.5\columnwidth}
\includegraphics[width=\columnwidth]{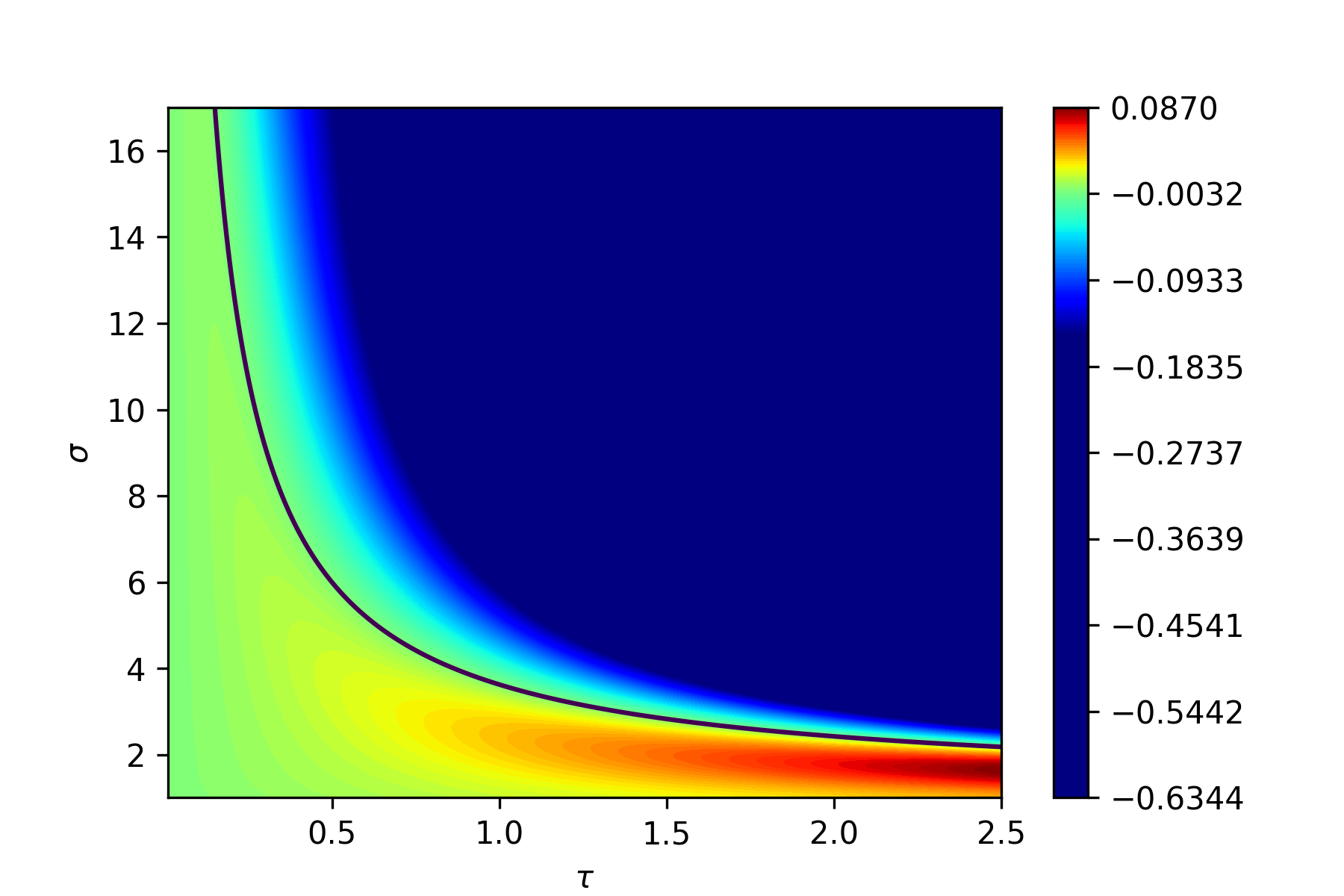}
\caption{$k=5$}\label{hm_5}
\end{subfigure}
\begin{subfigure}[H]{0.5\columnwidth}
\includegraphics[width=\columnwidth]{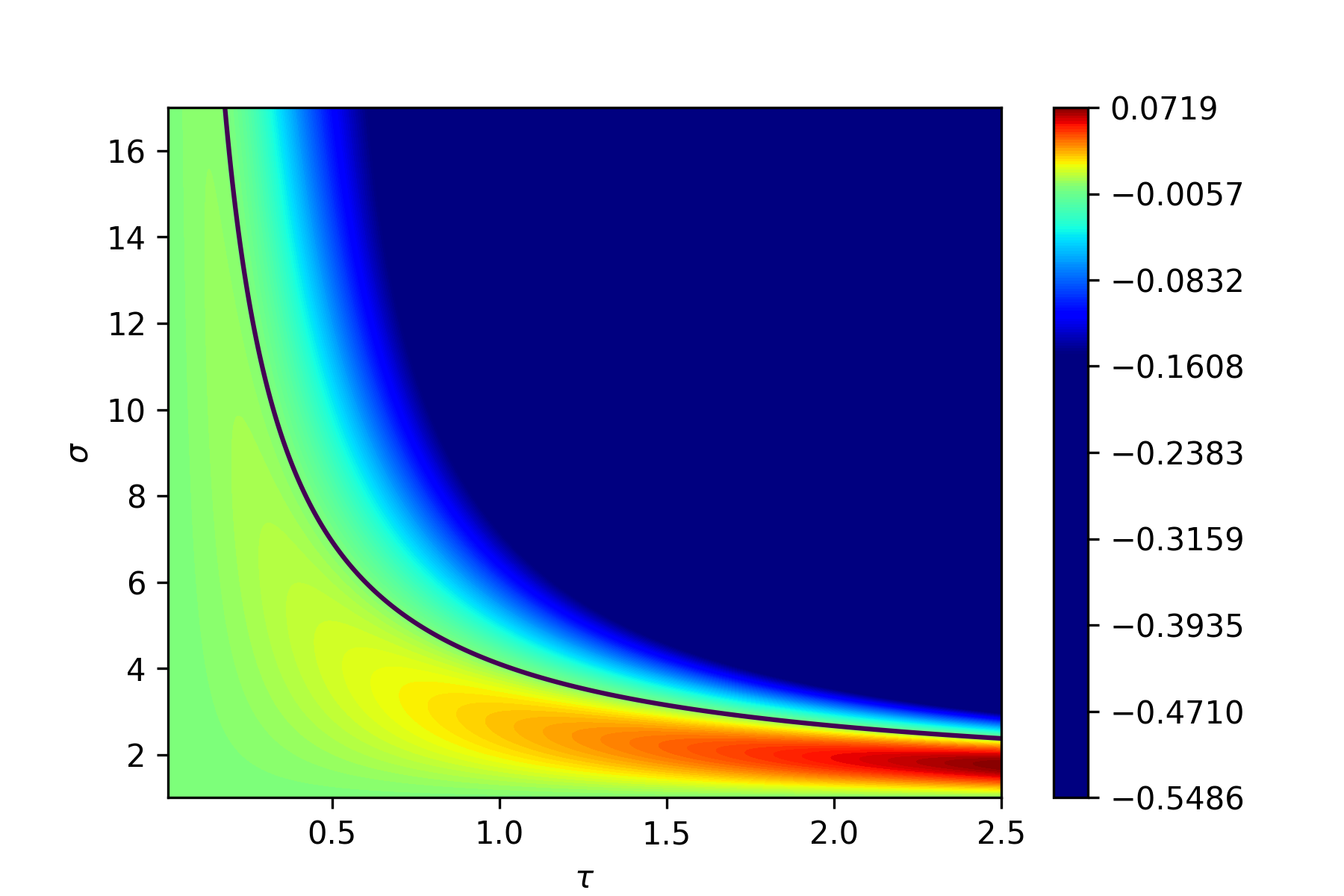}
\caption{$k=6$}\label{hm_6}
\end{subfigure}
\caption{Heatmaps of $\Gamma_k$ of QLLU-R in $(\tau, \sigma)$-plane}\label{figs:qllur_hm_sigma}
\end{figure}

\begin{figure}[H]
\begin{subfigure}[H]{0.5\columnwidth}
\includegraphics[width=\columnwidth]{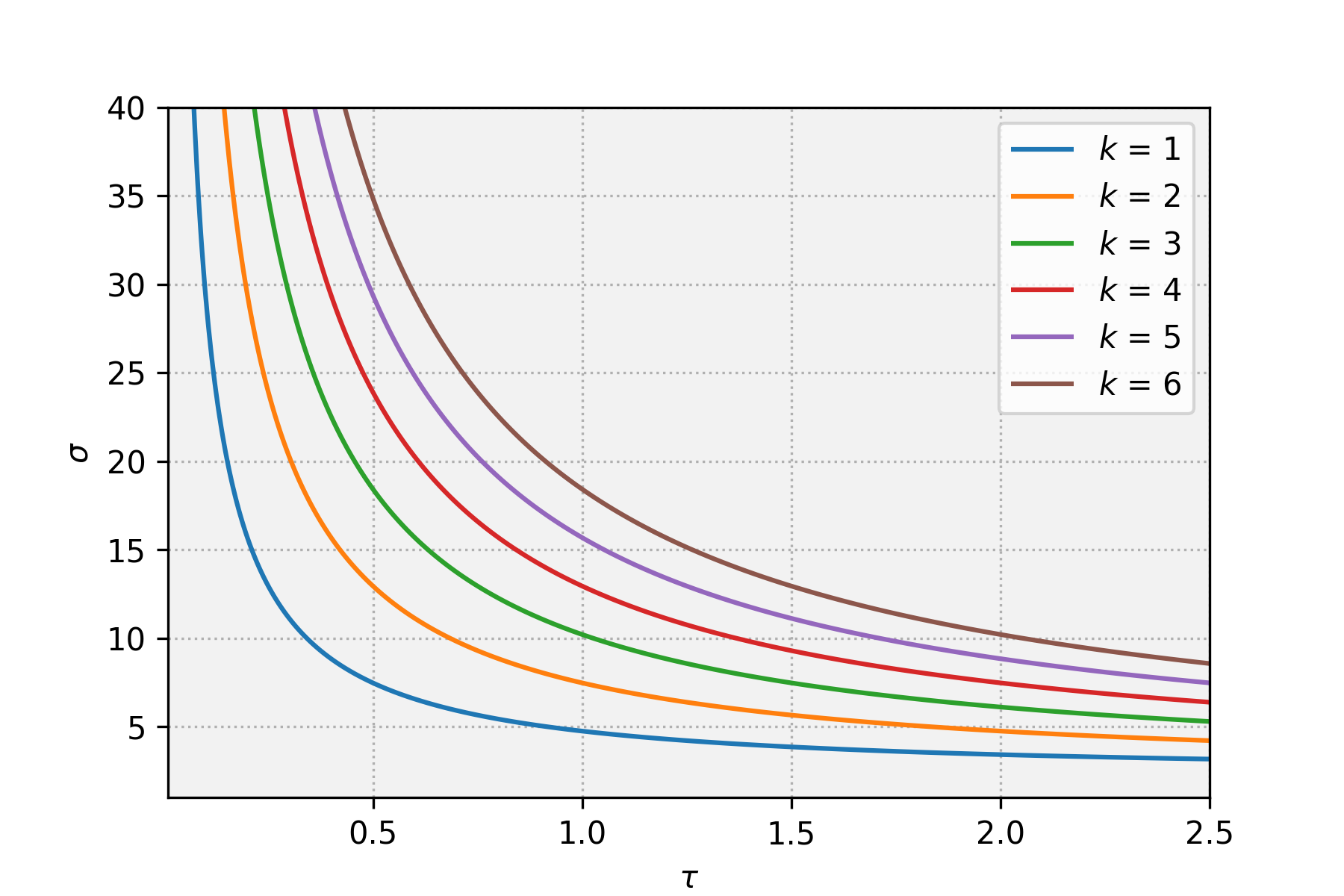}
\caption{CP-R}\label{cpr_cc}
\end{subfigure}
\begin{subfigure}[H]{0.5\columnwidth}
\includegraphics[width=\columnwidth]{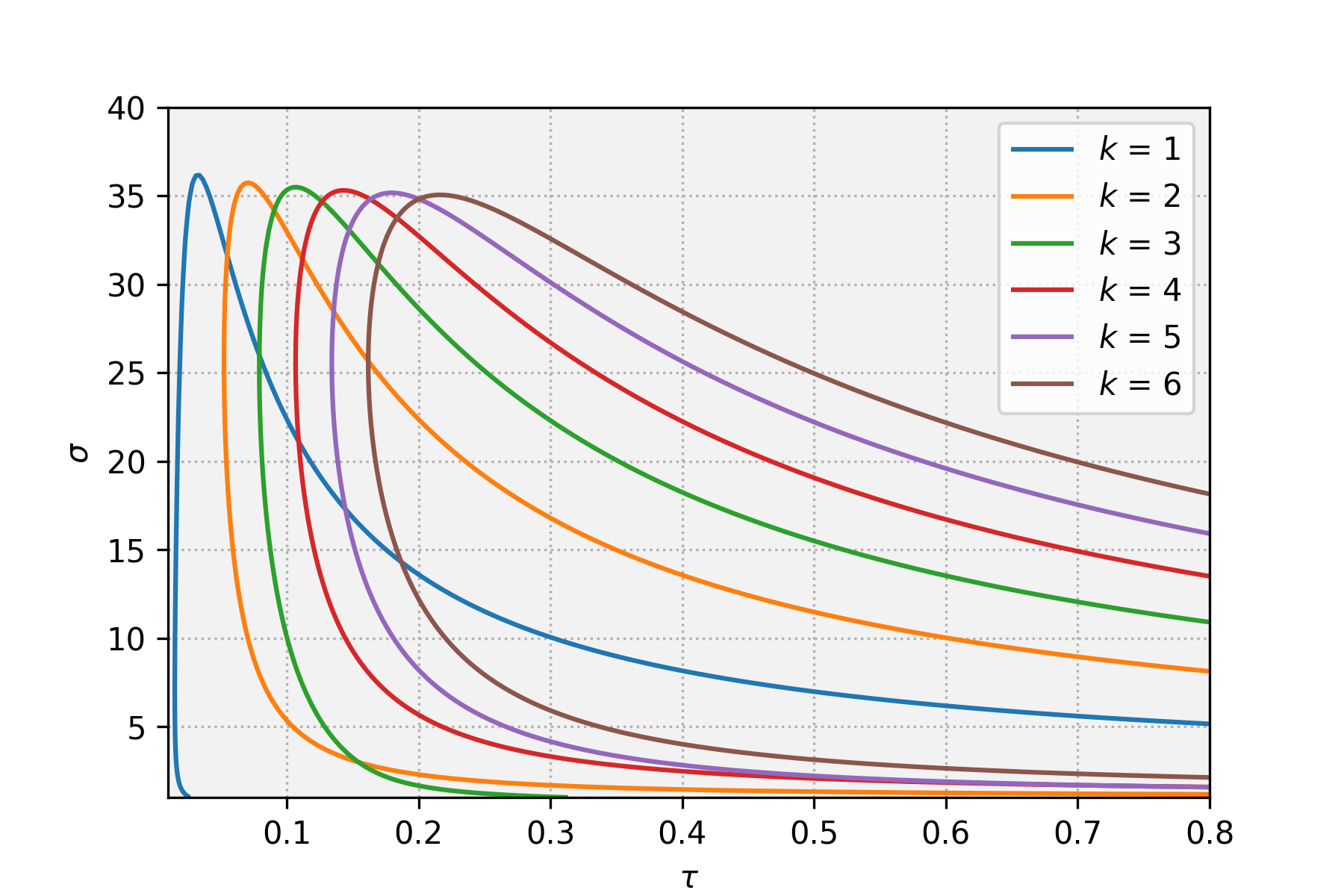}
\caption{CP-AD}\label{cpad_cc}
\end{subfigure}
\begin{subfigure}[H]{0.5\columnwidth}
\includegraphics[width=\columnwidth]{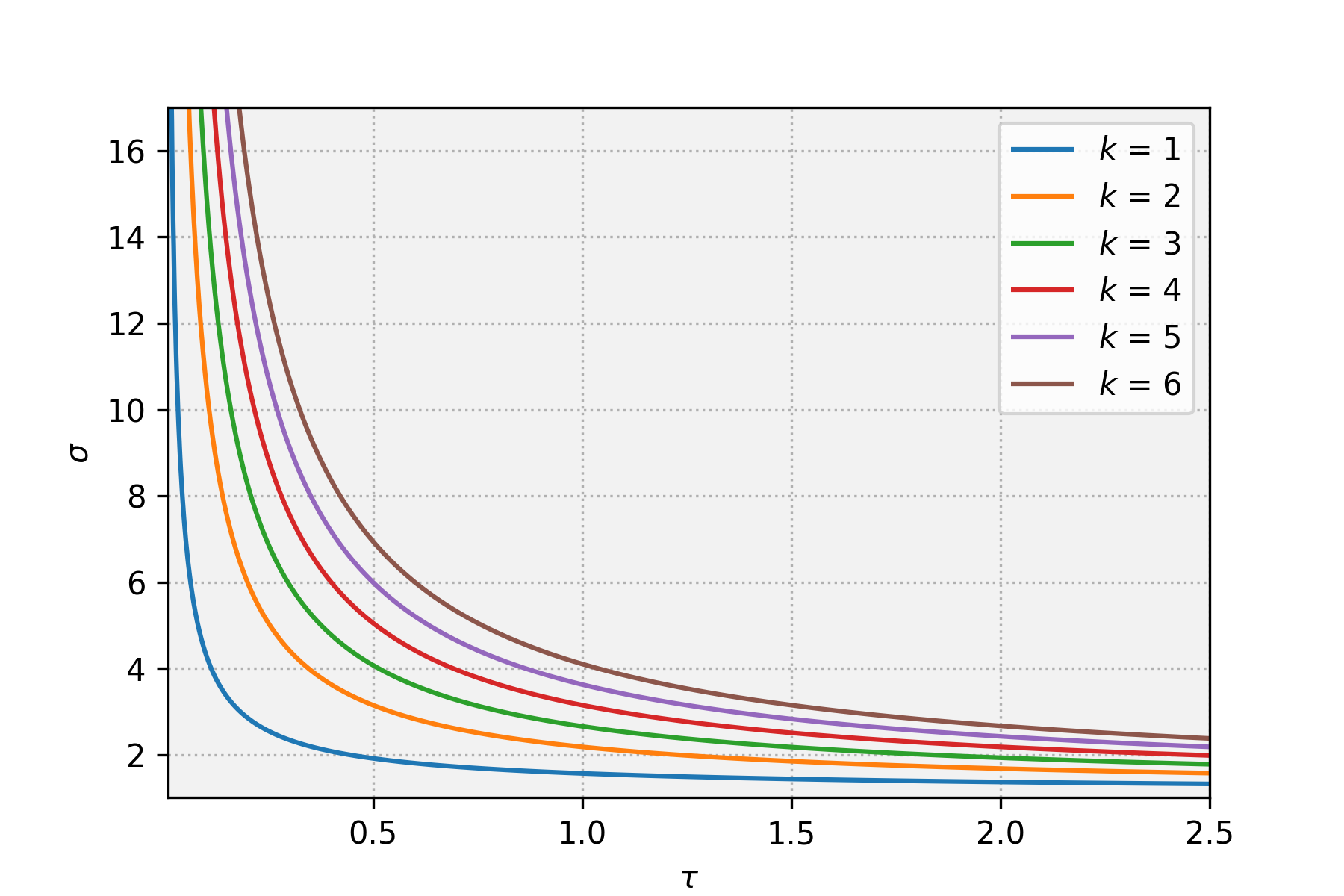}
\caption{QLLU-R}\label{qllur_cc}
\end{subfigure}
\begin{subfigure}[H]{0.5\columnwidth}
\includegraphics[width=\columnwidth]{sigma_contours.png}
\caption{QLLU-AD}\label{qlluad_cc_re}
\end{subfigure}
\caption{Critical curves for each model}
\end{figure}

Thus, the critical curves for each model are summarized as follows.\footnote{Fig.~\ref{qlluad_cc_re} is a redisplay of Fig.~\ref{figs:cc_sigma}.} See the critical curves for CP-R (Fig.~\ref{cpr_cc}) and QLLU-R (Fig.~\ref{qllur_cc}). As the absolute value of the spatial frequency increases, the corresponding critical curve appears to widen outward, thereby expanding the unstable region. We shall refer to this feature as inclusive. On the other hand, see the critical curves for CP-AD (Fig.~\ref{cpad_cc}) and QLLU-AD (Fig.~\ref{qlluad_cc_re}). There, the critical curves corresponding to different frequencies appear to shift to the right as the absolute value of the frequency increases. We shall refer to this feature as shifting. After all, whether the critical curves are inclusive or shifting does not depend on the difference between CP and QLLU, but solely on whether the dynamics is R or AD.

\subsection{Agglomeration shape}
\begin{figure}[H]
 \begin{subfigure}{0.5\columnwidth}
  \centering
  \includegraphics[width=\columnwidth]{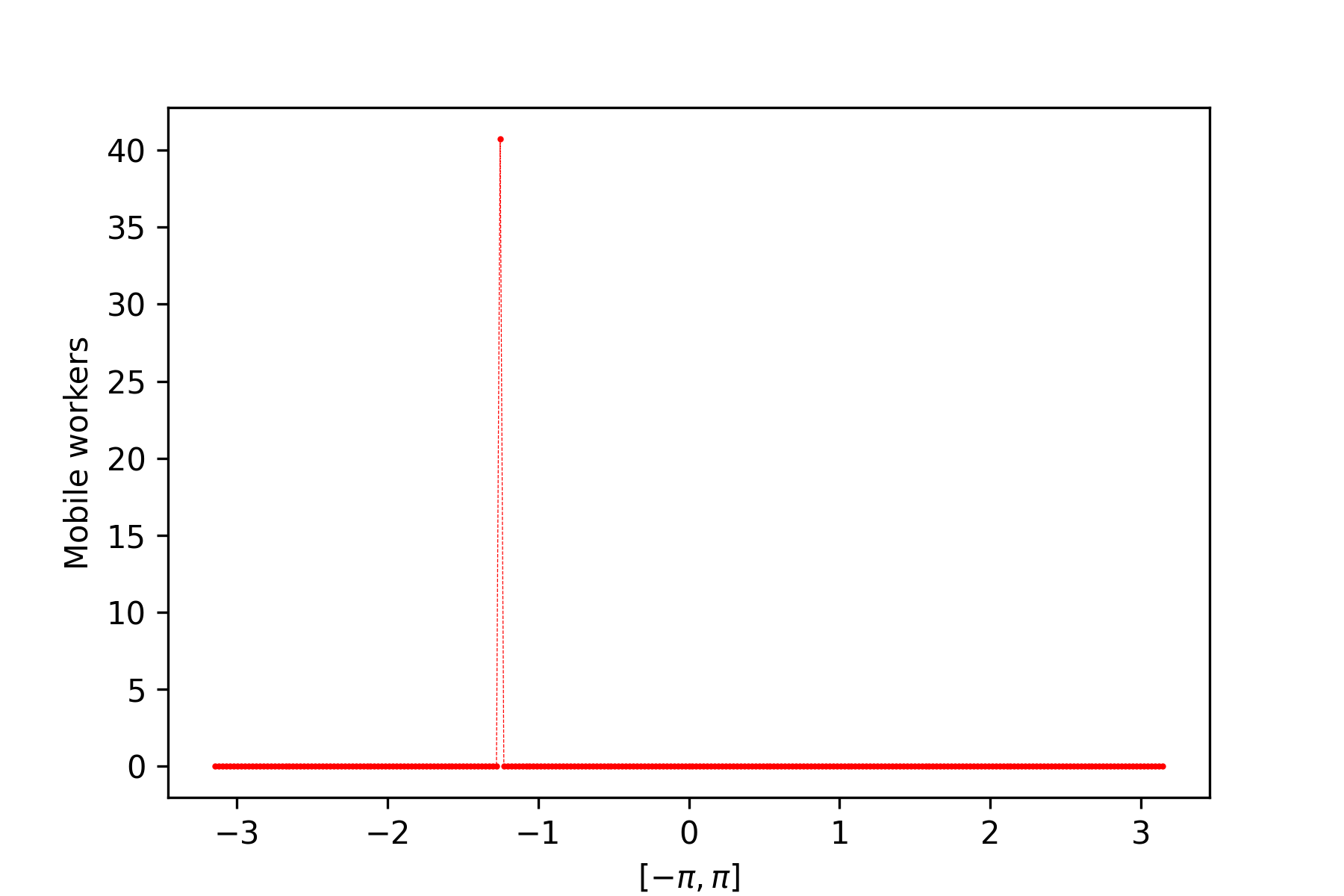}
  \caption{Mobile population density $\lambda^*$}
 \end{subfigure}
 \begin{subfigure}{0.5\columnwidth}
  \centering
  \includegraphics[width=\columnwidth]{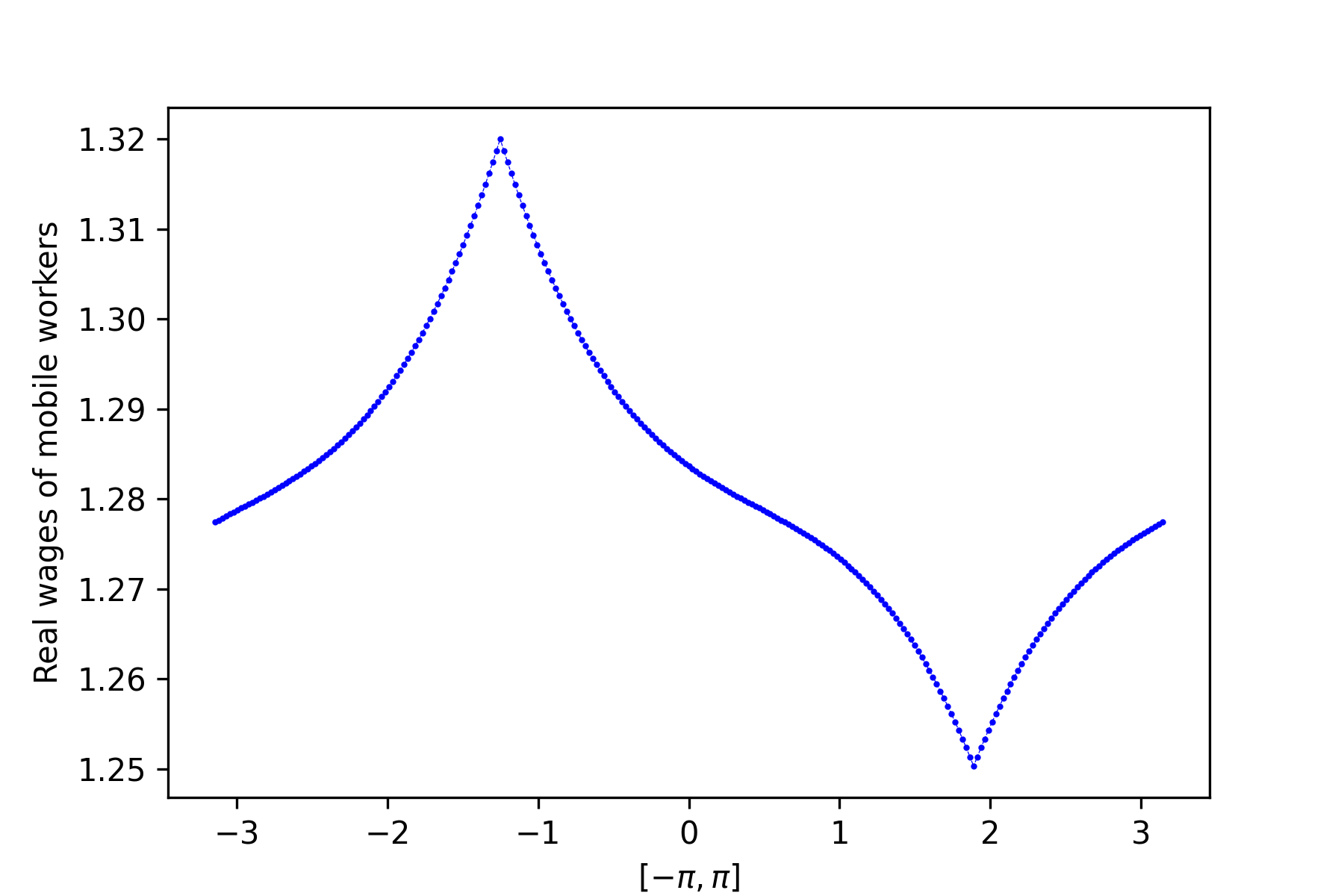}
  \caption{Real wage $\omega^*$}
 \end{subfigure}\\
 \caption{Numerical stationary solution to QLLU-R for $(\sigma,\tau)=(5.0, 0.05)$}
 \label{fig:qllur_one-city-tau}
\end{figure}

\begin{figure}[H]
 \begin{subfigure}{0.5\columnwidth}
  \centering
  \includegraphics[width=\columnwidth]{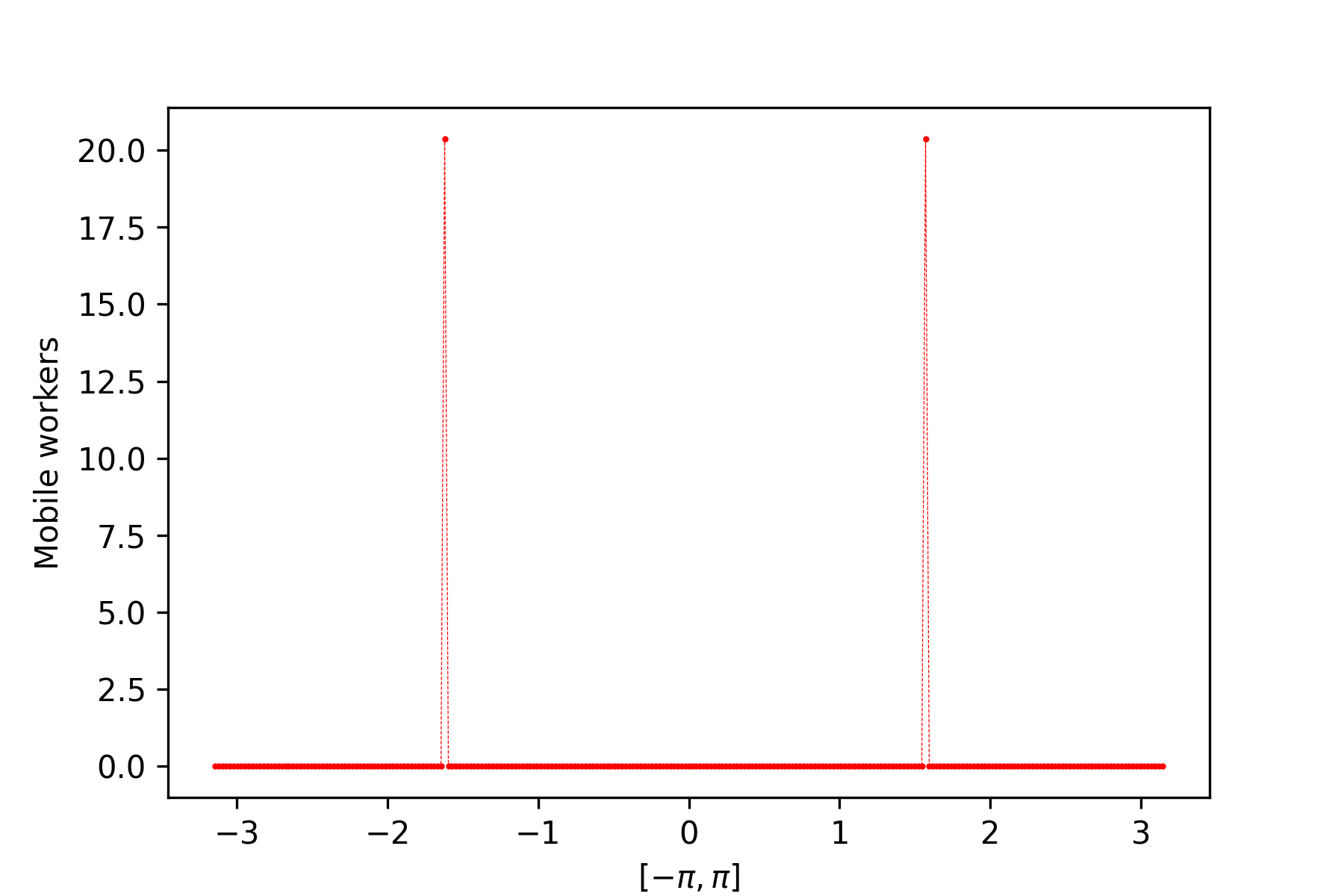}
  \caption{Mobile population density $\lambda^*$}
 \end{subfigure}
 \begin{subfigure}{0.5\columnwidth}
  \centering
  \includegraphics[width=\columnwidth]{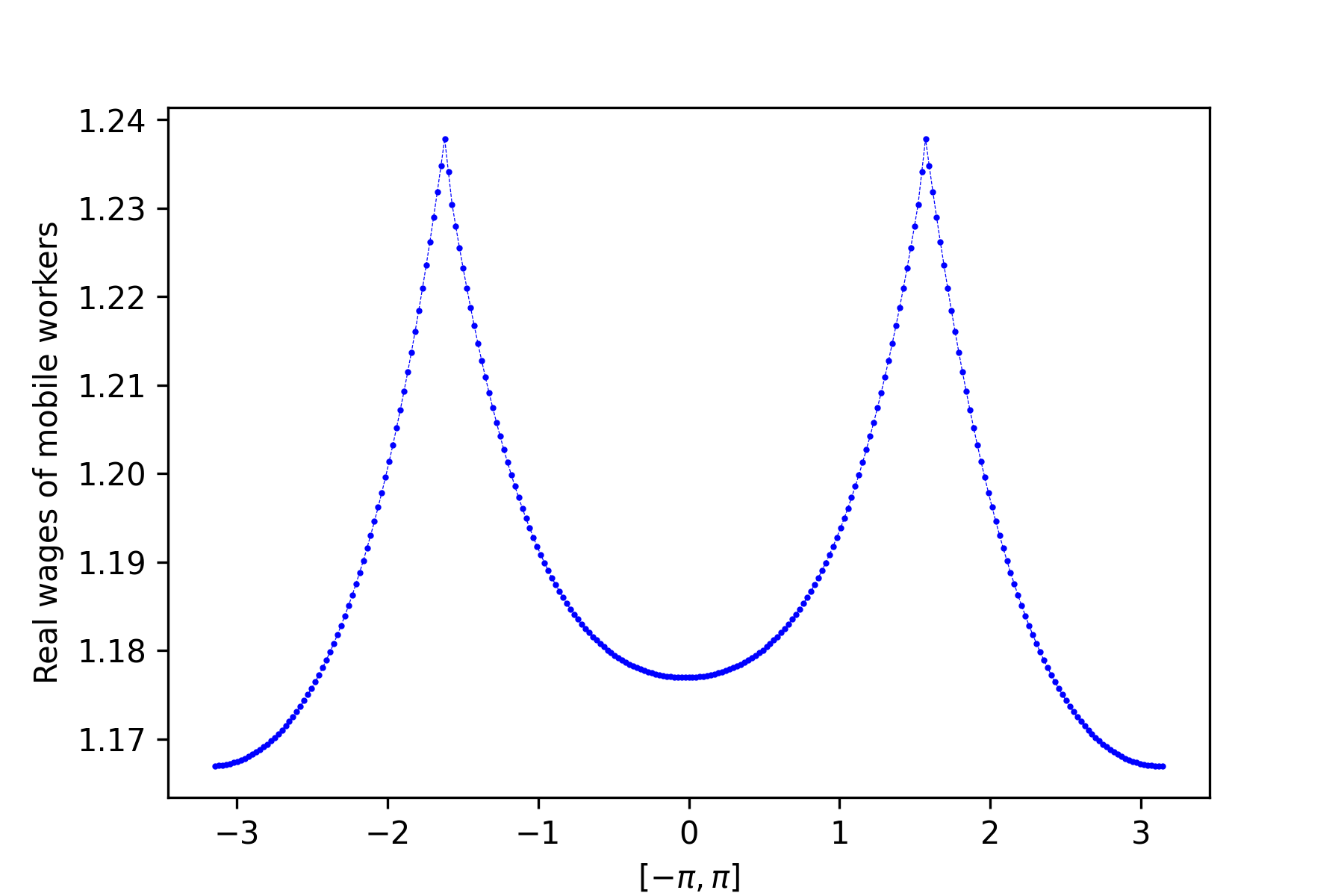}
  \caption{Real wage $\omega^*$}
 \end{subfigure}\\
 \caption{Numerical stationary solution to QLLU-R for $(\sigma,\tau)=(5.0, 0.15)$}
 \label{fig:qllur_two-city-tau}
\end{figure}

\begin{figure}[H]
 \begin{subfigure}{0.5\columnwidth}
  \centering
  \includegraphics[width=\columnwidth]{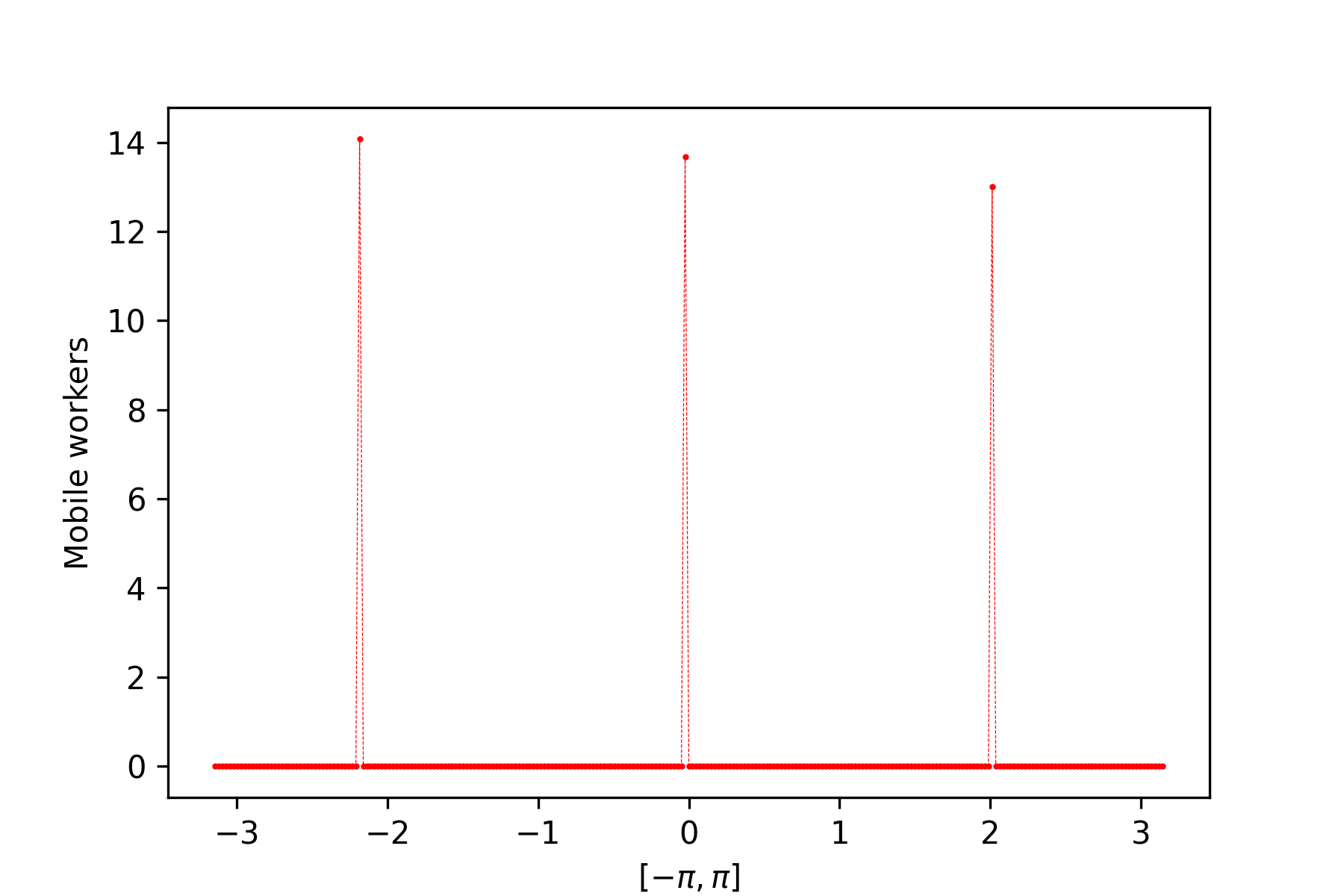}
  \caption{Mobile population density $\lambda^*$}
 \end{subfigure}
 \begin{subfigure}{0.5\columnwidth}
  \centering
  \includegraphics[width=\columnwidth]{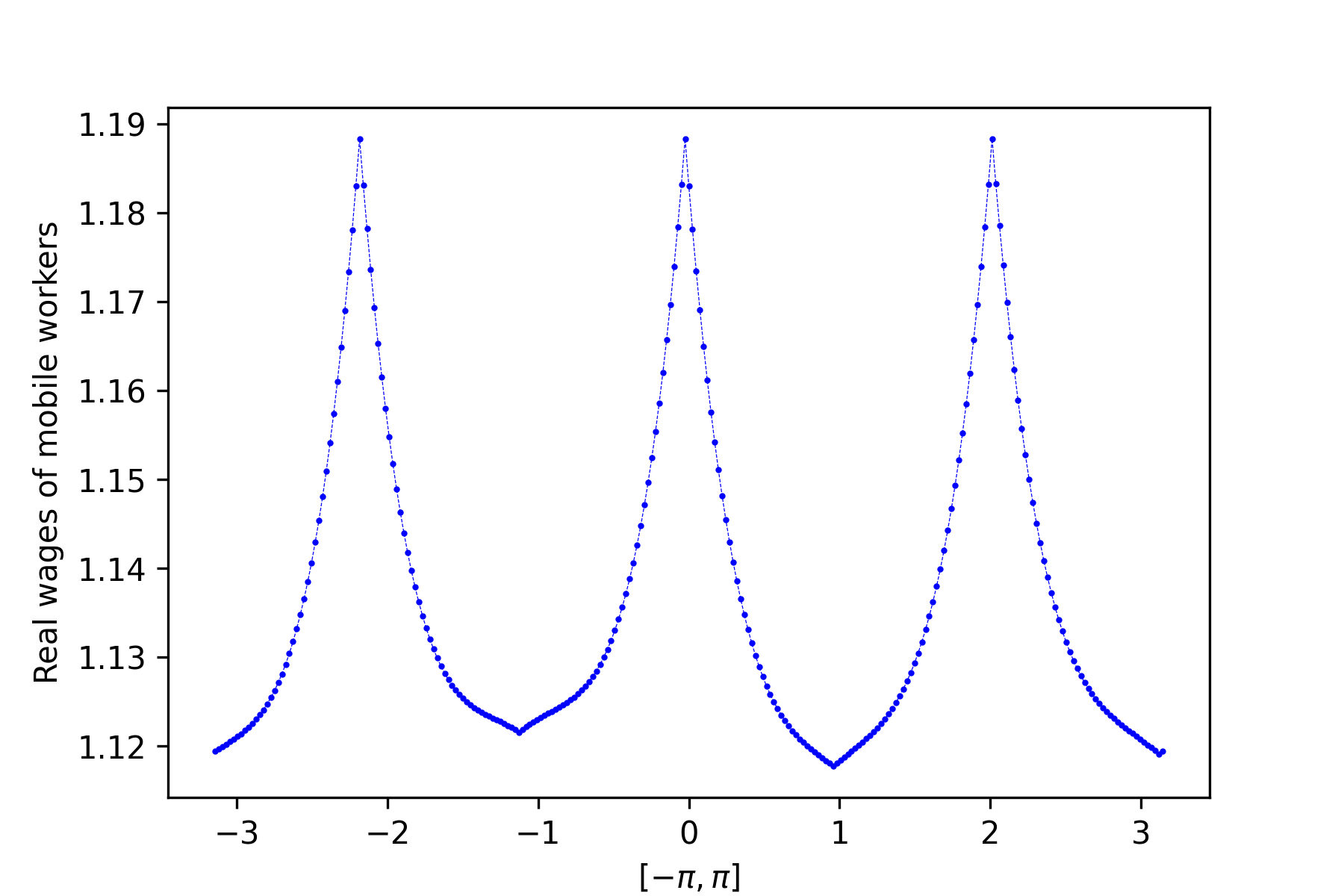}
  \caption{Real wage $\omega^*$}
 \end{subfigure}\\
 \caption{Numerical stationary solution to QLLU-R for $(\sigma,\tau)=(5.0, 0.25)$}
 \label{fig:qllur_three-city-tau}
\end{figure}

Next, we discuss the shape of the non-homogeneous stationary solution that is asymptotically reached by numerical solutions of each model. Numerical confirmation has shown that the time-evolving solutions of mobile population density to CP-R and QLLU-R asymptotically approach spiky, non-homogeneous stationary solutions with multiple spikes, which can be approximated mathematically by delta functions.\footnote{\citet{TabaEshi_explosion} analytically prove that a solution to CP-R converges to the delta function under certain assumptions.} The corresponding distribution of real wages also produces non-differentiable points in agglomerated regions. Figs.~\ref{fig:qllur_one-city-tau}-\ref{fig:qllur_three-city-tau} show examples of numerical stationary solutions of QLLU-R.\footnote{\citet[Setion 5, Figs.~1-12]{Ohtake2023cont}. The results shown in Figs.~\ref{fig:qllur_one-city-tau}-\ref{fig:qllur_three-city-tau} are newly computed for the present paper.} Similar convergence to spiky numerical stationary solutions is observed in CP-R (See \citet[Figs.~1-9]{OhtakeYagi_Asym}.). On the other hand, as already discussed, the solutions of mobile population density to QLLU-AD are not spiky, but converge to solutions having a mountainous shape with some spatial spread. The distribution of real wages also appears to be a smooth function. Although it is difficult to compute that for the non-homogeneous stationary solution of the CP-AD (which is one of the motivations for proposing the model in this paper), it appears to form a similar mountainous distribution rather than spiky one, according to \citet[Figs.~2-3]{MossMar2004}. In short, whether the shape of the nonhomogeneous stationary solution is spiky or mountainous does not depend on whether it is CP or QLLU, but solely on whether the dynamics is R or AD.

\begin{figure}[H]
 \begin{subfigure}{0.5\columnwidth}
  \centering
  \includegraphics[width=\columnwidth]{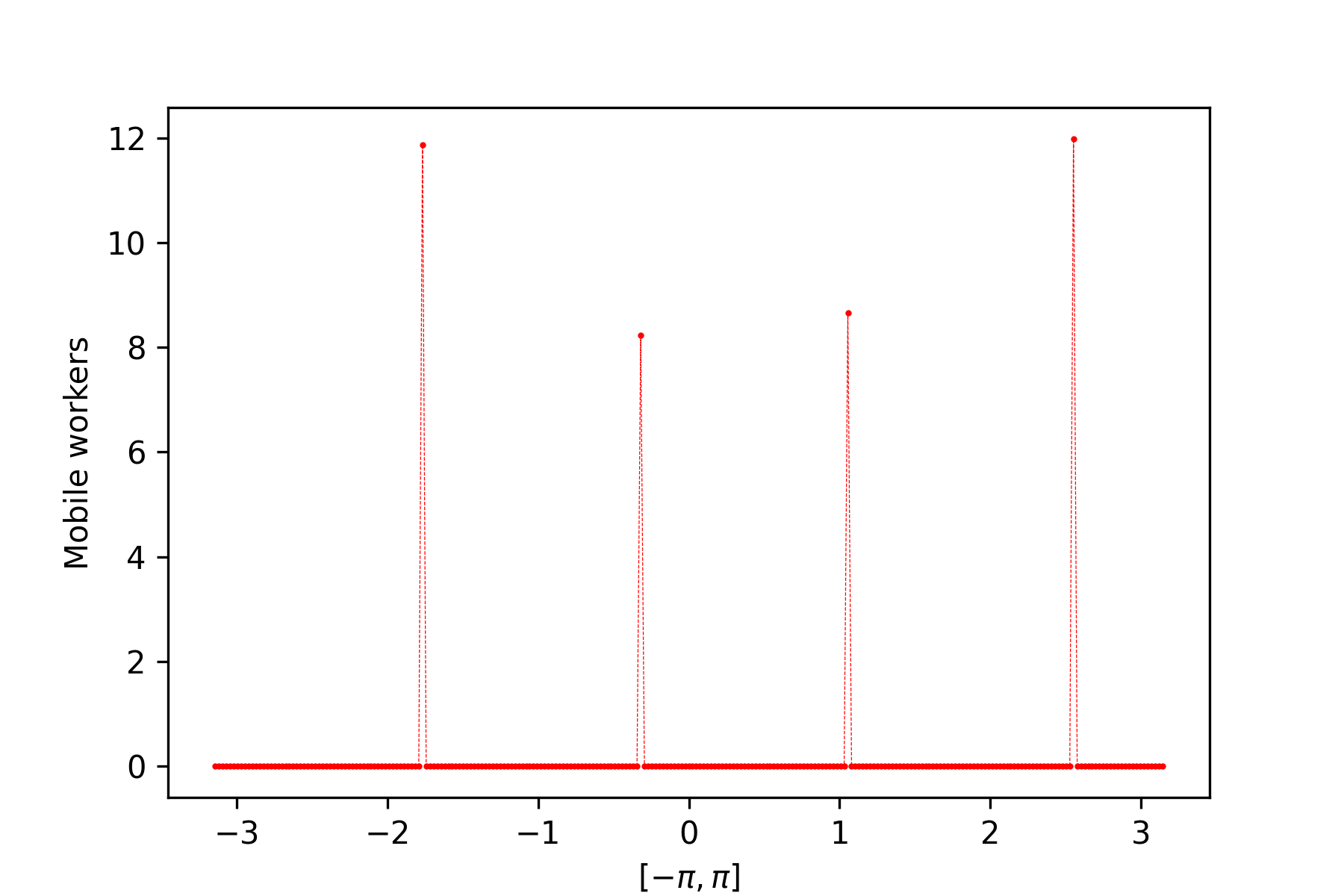}
  \caption{Mobile population density $\lambda^*$}
 \end{subfigure}
 \begin{subfigure}{0.5\columnwidth}
  \centering
  \includegraphics[width=\columnwidth]{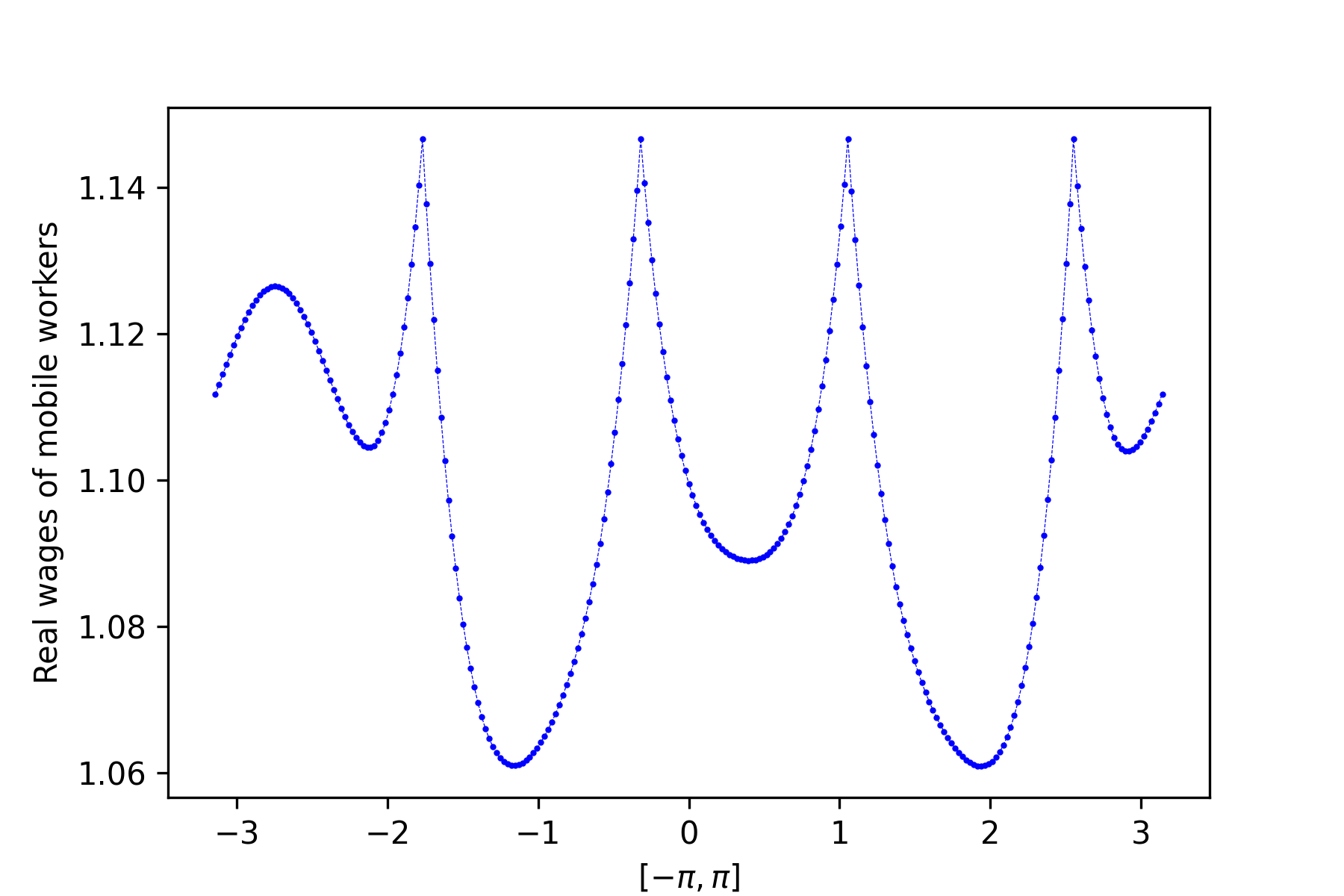}
  \caption{Real wage $\omega^*$}
 \end{subfigure}\\
 \caption{Numerical stationary solution to QLLU-R for $(\sigma,\tau)=(5.0, 0.35)$}
 \label{fig:qllur_four-city-tau}
\end{figure}

\begin{figure}[H]
 \begin{subfigure}{0.5\columnwidth}
  \centering
  \includegraphics[width=\columnwidth]{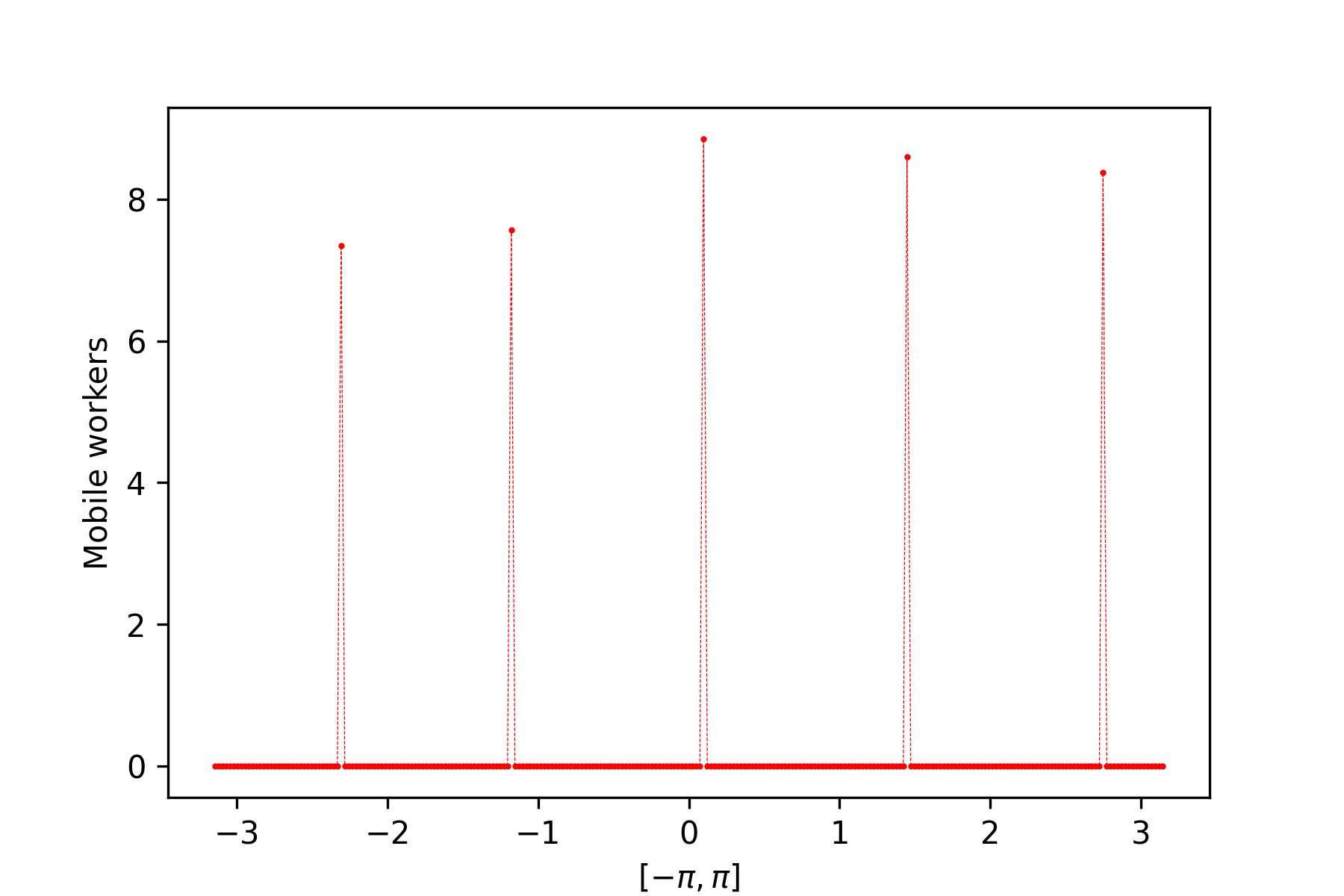}
  \caption{Mobile population density $\lambda^*$}
 \end{subfigure}
 \begin{subfigure}{0.5\columnwidth}
  \centering
  \includegraphics[width=\columnwidth]{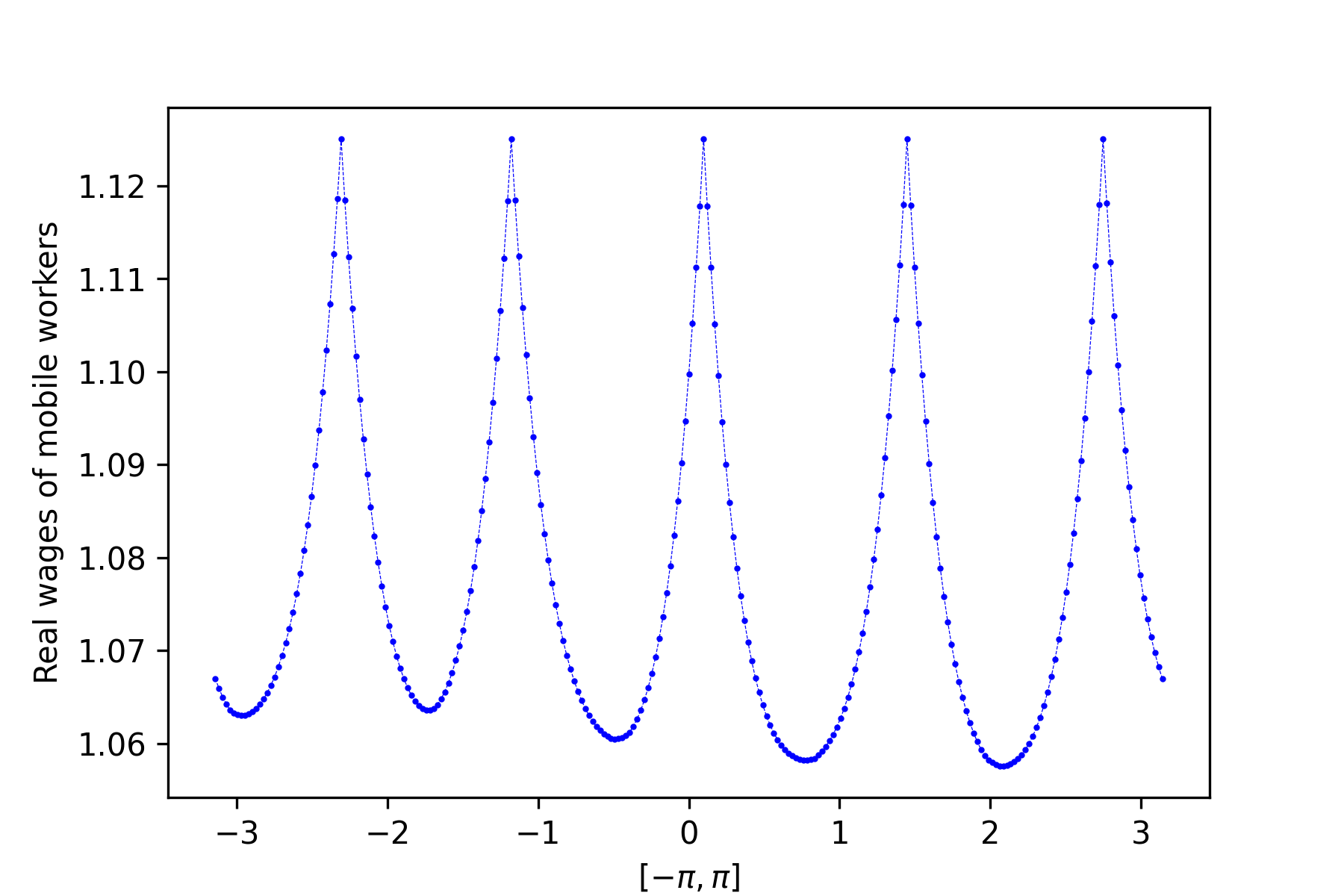}
  \caption{Real wage $\omega^*$}
 \end{subfigure}\\
 \caption{Numerical stationary solution to QLLU-R for $(\sigma,\tau)=(5.0, 0.37)$}
 \label{fig:qllur_five-city-tau}
\end{figure}

\begin{figure}[H]
 \begin{subfigure}{0.5\columnwidth}
  \centering
  \includegraphics[width=\columnwidth]{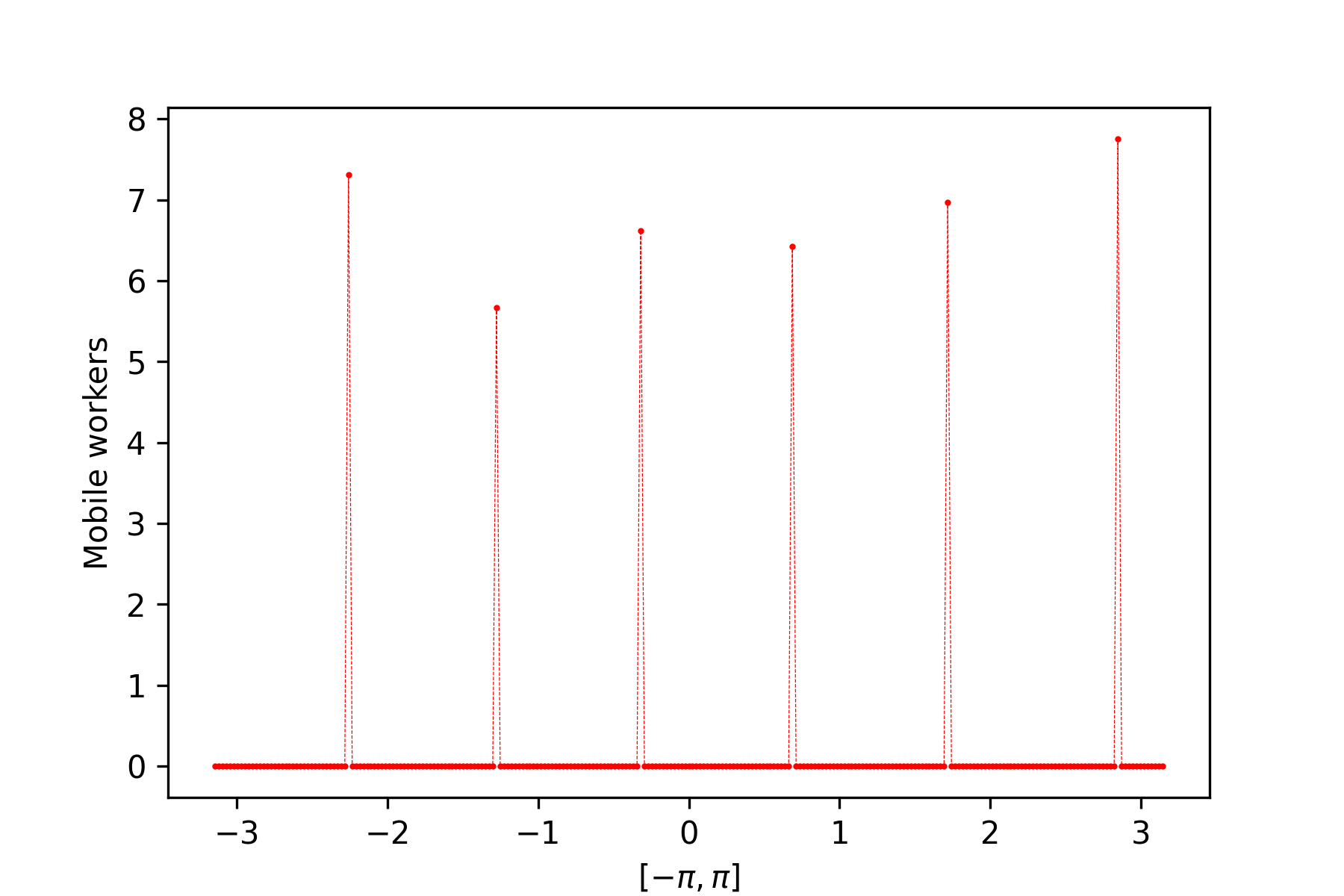}
  \caption{Mobile population density $\lambda^*$}
 \end{subfigure}
 \begin{subfigure}{0.5\columnwidth}
  \centering
  \includegraphics[width=\columnwidth]{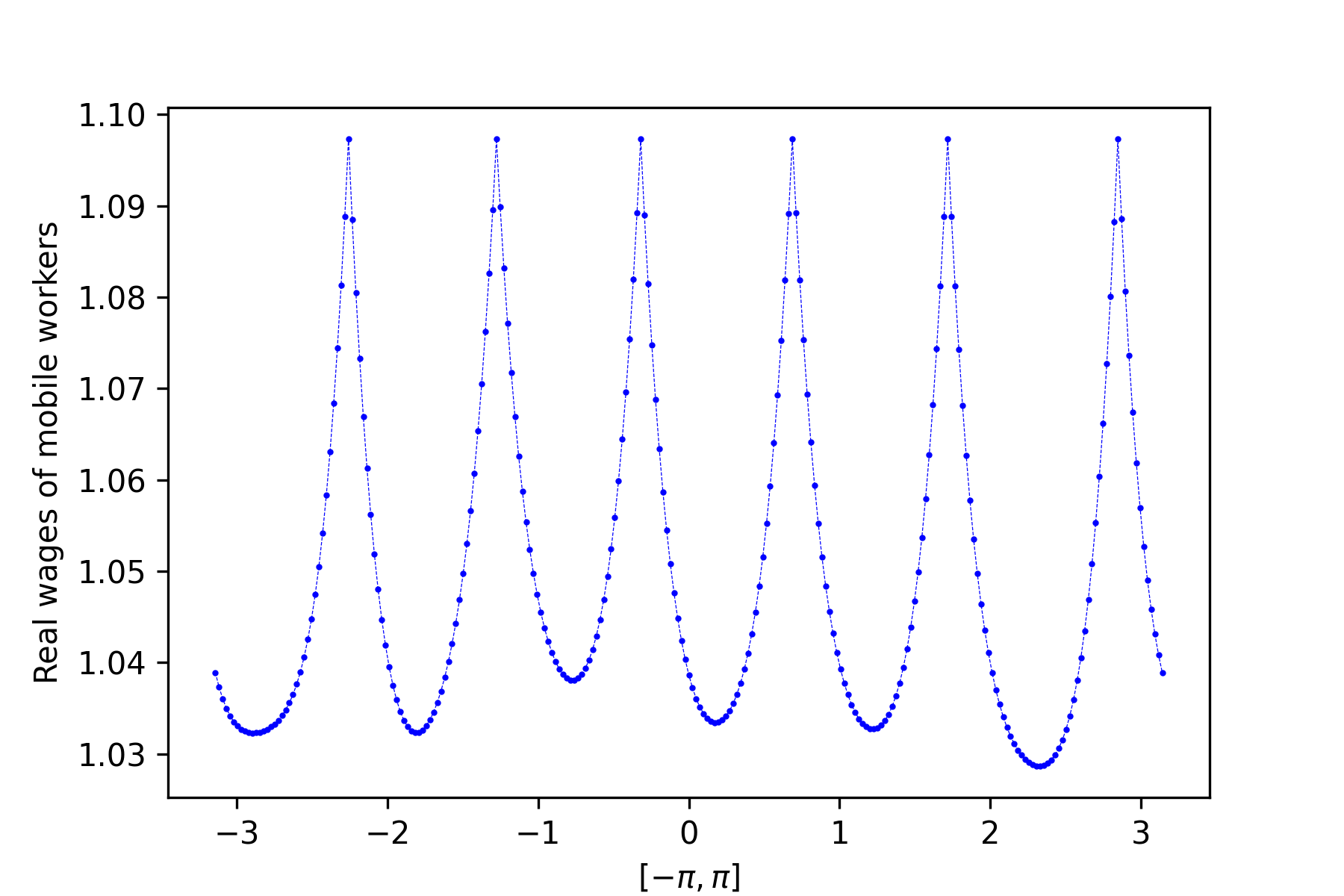}
  \caption{Real wage $\omega^*$}
 \end{subfigure}\\
 \caption{Numerical stationary solution to QLLU-R for $(\sigma,\tau)=(5.0, 0.45)$}
 \label{fig:qllur_six-city-tau}
\end{figure}

\section{Conclusion and discussion}\label{sec:concdisc}
The contribution of this study is to elucidate the role of advection-diffusion dynamics in spatial economic pattern formation. It has been investigated from two approaches: one is the eigenvalue analysis of the linearized equations, and the other is the numerical computation of solutions to the nonlinear model.

First, the eigenvalue analysis has revealed that the replicator dynamics and the advection-diffusion dynamics differ in the appearance of the critical curves on the parameter plane. In the replicator dynamics, the unstable area in the parameter plane tends to expand uniformly with the absolute value of the frequency of modes. On the other hand, in the presence of diffusion, the frequency dependence is not monotonic; overall, the unstable area appears to shift to the right in the parameter plane as the absolute frequency increases. This makes the homogeneous stationary solution stable when transport costs are sufficiently low. More interestingly, if $\sigma$ is not too small, the homogeneous stationary solution can become stable even at intermediate values of transport costs, or a cascade of instability and stability can occur as transport costs decrease. Furthermore, if $\sigma$ is sufficiently large, the homogeneous stationary solution is stable for any values of transport costs. It has been found that these differences arise exclusively from whether the dynamics is R or AD, and do not depend on whether the market equilibrium model is CP or QLLU.\footnote{\citet{Ohtake2025unique} recently conducted research on a model in which market equilibrium is given by the so-called Footloose Entrepreneur model and migration dynamics are given by the replicator equation. The results also support the findings of the present paper.}

Second, the large-time behavior of solutions to QLLU-AD has been computed to obtain the number, size, and agglomeration shape of urban areas. This can never be achieved by analyzing the linearized equations and requires specific computations of the large-time behavior of solutions to the original nonlinear system. When the homogeneous stationary solution is unstable, it has been found that lower transport costs lead to the formation of smaller numbers of large urban areas with larger population sizes. The same is true for an enhanced preference for variety, but given that transport costs are somewhat large. These are theoretically important in that they ensure that the findings from the linearized equations hold in the large-time equilibrium of the system. Regarding the agglomeration shape, it has been found that urban areas are not points but have some extent of geographic spread. This is in contrast to the continuous space CP model with replicator dynamics, where agglomeration is formed as perfect points (and thus mathematically is expressed as a superposition of delta functions).\footnote{For delta function-like agglomeration in continuous space CP models, see \citet{TabaEshi_explosion}, \citet{OhtakeYagi_Asym}, \citet{OhtakeYagi_point}, \citet{Ohtake2023cont}, and \citet{Ohtake2025agriculture}.} Furthermore, it has been suggested that the difference in the agglomeration shape is determined by whether the dynamics is replicator or advection-diffusion. After all, advection-diffusion, particularly diffusion, has significant effects on pattern formation in spatial economy, and the findings have some implications for comparing and evaluating the real-world validity of NEG models. While diffusion has the advantage of producing interesting patterns with real-world relevance, it has the weakness of being time-consuming for numerical computation. The model proposed in this paper can compensate for that weakness.

The insights gained from this study may help in understanding agglomeration phenomena in real-world spatial economies. In reality, complete agglomeration into tiny areas that can be considered points rarely occurs. Instead, it is more likely to form clusters, each with a particular spatial spread, which is referred to as the urban area in this paper. The present study suggests that such realistic spatial agglomerations can be explained mainly by the effect of diffusion. The possibility that agglomeration may be prevented by diffusion may have some policy implications for local economies that suffer from the loss of people and resources to large cities. A more microeconomic and empirical basis for the intensity of advection and diffusion, which may not be spatially uniform as assumed in this study, would provide more practical spatial economic simulations and valuable insights for regional policy-making.

\section{Appendix}\label{sec:app}

\subsection{Derivation of an equation of continuity}\label{subsec:derivationeqcont}

\citet{Moss} has derived the advection-diffusion dynamics from microscopic motions of workers in the one-dimensional case. Instead, a more macroscopic derivation is provided in the following Subsection \ref{subsec:derivationad} by using an equation of continuity in multi-dimensional spaces. We begin with the derivation of the equation of continuity, but this is to make this paper self-contained, and does not include any new ideas. Intuitive discussions are provided for one- and two-dimensional cases, but similar discussions can be easily found in textbooks on fluid mechanics and related fields.\footnote{In fact, the explanation here is just one and two-dimensional versions of \citet[pp.147-148]{Arf2012mathphys} in which they are concerned with three dimensions.}

For any time $t$ and location $x\in\Omega\subset\mathbb{R}^n$, where $n=1$, $2$, or $3$, consider a vector field that maps a vector $J(t, x)\in\mathbb{R}^n$ that represents the direction and magnitude of mobile population migration in unit time per unit volume there. The vector $J(t, x)$, called the flux density. Using the flux density, one can formulate the flow of mobile population into and out of a given volume of space. In deriving the equation of continuity, the specific form of $J$ is not specified, but is discussed in its abstract form. The advantage of the equation of continuity is that various phenomena can be expressed by changing the specific form of $J$ in different ways.

First, let us consider the case that $\Omega$ is $1$-dimensional. As in Fig.~\ref{fig:1dimcons}, set up an interval $[x-\frac{d x}{2}, x+\frac{d x}{2}]$ of width $dx$ anywhere on $\Omega$. The change in the mobile population within this interval in unit time can be calculated by counting the population entering and leaving at boundaries $x-\frac{d x}{2}$ and $x+\frac{d x}{2}$. The population inflow into (resp. outflow from) this interval at the boundary on $x-\frac{d x}{2}$ (resp. $x+\frac{d x}{2}$) is given by the flux density $J(t, x-\frac{d x}{2})\in\mathbb{R}$ (resp. $J(t, x-\frac{d x}{2})\in\mathbb{R}$). Therefore, the change in the mobile population within this interval is given by
\begin{equation}
-\left[J\left(t, x+\frac{d x}{2}\right) - J\left(t, x-\frac{d x}{2}\right)\right]
\fallingdotseq -\frac{\partial J}{\partial x}(t, x)d x,
\end{equation}
where the right-hand side is the first-order approximation. Hence, we have
\begin{equation}\label{1d_cons}
\frac{\partial\lambda}{\partial t}(t, x) + \frac{\partial J}{\partial x}(t, x)=0.
\end{equation}
This is the equation of continuity in one-dimensional space. 
\begin{figure}[H]
\centering
\begin{tikzpicture}
 \draw[name path=xaxis, ->, thin] (-3.5,0)--(3.5,0);
 \draw[name path=Lb, densely dotted] (-1.0, -0.7) -- (-1.0, 0.7)node[above]{$x-\frac{dx}{2}$};
 \draw[name path=Rb, densely dotted] (1.0, -0.7) -- (1.0, 0.7)node[above]{$x+\frac{dx}{2}$};
 \path[name intersections={of = xaxis and Lb, by ={Left}}];
 \path[name intersections={of = xaxis and Rb, by ={Right}}];
 \fill[gray!20!white] (-1, 0.01) -- (-1, 0.3) -- (1, 0.3) -- (1, 0.01); 
 \fill[gray!20!white] (-1, -0.01) -- (-1, -0.3) -- (1, -0.3) -- (1, -0.01); 
 \fill[red] (Left) circle (0.06) node[below left]{$J\left(t, x-\frac{dx}{2}\right)$};
 \fill[red] (Right) circle (0.06) node[below right]{$J\left(t, x+\frac{dx}{2}\right)$};
 \draw[red, ultra thick][->, >=stealth] (-1.0, 0.0) -- (-0.5, 0.0); 
 \draw[red, ultra thick][->, >=stealth] (1.0, 0.0) -- (1.5, 0.0); 
\end{tikzpicture}
\caption{one-dimensional space and flux density}
\label{fig:1dimcons}
\end{figure}

A similar idea can be applied to the case where $\Omega$ is two-dimensional. In this case, the $1$st-component (resp. $2$nd-component) of $J(t,x)\in\mathbb{R}^2$ is denoted by $J_1(t,x)$ (resp. $J_2(t,x)$). As shown in Fig.~\ref{fig:2dimcons}, consider a rectangular region consisting of line segments parallel to the $x_1$- and $x_2$-axes, each with lengths $d x_1$ and $d x_2$, respectively. The population inflow into (resp. outflow from) this rectangular region at the boundary parallel to the $x_2$-axis on $x_1-\frac{d x_1}{2}$ (resp. $x_1+\frac{d x_1}{2}$) in unit time is given by $J_1(t, x_1-\frac{d x_1}{2}, x_2)d x_2$ (resp. $J_1(t, x_1+\frac{d x_1}{2}, x_2)d x_2$). Thus, the total population inflow and outflow per unit time from the boundary parallel to the $x_2$-axis is given by 
\begin{equation}
-\left[J_1\left(t, x_1+\frac{d x_1}{2}, x_2\right) - J_1\left(t, x_1-\frac{d x_1}{2}, x_2\right)\right]d x_2
\fallingdotseq -\frac{\partial J_1}{\partial x_1}(t, x)d x_1d x_2,
\end{equation}
where the right-hand side is the first-order approximation. Similarly, the total population inflow and outflow per unit time from the boundary parallel to the $x_1$-axis is given by
\begin{equation}
-\left[J_2\left(t, x_1, x_2+\frac{d x_2}{2}\right) - J_2\left(t, x_1, x_2-\frac{d x_2}{2}\right)\right]d x_1
\fallingdotseq -\frac{\partial J_2}{\partial x_2}(t, x)d x_1d x_2,
\end{equation}
where the right-hand side is the first-order approximation. Hence, we obtain the equation of continuity in two-dimensional space
\begin{equation}\label{2d_cons}
\frac{\partial\lambda}{\partial t}(t, x) + \frac{\partial J_1}{\partial x_1}(t, x) + \frac{\partial J_2}{\partial x_2}(t, x)=0.
\end{equation}
\begin{figure}[H]
\centering
\begin{tikzpicture}
 \draw[name path=xaxis,->,thin] (0, 0) -- (7, 0); 
 \draw[name path=yaxis,->,thin] (0, 0) -- (0, 4); 
 \draw[name path=Lb, densely dotted] (1.5, 3) -- (1.5, 0)node[below]{$x_1-\frac{dx_1}{2}$}; 
 \draw[name path=Rb, densely dotted] (5.5, 3) -- (5.5, 0)node[below]{$x_1+\frac{dx_1}{2}$}; 
 \draw[name path=Tb, densely dotted] (5.5, 3) -- (0, 3)node[left]{$x_2+\frac{dx_2}{2}$}; 
 \draw[name path=Bb, densely dotted] (5.5, 1) -- (0, 1)node[left]{$x_2-\frac{dx_2}{2}$}; 
 \path[name intersections={of = xaxis and Lb, by ={x1}}]; 
 \path[name intersections={of = xaxis and Rb, by ={x2}}]; 
 \path[name intersections={of = yaxis and Bb, by ={y1}}]; 
 \path[name intersections={of = yaxis and Tb, by ={y2}}]; 
 \fill[fill=gray!20!white, densely dotted] (1.5, 1) -- (5.5, 1) -- (5.5, 3) -- (1.5, 3) -- cycle; 
 \draw[red, ultra thick][->, >=stealth] (1.5, 2.0) -- (2.0, 2.0); 
 \fill[red] (1.5, 2) circle (0.06) node[left]{$J_1\left(t, x_1-\frac{dx_1}{2}, x_2\right)$};
 \draw[red, ultra thick][->, >=stealth] (5.5, 2.0) -- (6.0, 2.0); 
 \fill[red] (5.5, 2) circle (0.06) node[below right]{$J_1\left(t, x_1+\frac{dx_1}{2}, x_2\right)$};
 \draw[blue, ultra thick][->, >=stealth] (3.5, 3.0) -- (3.5, 3.5); 
 \fill[blue] (3.5, 3) circle (0.06) node[above right]{$J_2\left(t, x_1, x_2+\frac{dx_2}{2}\right)$};
 \draw[blue, ultra thick][->, >=stealth] (3.5, 1.0) -- (3.5, 1.5); 
 \fill[blue] (3.5, 1) circle (0.06) node[below]{$J_2\left(t, x_1, x_2-\frac{dx_2}{2}\right)$};
\end{tikzpicture}
\caption{two-dimensional space and flux density}
\label{fig:2dimcons}
\end{figure}

By introducing 
\begin{align}
&\nabla:= \frac{\partial}{\partial x} \text{\hspace{21.5mm}when $n=1$}\\
&\nabla:= \left(\frac{\partial}{\partial x_1}, \frac{\partial}{\partial x_2}\right) \text{\hspace{5mm}when $n=2$}
\end{align}
according to the notation of vector analysis, the equations of continuity \eqref{1d_cons} and \eqref{2d_cons} can be written in a unified manner as
\begin{equation}\label{3cons}
\frac{\partial\lambda}{\partial t}(t, x) + \nabla\cdot J=0.
\end{equation}
It is quite correct to think that this can be true in three dimensions by the application of the same idea with $\nabla:= \left(\frac{\partial}{\partial x_1}, \frac{\partial}{\partial x_2},\frac{\partial}{\partial x_3}\right)$. However, the primary interest in spatial economics lies in one and two dimensions.\footnote{The derivation of the equation of continuity in three-dimensional space can be found in many physics textbooks. See for example \citet[pp.147-148]{Arf2012mathphys}.} 

\subsection{Derivation of the advection-diffusion equation}\label{subsec:derivationad}

As noted before, various models can be derived by linking the flux density $J$ to the migration behavior of mobile workers. The following procedure is similar to that used in \citet[Section 11.4]{Murray2002} to derive a reaction-diffusion-chemotaxis equation. Let mobile workers migrate in the direction of increasing real wages in space. Additionally, the size of the migrating population is assumed to be proportional to the size of the mobile population at a departure region. As a result, we have a vector field $J_{\text{adv}}$ such that
\[
J_{\text{adv}}(t,x) = a\lambda(t,x)\nabla\omega(t,x),~a>0,
\]
where $\nabla\omega(t,x)$ is the gradient of $\omega(t, x)\in\mathbb{R}^n$ with respect to $x$, that is, 
\[
\nabla\omega(t,x)=\left(\frac{\partial\omega}{\partial x_1},\frac{\partial\omega}{\partial x_2},\cdots,\frac{\partial\omega}{\partial x_n}\right).
\]
Meanwhile, mobile workers escape from a more populated region to a less populated region.\footnote{This term can be seen as the result of random migration due to idiosyncratic incentives, as in \citet{Moss}, or as an expression of behavior to avoid some congestion costs associated with agglomeration.} This can be expressed as a vector field $J_{\text{diff}}$ such that
\[
J_{\text{diff}}(t,x) = -d\nabla\lambda(t,x),~d>0,
\]
where $\nabla\lambda(t,x)\in\mathbb{R}^n$ is the gradient of $\lambda(t, x)$ with respect to $x$, that is,
\[
\nabla\lambda(t,x)=\left(\frac{\partial\lambda}{\partial x_1},\frac{\partial\lambda}{\partial x_2},\cdots,\frac{\partial\lambda}{\partial x_n}\right).
\]
Putting $J$ in \eqref{3cons} as $J = J_{\text{adv}} + J_{\text{diff}}$, we have
\[
\frac{\partial \lambda}{\partial t}(t,x) + a\nabla\cdot\left(\lambda(t,x)\nabla\omega(t,x)\right) = d\Delta\lambda(t,x),
\]
where $\Delta$ stands for the Laplace operator with respect to $x$,that is,
\[
\Delta\lambda=\sum_{i=1}^n\frac{\partial^2\lambda}{\partial x_i^2}.
\]

\subsection{Where does $Z_k$ come from?}\label{subsec:Zk}
A careful calculation shows that\footnote{The same calculation also appears in the analyses of other racetrack economy models as in \citet{Ohtake2023cont},  \citet{Ohtake2023city}, \citet{Ohtake2025agriculture}, and \citet{Ohtake2025unique}.}
\begin{equation}\label{inteikthemalpd}
\int_{-\pi}^\pi e^{iks}e^{-\alpha D(\rho r, \rho s)}\rho ds
= \frac{2\alpha \rho^2\left(1-(-1)^ke^{-\alpha \rho\pi}\right)}{k^2+\alpha^2 \rho^2}e^{ikr}.
\end{equation}
It is easy from \eqref{Gb} and \eqref{inteikthemalpd} to see that
\[
\ol{G}^{\sigma-1}\int_{-\pi}^\pi e^{iks}e^{-\alpha D(\rho r, \rho s)}\rho ds
= \frac{F}{\ol{\lambda}}Z_k e^{ik r}.
\]

\subsection{Eigenvalues of similar models}
In the following, the notations are the same as those used in this paper. For the CP model, the price index in the homogeneous stationary solution is given by
\begin{equation}
\ol{G}_{\rm CP} = \left[\frac{2\ol{\lambda}(1-e^{-\alpha\rho\pi})}{\alpha}\right]^{\frac{1}{1-\sigma}}.
\end{equation}

The eigenvalue of CP-R is given by \footnote{This is equivalent to \citet[An equation written as $J_n$ on p.85]{OhtakeYagi_Asym}.}
\[
\Gamma_k = v\ol{G}_{\rm CP}^{-\mu}
\left[
(1-\mu Z_k)\frac{-\frac{1}{\sigma}Z_k^2+\frac{\mu}{\sigma}Z_k}{1-\frac{\mu}{\sigma}Z_k-\frac{\sigma-1}{\sigma}Z_k^2}
+\frac{\mu Z_k}{\sigma-1}
\right].
\]

The eigenvalue of CP-AD is given by
\begin{align}
\Gamma_k &= 
\frac{k^2}{\rho^2}
\left[
a\ol{G}_{\rm CP}^{-\mu}\left(
(1-\mu Z_k)\frac{-\frac{1}{\sigma}Z_k^2+\frac{\mu}{\sigma}Z_k}{1-\frac{\mu}{\sigma}Z_k - \frac{\sigma-1}{\sigma}Z_k^2} + \frac{\mu Z_k}{\sigma-1}
\right)-d
\right].
\end{align}

The eigenvalue of QLLU-R is given by \footnote{This is equivalent to \citet[Eq. (43) on p.918]{Ohtake2023cont}.}
\begin{align}
\Gamma_k &= v\frac{\mu}{\sigma}\left[-\frac{\ol{\phi}+\ol{\lambda}}{\ol{\lambda}}Z_k^2 + \frac{2\sigma-1}{\sigma-1}Z_k\right].
\end{align}

\vspace{10mm}
\noindent
{\large {\bf Statements and Declarations}}

\vspace{3mm}
\noindent
{\small {\bf Competing Interests}:\\
The author has no relevant financial or non-financial interests to disclose.}

\vspace{3mm}
\noindent
{\small{\bf Ethical Approval}: Not applicable.}

\vspace{3mm}
\noindent
{\small{\bf Authors' contributions}: Single author (Kensuke Ohtake).}

\vspace{3mm}
\noindent
{\small{\bf Funding}:\\
The author declares that no funds, grants, or other support were received during the preparation of this manuscript.}

\vspace{3mm}
\noindent
{\small {\bf Availability of data and materials}:\\
All the source code for the numerical computations in this paper is available on \url{https://github.com/k-ohtake/advection-diffusion-neg}.}

\bibliographystyle{econ-aea}

\end{document}